\documentclass[12pt]{article}
\usepackage{latexsym}
\usepackage{epsfig,amssymb,euscript,slashed,euscript}
\usepackage{amsmath}
\usepackage{cite} 
\usepackage{bm,bbm}
\allowdisplaybreaks
\usepackage{array,calc,epsfig}
\usepackage[linktocpage]{hyperref}
\usepackage{pstricks}
\usepackage{bbm}
\usepackage{fancybox}
\usepackage{xcolor}

\def\be{\begin{equation}}
\def\ee{\end{equation}}
\def\bseq{\begin{subequations}}
\def\eseq{\end{subequations}}

\def\bea{\begin{eqnarray}}
\def\eea{\end{eqnarray}}
\newcommand\bbone{\ensuremath{\mathbbm{1}}}

\def\bseq{\begin{subequations}}
\def\eseq{\end{subequations}}

\numberwithin{equation}{section} 
\usepackage[]{graphicx}

\def\d {{\rm d}}

\def\cala         {{\cal A}}

\def\calb         {{\cal B}}
\def\calc         {{\cal C}}
\def\cald         {{\cal D}}

\def\calf         {{\cal F}}
\def\calg         {{\cal G}}

\def\calh         {{\cal H}}

\def\calj         {{\cal J}}
\def\calk         {{\cal K}}
\def\call         {{\cal L}}
\def\calm         {{\cal M}}
\def\caln         {{\cal N}}
\def\calo         {{\cal O}}
\def\calp         {{\cal P}}
\def\calq         {{\cal Q}}
\def\calr         {{\cal R}}
\def\cals         {{\cal S}}
\def\calt         {{\cal T}}
\def\calu         {{\cal U}}

\def\del          {\partial}
\def\delbar       {\bar\partial}
\def\ii           {{\rm i}}

\def\tr           {\mathop{\rm Tr}}
\def\Re           {{\rm Re\hskip0.1em}}
\def\Im           {{\rm Im\hskip0.1em}}

\def\sqr#1#2{{\vcenter{\vbox{\hrule height.#2pt
 \hbox{\vrule width.#2pt height#1pt \kern#1pt \vrule width.#2pt}\hrule
 height.#2pt}}}}
\def\square{%
  \mathop{\mathchoice{\sqr{12}{15}}{\sqr{9}{12}}{\sqr{6.3}{9}}{\sqr{4.5}{9}}}}

\def\d{\text{d}}

\def\slashchar#1{\setbox0=\hbox{$#1$}           
\dimen0=\wd0                                 
\setbox1=\hbox{/} \dimen1=\wd1               
\ifdim\dimen0>\dimen1                        
\rlap{\hbox to \dimen0{\hfil/\hfil}}      
#1                                        
\else                                        
\rlap{\hbox to \dimen1{\hfil$#1$\hfil}}   
/                                         
\fi}

\newcommand{\bC}{\mathbb{C}}

\newcommand{\bP}{\mathbb{P}}

\newcommand{\bR}{\mathbb{R}}
\newcommand{\bZ}{\mathbb{Z}}

\newcommand{\PE}{\mathrm{PE}}

\begin{document}
\font\cmss=cmss10 \font\cmsss=cmss10 at 7pt

\hfill
\begin{center}
{\Large \textbf{Semiclassics of three-dimensional SCFTs \\ \bigskip from holography}}
\end{center}

\vspace{6pt}
\begin{center}
{
Stefano Cremonesi$^1$, Stefano Lanza$^2$ and Luca Martucci$^3$
}

\vspace{1cm}
{\small $^1$ Department of Mathematical Sciences, Durham University, DH1 3LE Durham, UK\vspace{5pt}\\
$2$ 
Institute for Theoretical Physics, Utrecht University\\ Princetonplein 5, 3584 CE Utrecht, The Netherlands\vspace{5pt}\\
$^3$ Dipartimento di Fisica e Astronomia ``Galileo Galilei'', Universit\`a degli Studi di Padova\\
\& I.N.F.N. Sezione di Padova, Via F. Marzolo 8, 35131 Padova, Italy
}
\end{center}


\vspace{12pt}

{\abstract
\noindent
 We use holography to compute the large-$N$ effective field theory  along the moduli space of vacua of an infinite class of three-dimensional  $\caln=2$ SCFTs admitting a dual M-theory description. We focus in particular on toric models and show how the spectrum of large  $R$-charge SCFT chiral scalar operators corresponds to a set of explicit semiclassical solutions of our effective field theory, which describe bound states of backreacting giant gravitons and baryonic-like M5-branes. Our semiclassical description allows for a direct computation of the scaling dimensions of these operators and provides a starting point for a semiclassical investigation of the SCFT data in the large  $R$-charge sector. 
 We consider the models corresponding to the $Y^{12}(\mathbb{P}^2)$ and $Q^{111}$ Sasaki-Einstein spaces as explicit examples. 
 }

\vspace{1cm}


\thispagestyle{empty}

\newpage

\setcounter{footnote}{0}

\tableofcontents

\newpage



%

\section{Introduction and overview}

Supersymmetric quantum field theories typically admit a non-trivial moduli space $\calm$ of inequivalent vacua. One then expects the existence of an effective field theory (EFT) encoding the physical information relevant at low energy. It would be clearly important to have general control over these EFTs, but most of the times perturbative methods work only in a limited region of the moduli space, in which the microscopic theory is weakly coupled. Generically, absent extended supersymmetry or other strongly constraining symmetries, the structure of the EFT on most of the moduli space is thus inaccessible through ordinary quantum field theory techniques.

Superconformal field theories (SCFTs) also typically have a moduli space $\calm$ of inequivalent vacua and an associated EFT. Along the moduli space  the conformal symmetry is spontaneously broken, but it has been recently realized \cite{Hellerman:2017veg,Hellerman:2017sur,Hellerman:2018xpi,Hellerman:2021yqz} -- along the lines of \cite{Hellerman:2015nra,Monin:2016jmo}, see \cite{Gaume:2020bmp} for a review and more references -- that the EFT  can provide important information on more intrinsic CFT data 
beyond the spontaneously broken phase, such as scaling dimensions and correlation functions in the superconformal vacuum. More precisely, the EFT provides a concrete tool for investigating a sector of the CFT in which one or more operators have large ($R$-) charge.

In several known examples including those considered in the present paper, the SCFT is strongly coupled and the identification of the EFT may be problematic. For instance, for an SCFT arising at the IR fixed point of an ordinary UV supersymmetric field theory, the moduli space of the SCFT is obtained by zooming in the moduli space of its parent UV theory in a neighbourhood of its conformal invariant vacuum. In this region the UV theory is often in its maximally strongly coupled regime and no information on the EFT can be computed in perturbation theory at all.  

Holography offers an alternative route to a direct calculation of the EFT of  SCFTs admitting a large-$N$ expansion and  a dual gravitational description. This is the approach adopted in \cite{Martucci:2016hbu}, which considered an infinite class of four-dimensional $\caln=1$  SCFTs  admitting a dual type IIB description, the prototypical example being the Klebanov-Witten (KW) model \cite{Klebanov:1998hh,Klebanov:1999tb}. In this paper we apply the same  strategy to a large family of three-dimensional $\caln=2$ SCFTs engineered in M-theory, which generalise the more supersymmetric ABJM model \cite{Aharony:2008ug}. These SCFTs have a particularly rich dynamics, with peculiar features that are absent in their four-dimensional cousins: the generic presence of flavors \cite{Hohenegger:2009as,Gaiotto:2009tk, Hikida:2009tp,BeniniClossetCremonesi2010,Jafferis:2009th}; the freedom to gauge or not gauge abelian symmetries \cite{Witten:2003ya}; the relevance of monopole operators \cite{Aharony:2008ug,Gaiotto:2009tk,BeniniClossetCremonesi2010,Jafferis:2009th}; the possibility to turn on internal fluxes in the dual gravitational background \cite{Aharony:2008gk,Benishti:2009ky,Benini:2011cma}; the presence of  non-perturbative corrections to the moduli space \cite{Benishti:2010jn}.

These $\caln=2$ SCFTs  are obtained as IR fixed points of (possibly flavored) Yang-Mills/Chern-Simons quiver gauge theories engineered on stacks of M2-branes sitting at the tip of conical Calabi-Yau (CY) four-folds. For concreteness we will mostly restrict to toric models, although several results  extend to more general settings.  Due to the underlying $\caln=2$ superconformal symmetry, the SCFT moduli space $\calm$ must be a conical K\"ahler manifold.  As in the four-dimensional $\caln=1$ models, the complex structure of $\calm$ can be often understood within the field theory description -- see \cite{CremonesiMekareeyaZaffaroni2016} and references therein. In contrast, the K\"ahler potential  $\calk$  on $\calm$, which determines the two-derivative EFT  \cite{deBoer:1997kr,Aharony:1997bx,Intriligator:2013lca}, typically receives strong quantum corrections which are hard to calculate purely in field theory. 

For all these  models the moduli space $\calm$ include a geometric branch which admits a quite universal M-theory description in terms of M2-branes over resolved Calabi-Yau cones. Our main goals  will be: 
\begin{enumerate}
    \item[a)] the identification of a general method to compute the large-$N$  EFT, henceforth dubbed {\em holographic EFT},  at generic points of the  geometric moduli space starting from the holographic M-theory description; 
    
    \item[b)] to initiate a semiclassical investigation of the  large-charge chiral spectrum of this infinite class of  SCFTs starting from their  holographic EFTs, along the lines of   \cite{Hellerman:2017veg,Hellerman:2017sur,Hellerman:2018xpi,Hellerman:2021yqz}  -- see \cite{Loukas:2018zjh,Liu:2020uaz,delaFuente:2020yua} for other applications of the large-charge EFT approach to holographic models;
    
    \item[c)] to illustrate the general results of items a) and b) in the concrete models corresponding to Calabi-Yau cones over the Sasaki-Einstein $7$-folds $Y^{1,2}(\bP^2)$ and $Q^{111}$.
    
\end{enumerate}
Since the paper is quite long and unavoidably technical, we will reserve the rest of this first section to a qualitative description of our approach and of our main results.


\subsection{Holographic EFT: the general idea}\label{sec:HEFT_general_idea}

In this paper we focus on three-dimensional SCFTs whose  superconformal vacuum  is dual to an AdS$_4\times Y$ background in M-theory, where $Y$ is a Sasaki-Einstein seven-dimensional manifold and the AdS$_4$ space supports $N$ units of four-form flux $F_4$. This background can be interpreted as the near-horizon geometry of $N$ M2-branes at the tip of the eight-real-dimensional Calabi-Yau cone $C(Y)$ over $Y$.
 \begin{figure}[t!]
  \centering
    \includegraphics[width=0.7\textwidth]{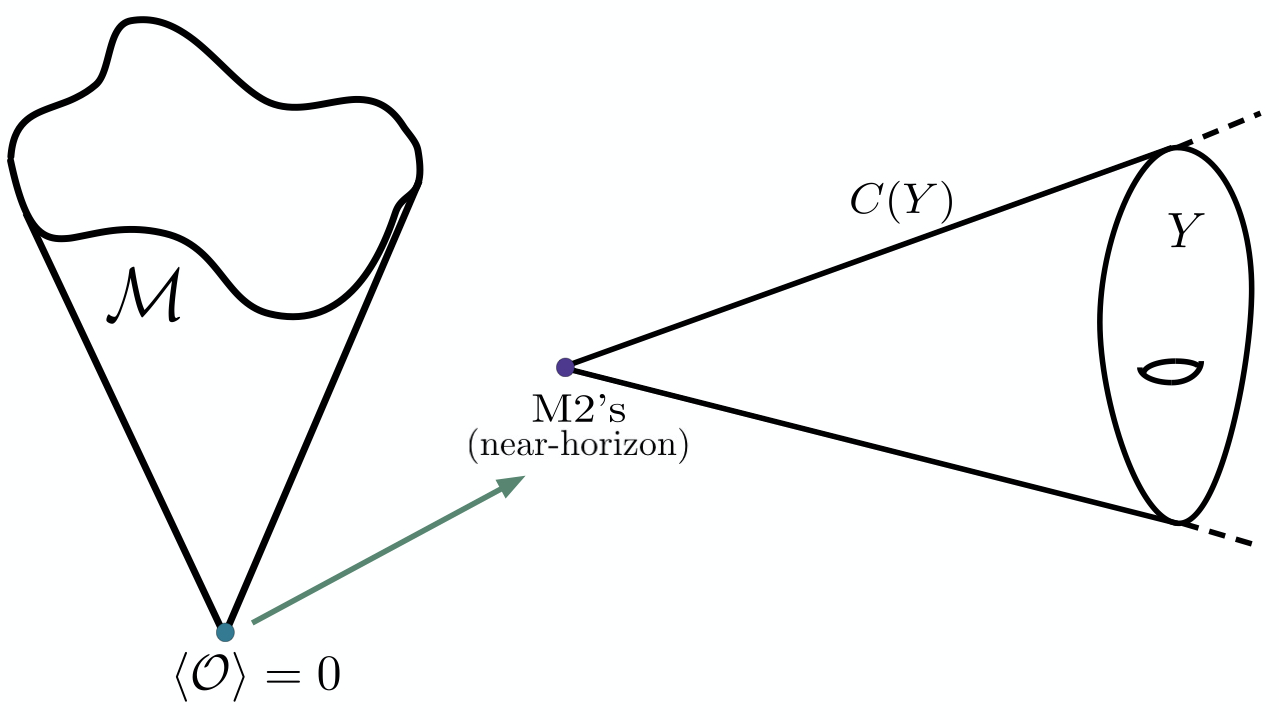}
      \caption{\footnotesize Superconformal vacuum as origin of the moduli space (on the left) and its dual configuration of M2-branes at the tip of a Calabi-Yau cone, leading to an AdS$_4$ geometry in the near-horizon limit. }\label{fig:MS1}
\end{figure}
The superconformal vacuum  corresponds to the `origin' of the moduli space, at which all operators but the identity have vanishing VEV. Because of the superconformal symmetry, the moduli space $\calm$ must be conical  as on the left-hand side of Figure \ref{fig:MS1}, and  the superconformal vacuum sits at the tip of this cone.

\begin{figure}[t!]
  \centering
    \includegraphics[width=0.7\textwidth]{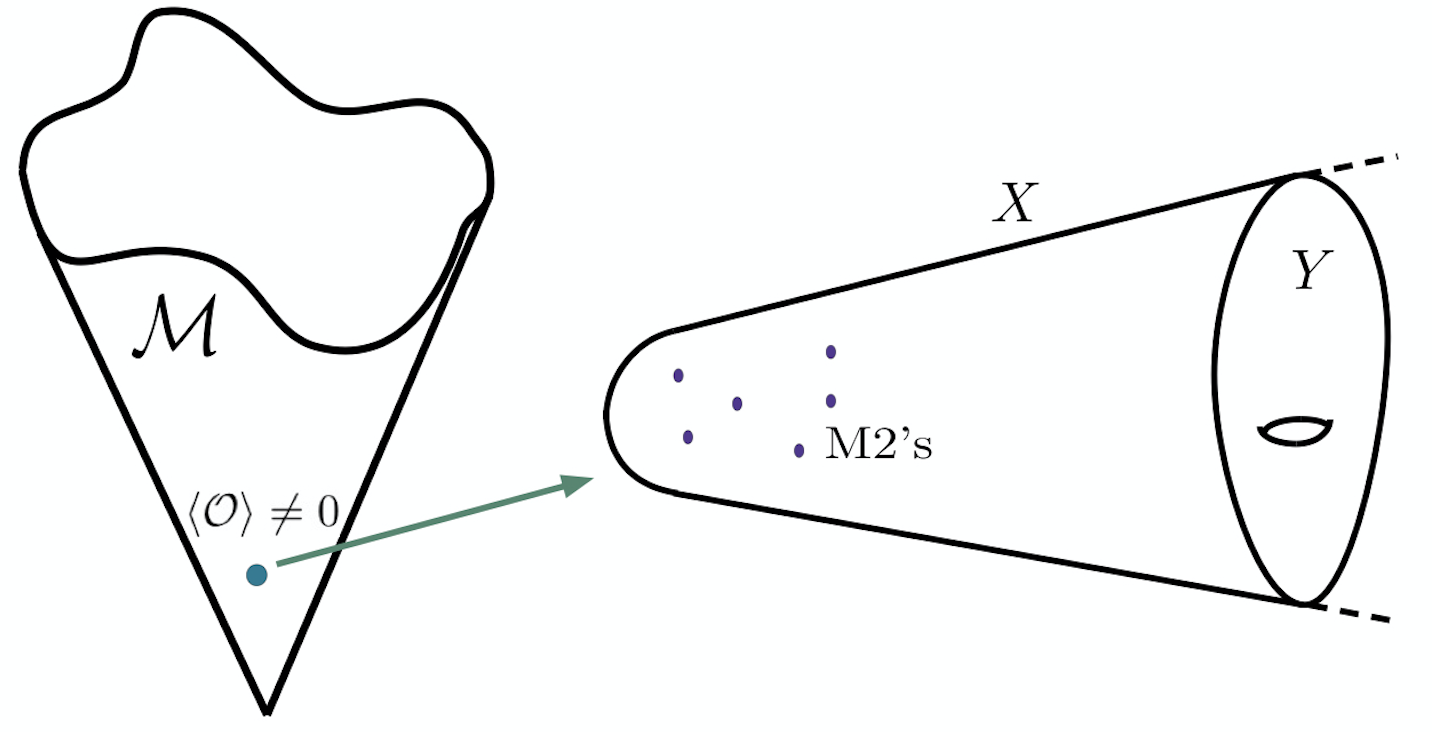}
      \caption{\footnotesize Generic vacuum  on the moduli space (on the left), which spontaneously breaks the conformal symmetry,  and its dual asymptotically AdS geometry (on the right).}\label{fig:MS2}
\end{figure}

Different points of the  moduli space $\calm$ correspond to different vacua of the SCFT, in which some  operators $\calo$ with positive scaling dimension acquire a VEV $\langle\calo\rangle\neq 0$. There the conformal symmetry is spontaneously broken and we expect the existence of a weakly coupled low-energy EFT for the moduli.\footnote{This EFT may break down at special points, or rather rays, of $\calm$, at which the conformal  breaking is partial, in the sense that it triggers an RG flow to another interacting SCFT, weakly coupled to some remaining moduli
 through irrelevant operators.}       
Holographically, moving away from the origin of $\calm$ corresponds to  deforming the AdS M-theory background as in \cite{Klebanov:1999tb} for the KW model (see also \cite{Martelli:2008cm,Benishti:2010jn}), by  modifying the cone $C(Y)$ into a resolved Calabi-Yau space $X$  and distributing the M2-branes along $X$, keeping fixed the AdS asymptotics -- see Figure \ref{fig:MS2}. The {\em geometric} moduli space of the SCFT corresponds to the moduli space of this family of backgrounds and, in general, may constitute only a branch of the entire SCFT moduli space. In this paper we will ignore other possible branches of the moduli space and then $\calm$ will in fact refer only to the geometric one.

The  EFT of the SCFT along $\calm$ can in principle be computed from the M-theory description.
One should impose appropriate boundary conditions on the M-theory fields and expand them along the internal space in modes with different three-dimensional masses. The massless normalizable modes are the dynamical moduli entering the EFT, the details of which can in principle be computed by integrating out the massive modes. Working at two-derivative level in eleven-dimensional supergravity, one can obtain the leading large-$N$ contribution to the two-derivative three-dimensional holographic EFT. Perturbative contributions coming from higher derivative terms, Planckian M-theory states and loop corrections are then expected to be suppressed by powers of $1/N$.  

\begin{figure}[t!]
  \centering
    \includegraphics[width=0.58\textwidth]{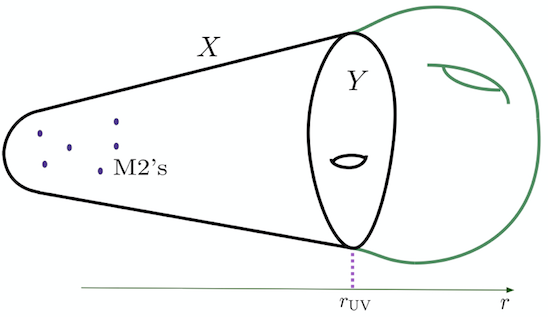}
      \caption{\footnotesize The holographic theory as  local geometry of a warped compactification.}\label{fig:compCY}
\end{figure}

We will however follow an alternative route to determine this holographic EFT: we will obtain it from a rigid limit of  the effective supergravity  for   warped M-theory compactifications to three-dimensions derived in  \cite{Martucci:2016pzt}. Indeed, one may think of the M-theory vacua of Figure \ref{fig:MS2} as local geometries of the warped compactifications introduced in \cite{Becker:1996gj}, see Figure \ref{fig:compCY}. 
We may think of $X$ as cut off at some 
very large value of the radial coordinate $r_{\rm UV}$ and then completed into a compact space.\footnote{We will ignore the tadpole conditions, which relate the number of M2-brane and $G_4$ flux quanta to the geometry of the compact space, since they are irrelevant for us.} 
The M-theory modes that extend to the compact space  constitute some hidden `Planckian' sector which includes gravity and couples to  the dynamical sector of modes localised in the throat, and then to the SCFT,  via $1/M_{\rm P}$ suppressed operators. The decompactification $r_{\rm UV}\rightarrow\infty$ limit  is equivalent to the rigid $M_{\rm P}\rightarrow\infty$ limit  which  decouples the Planckian sector, thus recovering  
the holographic dual to the pure SCFT. 

This is the basic idea behind our derivation of the   holographic EFT, which was already applied in \cite{Martucci:2016hbu} to ${\rm AdS}_5/{\rm CFT}_4$ models. By adapting the results of \cite{Martucci:2016hbu}  to the  ${\rm AdS}_4/{\rm CFT}_3$ context, we will determine the K\"ahler  potential $\calk(\phi,\bar\phi)$ which describes  the large-$N$ dynamics of a set of chiral moduli $\phi_A$ parametrizing the geometric moduli space $\calm$. In the M-theory viewpoint, these include both the M2-brane positions and the bulk moduli.
As we will discuss, in the presence of (perturbative) $U(1)$ symmetries (\emph{e.g.} bulk axionic symmetries or  geometric toric symmetries)  it is often convenient to adopt a dual description of the EFT in which part of the chiral multiplets are traded for vector multiplets. In this paper we restrict to solutions which do not support internal $G_4$ flux, but our methods can be readily extended to flux backgrounds. We leave the investigation of such models to future work.

If the internal space $X$ contains compact six-cycles, that is $b_6(X)\neq 0$, non-perturbative corrections generated by Euclidean M5-branes instantons are expected. 
In particular, as already pointed out in \cite{Benishti:2010jn}, a dynamically generated superpotential in the EFT  may lift (part of) the moduli space.  This aspect will be investigated in a follow-up  \cite{LucaStefano}, while in this paper we will focus on the perturbative contribution and often restrict to the $b_6(X)= 0$ case.

\subsection{Chiral operators as EFT monopoles}

Having derived the holographic EFT from M-theory, we will test it by computing the spectrum of `heavy' chiral operators and the corresponding large  ($R$-) charge and scaling dimensions. The idea of looking at  sectors of operators with large quantum numbers was introduced in the seminal papers \cite{Berenstein:2002jq,Gubser:2002tv}, which have inspired several subsequent developments. Here we will adopt the general philosophy of \cite{Hellerman:2017veg,Hellerman:2017sur,Hellerman:2018xpi}, in which information on the large $R$-charge sector is obtained from BPS semiclassical states of the EFT -- see also  \cite{Hellerman:2015nra,Bourget:2018obm,Beccaria:2018xxl,Grassi:2019txd,Hellerman:2020sqj,Sharon:2020mjs,Cuomo:2022xgw} for other works on  the large charge sectors of supersymmetric models related to the EFT approach. We will show how to do so for general toric models, deriving some universal results. In particular, we will see that the semiclassical states are more naturally described in the dual EFT picture in which all chiral fields are traded for vector multiplets. This dual description is related to the symplectic description of the moduli space, which 
is a toric K\"ahler cone itself (up to a quotient by the symmetric group) \cite{abreu2000kahler}. In the vector multiplet formulation of the EFT, the semiclassical states are BPS monopole solutions which correspond, via  the usual state-operator correspondence, to chiral monopole operators. These EFT operators are in one-to-one correspondence with chiral operators of the SCFT, which can in turn be identified with appropriate chiral operators in the UV quiver gauge theory. In particular, the charges of the EFT monopoles correspond to the toric and Betti charges of the SCFT operators.  
  
We will apply these general results to two concrete models, corresponding to  the Sasaki-Einstein spaces $Y^{12}(\mathbb{P}^2)$ and $Q^{111}$. The cones over $Y^{12}(\mathbb{P}^2)$ and $Q^{111}$ admit  resolutions with explicitly known metrics \cite{Martelli:2007mk,Cvetic:2000db,Cvetic:2001ma,Benishti:2010jn}. This will allow us to explicitly compute the holographic EFT and, from it, the scaling dimensions of the scalar chiral spectrum. These results will then be matched with expectations in the dual SCFTs, which can be identified with IR fixed points of  UV
field theories obtained by applying the S operation of \cite{Kapustin:1999ha,Witten:2003ya} to certain
quiver Chern-Simons theories  \cite{BeniniClossetCremonesi2010,Jafferis:2009th}.

The holographic derivation of these scaling dimensions is not new, of course. Indeed, our heavy semiclassical states  provide a three-dimensional low-energy  description of  bound states of giant gravitons and baryonic branes. However, unlike previous treatments, not only does our EFT approach systematically produce \emph{all} such consistent  brane configurations with the appropriate quantised charges and scaling dimensions, but it also describes their backreaction in a controlled low-energy regime. For instance, we will see how the backreaction of the usual wrapped branes on AdS$_4\times Y$ which are dual to baryonic-like operators does in fact dynamically resolve the underlying cone $C(Y)$. 

We stress that the EFT framework provides a  direct identification between these brane configurations and the dual SCFT operators, which are usually linked via indirect arguments.
 Our results then provide a novel starting point for investigating the ``heavy" sector, possibly carrying Betti charges, of the SCFTs. Indeed, even though in this paper we will restrict to investigating the BPS sector, the EFT provides a natural starting point to study also the non-BPS sector as in \cite{Hellerman:2017veg} and to compute correlation functions \cite{Monin:2016jmo,Hellerman:2018xpi}.

Finally, we will also comment on the presence of massive $\frac12$-BPS charged particles in the holographic EFT, which can provide additional  semiclassical information on the sector of `spinning' charged operators as in \cite{Cuomo:2017vzg,Cuomo:2019ejv}.

We remark that our results can be easily adapted to the AdS$_5$/CFT$_4$ context in type IIB string theory. We will come back to this in future work.

\subsection{Structure of the paper}

The rest of the paper is organised as follows. 
Section \ref{sec:FT} introduces the two field theory models that we will use as our main examples in later chapters: these are the worldvolume theories on M2-branes probing the (resolved) cones over the Sasaki-Einstein $7$-folds $Y^{1,2}(\bP^2)$ and $Q^{111}$. We discuss the field theories associated to different choices of boundary conditions for the Betti multiplets in the bulk, and their spectra of chiral operators. Section \ref{sec:prelrem} discusses the holographic duals of these field theories in M-theory (or rather, 11-dimensional supergravity plus M2-branes). Section \ref{sec:G4=0} describes the holographic EFT of these supergravity backgrounds, with vanishing four-form flux. Section \ref{sec:Soperation} discusses the effect of the $S$ operation of \cite{Kapustin:1999ha,Witten:2003ya} on the holographic EFT. Section \ref{sec:toric} investigates in more detail the holographic EFT of toric models in terms of vector and linear multiplets, making connection with the symplectic formulation of toric varieties. Section \ref{sec:chiralop} studies the effective chiral operators of the holographic EFT, making connection with the holomorphic description of toric varieties. In section \ref{sec:semi} we construct semiclassical magnetically charged solutions of the holographic EFT on $\bR\times S^2$, which are mapped under the state/operator correspondence to 't Hooft monopole operators in the holographic EFT on $\bR^3$.
Section \ref{sec:effM5} discusses the M-theory interpretation of these states/operators, as bound states of AdS$_4$ giant gravitons and baryonic M5-branes. The holographic EFT description fully incorporates the backreaction of these branes and realizes charge quantization directly, with no need to geometrically quantize the classical configuration space of these branes. We also briefly discuss charged BPS particles, which are realised as M2-branes wrapping effective curves, from the viewpoint of the holographic EFT. Finally, in sections \ref{sec:Y12model} and \ref{sec:Q111model} we apply the general theory developed in previous sections to the resolved cones over $Y^{1,2}(\bP^2)$ and $Q^{111}$ respectively, and match the chiral operators of the holographic EFT with chiral operators of the microscopic SCFT on the worldvolume of M2-branes probing the geometry.
We include a number of appendices covering our conventions, a brief review of the conical K\"ahler structure of the moduli space of vacua of three-dimensional $\caln=2$ SCFTs, and details of geometric and field theory calculations for the $Y^{1,2}(\bP^2)$ and $Q^{111}$ models.


\section{A field theory appetizer}
\label{sec:FT}

We start by introducing two three-dimensional quiver Chern-Simons theories which flow to IR SCFTs with a dual eleven-dimensional supergravity description in the large-$N$ limit. These theories have a geometric branch of the moduli space and a corresponding EFT which we will explicitly derive  in sections \ref{sec:Y12model} and \ref{sec:Q111model} by applying the general results of  sections \ref{sec:prelrem}--\ref{sec:effM5}. The purpose of this section is to provide a concrete idea of the kind of SCFTs that can be  studied holographically by means of the general results derived  in sections \ref{sec:prelrem}--\ref{sec:effM5}.        

 We will start by reviewing the parent $U(N)\times U(N)$ flavoured ABJM quivers introduced in \cite{BeniniClossetCremonesi2010,Jafferis:2009th}. We will actually be interested in a variation of these models obtained by applying the S operation \cite{Kapustin:1999ha,Witten:2003ya} to combinations of topological and flavor $U(1)$ symmetries. As we will discuss in the following sections, this choice allows for additional  directions in the moduli space and corresponds, on the dual M-theory side,  to the choice of specific boundary conditions for the corresponding supergravity fields \cite{Klebanov:1999tb,Witten:2003ya,Benishti:2010jn}.

\subsection{The alternate \texorpdfstring{$Y^{1,2}(\bP^2)$}{Y12(P2)} quiver}
\label{sec:Y12quiver}

\begin{figure}[t]
\centering
	\includegraphics[scale=0.6]{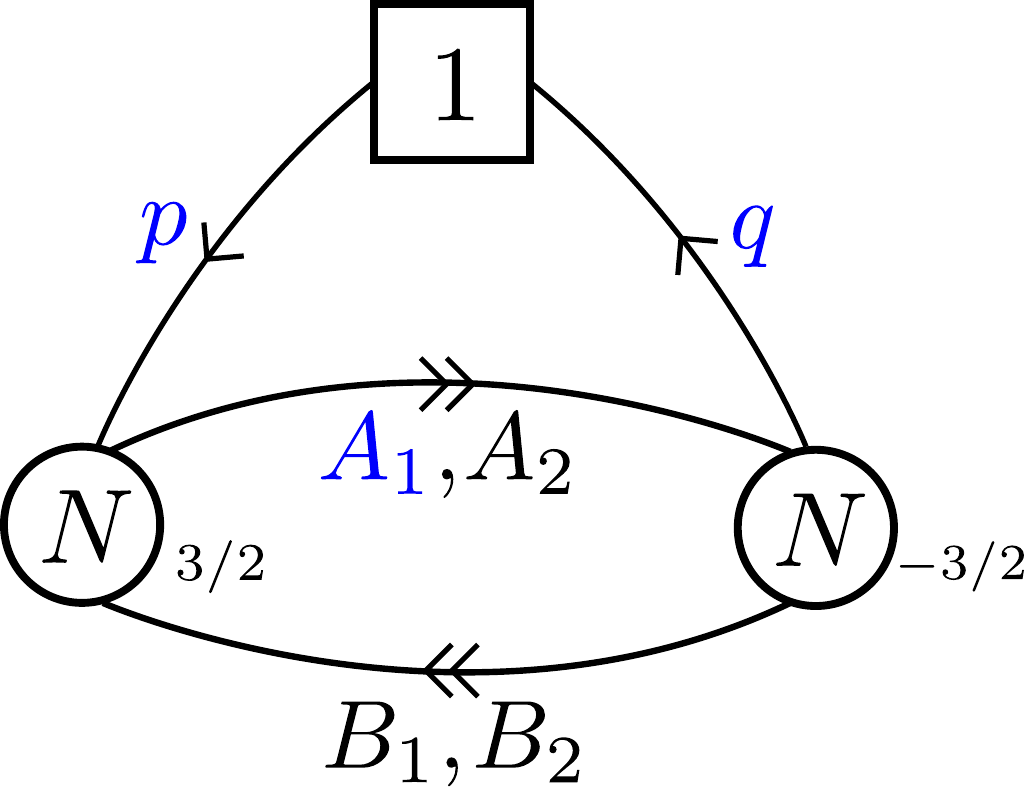}
\caption{Quiver diagram for the worldvolume theory of $N$ M2-branes probing the cone over $Y^{1,2}(\bP^2)$. Coloured fields are present in an extra superpotential term, here $\Delta W = p A_1 q$, in addition to the standard ABJM superpotential.}
	\label{fig:quiverY12}
\end{figure}

Let us start with the worldvolume gauge theory on $N$ regular M2-branes probing the cone over the Sasaki-Einstein 7-fold $Y^{1,2}(\bP^2)$ \cite{Gauntlett:2004hh,Martelli:2008rt} that was derived in \cite{BeniniClossetCremonesi2010} by reducing the system to type IIA string theory.
\footnote{See \cite{Cheon:2011th,Cheon:2011vi,JafferisKlebanovPufuEtAl2011} for exact tests of this duality beyond the moduli space of vacua. An alternative UV gauge theory was proposed in \cite{Benini:2011cma} based on a different type IIA reduction. The two UV gauge theories are expected to be IR dual.}
It is a three-dimensional $\caln=2$ flavoured version of the ABJM theory \cite{Aharony:2008ug}: a $U(N)_{3/2}\times U(N)_{-3/2}$ Chern-Simons matter theory, with vector multiplets $V_1$ and $V_2$, bifundamental matter fields $A_{1,2}\in (N,\overline{N})$ and $B_{1,2}\in (\overline{N},N)$, and fundamental flavours $p \in (\overline{N},1)$ and $q \in (1,N)$. See Fig.~\ref{fig:quiverY12} for the associated quiver diagram. \footnote{One can add a fractional M2-brane to the previous configuration: this has the effect of changing the gauge group of the quiver to $U(N+1)_{3/2}\times U(N)_{-3/2}$ and of turning on a nontrivial torsion $4$-form flux in the gravity dual \cite{Cremonesi2011}. We will not consider this modification further in this paper.}
The matter fields interact through a superpotential
\be\label{W_Y12}
W=\tr(A_1B_1A_2B_2-A_1B_2A_2B_1)+ p A_1 q\,.
\ee
In order to make the theory weakly coupled in the UV, we can add Yang-Mills kinetic terms $-\frac{1}{4g^2_i}\int\d^2\theta\,\tr W_iW_i$ for the two $U(N)$ vector multiplets $V_1$ and $V_2$, with dimensionful Yang-Mills couplings $g_i$. 

In the following we will make use of the topological conserved current superfield $\calj_{\rm T}$ associated with the `baryonic' $U(1)_{\rm B}$ gauge symmetry 
\be\label{Y12top}
\calj_{\rm T}\equiv \frac1{2\pi}\Sigma_{\rm B}\equiv \frac{\ii}{8\pi}D\bar D V_{\rm B} \quad~~~~~\text{with}\quad V_{\rm B}\equiv \tr V_1-\tr V_2\,.
\ee  
 Here and in the following we  mostly adopt the conventions of \cite{Intriligator:2013lca}, up to some minor differences -- see appendix \ref{app:3dsuperspace} for more details.

The normalization of $V_{\rm B}$ has been chosen so that the corresponding field-strength $F_{\rm B}$ obeys the quantization condition $\int F_{\rm B}\in 2\pi\bZ$.

The model that we will be interested in is obtained  from the above model by ``ungauging'' (or freezing) the baryonic $U(1)_{\rm B}$ through a so-called S operation \cite{Kapustin:1999ha,Witten:2003ya}. The S operation is performed by first coupling the topological symmetry to a background $U(1)$ vector multiplet 
$\cala$ through the supersymmetric BF-term  
\be\label{Bc}
\int \d^4\theta \calj_{\rm T}\cala=
\frac{1}{2\pi}\int \d^4\theta \Sigma_{\rm B} \cala=\frac{1}{2\pi}\int \d^4\theta \,V_{\rm B}\Sigma\,,
\ee
where $\Sigma\equiv \frac\ii4 D\bar D\cala$ is the linear multiplet associated to the new vector multiplet $\cala$, and then making $\cala$ dynamical.
The overall effect of this S operation is to introduce a dynamical FI parameter for the vector multiplet $V_{\rm B}$. 

Since $\cala$ appears only linearly, it may be integrated out exactly. This would impose the constraint $\calj_{\rm T}=0$, which forces $V_{\rm B}$ to be flat, leaving a \emph{global} $U(1)_{\rm B}$ symmetry which is embedded in the original gauge group as follows:
\be\label{baryonicsymm}
e^{\ii\theta_{\rm B}}\in U(1)_{\rm B}\quad \mapsto\quad (e^{\frac{\ii}{2N}\theta_{\rm B}}\bbone_N,e^{-\frac{\ii}{2N}\theta_{\rm B}}\bbone_N)\in U(N)\times U(N)\,.
\ee
Let us denote by $\calj_{\rm B}$ the corresponding conserved current multiplet. 

In fact, the S operation offers a natural alternative description of these quivers, which will turn out to be more useful in the following: instead of integrating out $\cala$ from the outset, we
keep it dynamical. Then the UV gauge symmetry $U(N)\times U(N)$ is preserved and is extended by an additional $U(1)_\cala$ factor. In this alternative but equivalent formulation, the Gauss law for the flat vector multiplet $V_{\rm B}$ implies that the conjugate conserved baryonic current $\calj_{\rm B}$  is redundant and is dynamically identified with the topologically conserved current associated to the $\cala$ vector multiplet:
\be\label{Y12Tcur}
\calj_{\rm B} = \frac{1}{2\pi}\Sigma\,.
\ee 
We will denote by $U(1)_{\rm T}$ the  topological symmetry associated to the $U(1)_\cala$ gauge symmetry, which has  conserved current $\frac{1}{2\pi}\Sigma$.

It is useful to provide yet another description of the microscopic theory, which is obtained by dualizing the linear multiplet $\Sigma$ to a chiral multiplet $\rho$. For clarity, let us add
an irrelevant kinetic term 
\be\label{kintilde}
-\frac1{e^2}\int \d^4\theta\,\Sigma^2
\ee
and eventually consider  the low-energy limit $e\rightarrow\infty$. The dualization is then performed as usual \cite{Lindstrom:1983rt,deBoer:1997kr,Intriligator:2013lca}. First replace the sum of \eqref{Bc} and \eqref{kintilde} with 
\be\label{VBrho}
-\frac1{e^2}\int \d^4\theta\,\Sigma^2-\int\d^4\theta \Big(\chi+\overline\chi-\frac1{2\pi}V_{\rm B}\Big)\Sigma\,,
\ee
 where $\Sigma$ is now an unconstrained real superfield and the chiral superfield $\chi$ is a  Lagrange multiplier, whose equation of motion enforces that $\Sigma$ is a linear superfield. To preserve gauge invariance under $V_{\rm B}\rightarrow V_{\rm B}-\ii(\Lambda_{\rm B}-\overline\Lambda_{\rm B})$, we must accordingly shift
 \be\label{rhoBgauge}
\chi\rightarrow  \chi-\frac{\ii}{2\pi}\Lambda_{\rm B}\, .
 \ee
Then $e^{-2\pi \chi}$ is a single-valued chiral superfield with (gauged) baryonic charge $1$.\footnote{Note that $\chi$ has periodicity $\Im\chi\simeq \Im\chi+1$ by construction.} 
Integrating out $\chi$ from \eqref{VBrho} gives back the original theory in terms of the vector multiplet $\cala$. Instead, integrating out $\Sigma$ imposes the identification 
\be\label{SVR}
\frac{2}{e^2}\Sigma =\frac1{2\pi}V_{\rm B}-(\chi+\bar\chi) 
\ee
and substituting in \eqref{VBrho} gives 
\be
\frac{e^2}2\int\d^4\theta \Big(\chi+\overline\chi-\frac{1}{2\pi}V_{\rm B}\Big)^2\,.
\ee
In this formulation the theory has an additional rigid symmetry, which shifts the imaginary part of $\chi$ by a constant and leaves all the other fields unchanged. This can be identified with the above topological $U(1)_{\rm T}$, under which $e^{-2\pi\chi}$ has charge $1$. From \eqref{SVR} it is clear that in the low-energy limit $e\rightarrow \infty$ one must impose  $V_{\rm B}=2\pi(\chi+\overline\chi)$, so that $V_{\rm B}$ is flat, the baryonic gauge symmetry is ungauged and can be identified with $U(1)_{\rm T}$. We anticipate that, in the dual M-theory description, the baryonic symmetry of the $Y^{12}(\bP^2)$ will be identified with a Betti symmetry.

\subsubsection{Moduli space}\label{subsec:moduli_Y12}

The quiver gauge theory introduced in the previous section flows to a strongly interacting $\caln=2$ SCFT with a non-trivial moduli space of vacua. This contains a `geometric' branch, whose structure can be understood from a semiclassical analysis of the above UV quiver theory  as long as the mass scale set by the VEV is much larger than the running  $g_i^2$ at that scale. (This condition is not needed if one focuses on holomorphic data, which are insensitive to the gauge coupling, as we will do in section \ref{subsec:chiral_Y12}.)
The point of keeping $\Sigma$  in the UV quiver introduced above is that it naturally enters the description of the moduli space. One can  follow the semiclassical analysis of \cite{Benini:2011cma} almost verbatim, with the only difference that the {\em bare} FI parameters $\xi_1=-\xi_2$ therein are replaced by the dynamical scalar field $\sigma\equiv \Sigma|_{\theta=\bar\theta=0}$, up to some numerical constant. The result is as follows.  

Along the geometric moduli space, the $U(N)\times U(N)$ quiver gauge symmetry is generically broken to the maximal torus $U(1)^N$ of the diagonal $U(N)$ subgroup. Let us denote by $s_{I}$ ($I=1,\ldots,N$)  the $N$ scalars of the  low-energy $U(1)$ vector  multiplets $V_I$. Furthermore, $p=q=0$ and the bifundamental matter solving the F-flatness conditions can be written in the form
\be\label{ABvevs}
\langle A_{i}\rangle=\left(\begin{array}{cccc} 
a_{i1} &0 &\ldots & 0\\
0& a_{i2}  &\ldots & 0\\
0 & 0 & \ddots & \vdots\\
0 &0 &\ldots & a_{iN}
\end{array}\right)\,,\quad \langle B_{i}\rangle=\left(\begin{array}{cccc} 
b_{i1} &0 &\ldots & 0\\
0& b_{i2}  &\ldots & 0\\
0 & 0 & \ddots & \vdots\\
0 &0 &\ldots & b_{iN}
\end{array}\right)
\ee
  One still needs to impose the D-flatness conditions, which must take into account  the one-loop corrections to the effective CS levels, as in \cite{BeniniClossetCremonesi2010,Jafferis:2009th,Benini:2011cma}. As a result, one gets  
\be\label{conifoldfib}
|a_{1I}|^2+|a_{2I}|^2-|b_{1I}|^2+|b_{2I}|^2=\xi_{\rm eff}(s_I,\sigma)
\ee
with
\be
\xi_{\rm eff}(s_I,\sigma)=-\sigma +\frac{3}{2}s_{I}-\frac12 |s_{I}|
\ee

 Additional directions in the moduli space are obtained by dualizing the low-energy $U(1)^N$ photons into axions. Supersymmetrically, this  corresponds to a dualization of the  $N$ low-energy $U(1)$ vector multiplets $V_I$ into  chiral multiplets $t_I$, completely analogous to the dualization of the vector multiplet $\cala$ into the chiral multiplet $\chi$ discussed in subsection \ref{sec:Y12quiver}. In particular, the
$U(1)^N$ photons are dual to $N$ axions $\varphi_I=\Im t_I$ -- in this paper we sloppily use the same symbol for a chiral superfield and its lowest component -- each parametrizing a circle $U(1)_{\text{\tiny M}}$.  There is a residual $S_N$ gauge symmetry which permutes the $N$ sets of fields $(a_{iI},b_{iI},t_I)$.
One then obtains  a double fibration. Namely, we have $N$ symmetrized copies of the resolved conifold times $U(1)_{\text{\tiny M}}$ fibered over $\mathbb{R}$, which are in turn fibered over $\mathbb{R}$ parametrized by $\sigma$. By trading $\sigma\equiv \Sigma|_{\theta=\bar\theta=0}$ for the dual $\Re\chi$ and adding the dual axion $\Im\chi$, one gets the complete description of the semiclassical moduli space in terms of chiral coordinates.

Following \cite{Aganagic:2009zk,BeniniClossetCremonesi2010,Jafferis:2009th,Benini:2011cma},  each of these $N$ copies can be identified with  the Calabi-Yau four-fold $X$ which is obtained by resolving the cone over $Y^{12}(\mathbb{P}^2)$. This can be described in terms of a $U(1)$ gauged linear sigma model with five complex homogeneous coordinates $Z_A$, $A=1,\ldots , 5$ (so that $d=5$ in the formulas of section \ref{sec:toric}), with charges \be\label{Y12glsm0}
\begin{array}{c|ccccc|r} 
 & Z_1 & Z_2 & Z_3  & Z_4 & Z_5  &{\rm FI} \\
\hline
U(1)& 1 & 1 & 1 & -2 &  -1   & \sigma\\ 
\hline
U(1)_{\text{\tiny M}} &0&0&1& -1 & 0& s_I
\end{array}
\ee
In the second line we have also indicated the choice of $U(1)_{\text{\tiny M}}$ circle action along which one needs to reduce in order to go back to the conifold \eqref{conifoldfib}, as in \cite{Aganagic:2009zk}.
In this description $\sigma$ represents the resolution parameter of the fourfold, which spans
the union of two K\"ahler cones, according to its sign. This is precisely the description that one gets starting from the dual M-theory model, which will be discussed in more detail in section \ref{sec:Y12model}.  

We emphasize that the above description is valid for vacua which are far away from the origin of the quiver moduli space and at which the quiver does not flow to a SCFT. It is then expected to capture only part of the information on the SCFT moduli space, which includes its holomorphic description. In particular, it tells us that the EFT should contain  $4N$ chiral multiplets $z^i_I$, whose lowest components parametrize the position in the above Calabi-Yau $X$, plus the linear multiplet $\Sigma$, or alternatively a dual (gauge invariant) chiral multiplet $\chi$. This motivates the above UV formulation in terms of $\Sigma$. On the other hand, even in the large-$N$ limit, we expect strong quantum corrections to the D-term sector of the EFT, which may be obtained by integrating out the massive fields in the UV quiver. We will derive this EFT (at the two-derivative level) from the dual M-theory description in section \ref{sec:Y12model}, following the general procedure discussed in sections \ref{sec:prelrem}-\ref{sec:effM5}.

\subsubsection{Chiral operators} \label{subsec:chiral_Y12}

Let us define as `mesonic' the chiral operators which are neutral under the $U(1)_{\rm T}$ symmetry. 
In the formulation with the dynamical $\cala$ vector multiplet, these operators do not carry any $U(1)_\cala$ monopole charge. The VEVs of these operators parametrize the geometric moduli space introduced in the above subsection. These operators were studied in \cite{BeniniClossetCremonesi2010,Jafferis:2009th} in the abelian case, and in  \cite{CremonesiMekareeyaZaffaroni2016} in the non-abelian case. A subset of chiral mesonic operators can be constructed by taking gauge invariant combinations of products of basic bulding blocks $A_iB_j$. In addition there are dressed monopole operators which are invariant under the $U(N)\times U(N)$ gauge group \cite{CremonesiMekareeyaZaffaroni2016,BeniniClossetCremonesi2010}. To construct a gauge invariant non-abelian dressed monopole operator, one starts from a \emph{bare} monopole operator with equal magnetic charges ${\mathfrak q}=(q_1,\ldots,q_N; q_1,\ldots,q_N)$ under the two gauge groups. The magnetic charge ${\mathfrak q}$ defines an embedding of $U(1)$ inside the $U(N)\times U(N)$ gauge group, and a supersymmetric bare monopole operator is introduced by prescribing a Dirac monopole singularity for the vector multiplet of the embedded $U(1)$ in the path integral. The monopole boundary condition induces an adjoint Higgsing of the gauge group to $U(1)^N \times U(1)^N$ for generic charges, or to $\prod_\alpha U(r_\alpha)\times \prod_\alpha U(r_\alpha)$ if the magnetic charges $q_I$ are equal in blocks of size $r_\alpha$, leading to a gauge symmetry enhancement. Due to the Higgsing, all the fundamental and antifundamental matter, as well as some of the bifundamentals, gain an effective mass, so that the residual gauge theory of light modes in the monopole background is a tensor product of decoupled $U(r_\alpha)\times U(r_\alpha)$ quivers. Integrating out the heavy modes leads to a quantum correction to the Chern-Simons levels, which in turn contributes to the electric charge of the bare monopole operators. 

To construct chiral gauge invariant operators of the residual gauge group $\prod_\alpha U(r_\alpha)\times \prod_\alpha U(r_\alpha)$, the bare monopole operator must be dressed with some of the massless bifundamentals.\footnote{If the magnetic charges are different (modulo permutations) for the two $U(N)$ gauge factors, the bare monopole operator cannot be dressed into a gauge invariant chiral operator.} Finally, one averages over the action of the Weyl group $S_N \times S_N$ to form a fully gauge invariant operator.

The geometric moduli space of the $U(N)\times U(N)$ theory on the worldvolume of $N$ M2-branes is the $N$-th symmetric product of the geometric moduli space of the $U(1)\times U(1)$ theory on a single M2-brane, as was shown using Hilbert series in \cite{CremonesiMekareeyaZaffaroni2016}. We can therefore set $N=1$ first, and rely on the results of \cite{BeniniClossetCremonesi2010}. We will use lower case letters to denote bifundamental fields and monopole operators in the abelian theory. The basic bare monopole operators $t$ and $\tilde t$ with magnetic charges ${\mathfrak q}=(1;1)$ and $(-1;-1)$ respectively have electric charges $(-1;1)$ and $(2;-2)$ under the $U(1)\times U(1)$ gauge group, and obey the quantum relation 
\be
t\tilde t= a_1~.
\ee 
Since $a_1$ is generated by the bare monopole operators, one can use $t,\tilde t, a_2, b_1, b_2$ to construct independent gauge invariant operators.  
The connection with the toric description \eqref{Y12glsm0} of the corresponding Calabi-Yau four-fold is obtained by identifying  
\be\label{quivertoric0}
Z_1\equiv b_{1}\quad,\quad Z_2\equiv b_{2}\quad,\quad Z_3\equiv t\quad,\quad Z_4\equiv \tilde t\quad,\quad Z_5\equiv a_2\,.
\ee

In the non-abelian theory, a convenient basis that generates the whole set of gauge invariant monopole operators is obtained starting from the monopole operators $T^{(r)}$ and $\tilde T^{(r)}$ of magnetic charges ${\mathfrak q}=\pm 1_{r}\equiv((\pm 1)^r,0^{N-r}; (\pm 1)^r,0^{N-r})$ for $r=1,\dots,N$.
\footnote{An equivalently good basis uses monopole operators of magnetic charges ${\mathfrak q}=(\pm r,0^{N-1}; \pm r,0^{N-1})$ with $r=1,\dots,N$, corresponding to the operators $(T^{(1)})^r$ and $(\tilde T^{(1)})^r$ in the residual gauge theory with gauge group $(U(1)\times U(N-1))^2$. In the parlance of symmetric polynomials, this choice corresponds to power sums, while the choice made in the text corresponds to `elementary' symmetric polynomials. Hilbert series calculations show that dressed power sum monopole operators and dressed elementary symmetric polynomial monopole operators with the same quantum numbers differ by products of dressed monopole operators of lower $R$-charge. This extends to any choice of basis of symmetric polynomials. All statements about dressed monopole operators in the main text are to be considered modulo mixing with products of lower dressed monopole operators with the same quantum numbers.
\label{sym_polys}}
These operators transform in the representations  $(\det\overline\square,1;\det\square,1)$ and $((\det\square)^2,1;(\det\overline\square)^2,1)$ under the residual gauge group $(U(r)\times U(N-r))\times (U(r)\times U(N-r))$, and satisfy the quantum relation 
\be 
T^{(r)} \tilde T^{(r)} = \det\nolimits_r A_1^{(r)}
\ee
in the residual gauge theory, where $A_1^{(r)}$ denotes the block of $A_1$ in the bifundamental representation of $U(r)\times U(r)$ and $\det\nolimits_r$ the determinant of an $r\times r$ matrix (we use $\det=\det\nolimits_N$ unless explicitly stated). Since $T^{(r)}$ is charged under the gauged baryonic symmetry of $U(r)\times U(r)$, in order to build a chiral gauge invariant operator under the residual gauge group it must multiply an operator of opposite baryonic charge in the $U(r)\times U(r)$ residual gauge theory, such as $\det\nolimits_r (A)^{(r)}$, $\det\nolimits_r (ABA)^{(r)}$, $\det\nolimits_r (A)^{(r)} \cdot \tr{(AB)^{(r)}}$ \emph{etc.},  with arbitrary $SU(2)$ indices. 
Finally, this operator must be symmetrized with respect to the broken Weyl group $(S_N/(S_r \times S_{N-r}))^2$ to form a fully gauge invariant operator. 

It was shown in appendix A of \cite{CremonesiMekareeyaZaffaroni2016} at the level of Hilbert series that the geometric moduli space of the theory on $N$ M2-branes is the $N$-th symmetric product of the theory on a single M2-brane.
\footnote{This assumes a conjecture of \cite{Forcella:2007wk} that baryonic generating functions of theories on multiple D3-branes are obtained by symmetrizing the baryonic generating functions of theories on a single D3-brane.}
The proof was detailed for ABJM \cite{Aharony:2008ug}, but it is a  straightforward exercise to generalise it to all flavoured ABJM theories. This result is expected from brane considerations, but is not obvious from purely field-theoretic  considerations and has not been worked out explicitly beyond counting techniques. In particular, it implies that the gauge invariant chiral operators of the non-abelian $U(N)\times U(N)$ gauge theory are in one-to-one correspondence with the gauge invariant chiral operators of $N$ copies of the gauged linear sigma model \eqref{Y12glsm0} of the Calabi-Yau fourfold, supplemented by a discrete $S_N$ gauge symmetry that permutes the $N$ copies. We will use this crucial result in section \ref{sec:Y12model} to match chiral operators in the holographic EFT to chiral operators in the microscopic quiver gauge theory that have been discussed in this section.

We can now consider monopole operators for the vector multiplet $\cala$. Let us first consider bare monopole operators $\calt_n$ of magnetic charge $n$ under the $U(1)_\cala$ gauge symmetry, which have charge $n$ under $U(1)_{\rm T}$. Since no matter is charged under $U(1)_\cala$, the classical relation $\calt_n=\calt^n$ holds for all $n \in \bZ$, and the operator $\calt\equiv \calt_1$ is invertible.
In the dual description in terms of the chiral field $\chi$, we can identify $\calt$ with $e^{-2\pi \chi}$ and $\calt_n$ with $e^{-2\pi n\chi}$.
As is clear from \eqref{rhoBgauge}, the bare monopole operators $\calt_n$ have charge $n$ under the baryonic gauge  symmetry $U(1)_{\rm B}$. They must then be `dressed' by other operators that have a net charge $-n$ under $U(1)_{\rm B}$. The simplest possibilities are
\be\label{UVbaryons}
  \begin{aligned}
 \calb_{{\mathfrak m}_3,{\mathfrak m}_4}&= \calt~  [T^{(N-\mathfrak r)} (\det\nolimits_{\mathfrak r} B^{(\mathfrak r)})_{{\mathfrak m}_3,{\mathfrak m}_4}]|_{\rm{sym}}
 \quad~~~~~{(\mathfrak{m}_3,{\mathfrak m}_4 \ge 0,~ {\mathfrak r}= {\mathfrak m}_3+{\mathfrak m}_4\leq N)}\\
 \tilde\calb_{1}&=\calt^{-1}~\det  A_2 \\
 \tilde\calb_{2}&= \calt^{-2}~\tilde T^{(N)} 
  \end{aligned}
  \ee
 where we used the shorthand $(\det B)_{a_1,a_2}$ for the dibaryon built out of $a_1$ factors of $B_1$ and $a_2$ factors of $B_2$, which has isospin $I_3 = (a_1-a_2)/2$ under the $SU(2)$ global symmetry that acts on the $B_a$ bifundamentals. (We identified $a_1\equiv \mathfrak{m}_3$ and $a_2\equiv \mathfrak{m}_4$ for later purposes.) The bare monopole operator $T^{(N-\mathfrak r)}$ in the first line breaks the gauge group to $(U(N-\mathfrak r) \times U(\mathfrak{r}))^2$, with the monopole charged under the centre of the $U(N-\mathfrak r)$ factors and the dibaryon charged under the centre of the $U(\mathfrak r)$ factors, and the subscript sym denotes the symmetrization under the (broken) Weyl group.

We note that as ${\mathfrak m}_3$ and ${\mathfrak m}_4$ are varied at fixed $N$, the dibaryonic operators in the first line of \eqref{UVbaryons} span the totally symmetric rank $N$ irreducible representation $[N,0]$ of an enhanced $SU(3)$ symmetry, which contains the manifest $SU(2)$ symmetry acting on the $B$ bifundamentals as well as a linear combination of the other $U(1)$ symmetries. In the large $N$ limit, one can use $F$-extremization \cite{JafferisKlebanovPufuEtAl2011} or equivalently volume minimization \cite{MartelliSparksYau2006} to find the scaling dimensions of these operators (see also appendix A of \cite{Cremonesi2011}): \be\label{dimensions_Y12_FT}
\Delta(\calb_{{\mathfrak m}_3,{\mathfrak m}_4}) \simeq 0.4505 N~, \qquad \Delta(\tilde \calb_1)\simeq 0.4136 N~, \qquad \Delta(\tilde \calb_2)\simeq 0.2349 N~.  
\ee

We can be more explicit about the operators if we assume that the magnetic charges ${\mathfrak q}=(q_1,\ldots,q_N; q_1,\ldots,q_N)$ for $U(N)\times U(N)$ are generic and break it to its maximal torus.
\footnote{See also appendix \ref{appsub:counting_Y12} for a more detailed count of chiral operators according to their charges under the toric symmetries.} This is the holomorphic counterpart of the semiclassical discussion of the geometric moduli space in section \ref{subsec:moduli_Y12}.
The massless modes in such a generic monopole background are described by $N$ copies of an effective abelian ABJM quiver with gauge group $U(1)\times U(1)$ and effective Chern-Simons levels $k^+=(1;-1)$ and $k^-=(2;-2)$. 
We label by $t_I$ and $\tilde t_I$ the monopole operators of magnetic charge $(\pm 1;\pm 1)$ of the $I$-th abelian ABJM quiver in the residual gauge theory, which have electric charges $- k^+$ and  $- k^-$ respectively, and by $a_{1,I}$, $a_{2,I}$, $b_{1,I}$, $b_{2,I}$ the bifundamentals in the same abelian quiver. There is a quantum relation $t_I \tilde t_I = a_{1,I}$ for all $I$, which can be used to eliminate products of $t_I$ and $\tilde t_I$ in favour of $a_{1,I}$ and vice versa.
\footnote{We have assumed that non-abelian diagonal monopole operators factorize into abelian diagonal monopole operators. This is not a general property of monopole operators, but is consistent with the quantum numbers of \emph{diagonal} monopole operators in flavoured ABJM theories.} 

Then we can write the general chiral operators as
\be\label{chiralopsFT_Y12}
\begin{split}
\calo_{{\bf \mathfrak q}, n;\,\bf\alpha, \beta}&= \calt^n \left(\prod_{I}\left[(t_I)^{\frac{q_I}{2}+\frac{|q_I|}{2}} (\tilde t_I)^{-\frac{q_I}{2}+\frac{|q_I|}{2}} (a_{1,I})^{\alpha^I_1} (a_{2,I})^{\alpha^I_2} (b_{1,I})^{\beta^I_1} (b_{2,I})^{\beta^I_2}\right]\right)_{\rm sym}~\\
&= \calt^n \left(\prod_{I}\left[(t_I)^{\frac{q_I}{2}+\frac{|q_I|}{2}+\alpha_1^I} (\tilde t_I)^{-\frac{q_I}{2}+\frac{|q_I|}{2}+\alpha_1^I} (a_{2,I})^{\alpha^I_2} (b_{1,I})^{\beta^I_1} (b_{2,I})^{\beta^I_2}\right]\right)_{\rm sym}~,
\end{split}
\ee 
where the powers $\alpha_i^I$ and $\beta_i^I$ of the bifundamentals are non-negative integers which satisfy the $U(1)\times U(1)$ gauge invariance constraint
\be\label{gauge_inv_Y12b}
\alpha^I_1+\alpha^I_2 - \beta^I_1 - \beta^I_2 = \xi_{\rm eff}(q_I,n) \equiv -n + \frac{3}{2}q_I - \frac{1}{2}|q_I|~
\ee
for each $I$. The subscript $\rm sym$ denotes symmetrization over the Weyl group. The operators \eqref{chiralopsFT_Y12} are `mesonic' if $n=0$ and `baryonic' if $n\neq 0$. 
 
The above discussion should be adjusted if some $q_I$ are equal, leading to a non-abelian factor in the residual gauge group. Then a product of bifundamentals charged under the Cartan subgroup of each non-abelian factor should be replaced by an appropriate dibaryonic operator for the same non-abelian factor, as explained in \cite{CremonesiMekareeyaZaffaroni2016} and briefly reviewed in appendix \ref{appsub:counting_Y12}. For instance, $t_1 a_{1,1} t_2 a_{1,2}$ should be replaced by $T^{(2)} \det_2 A_1^{(2)} = T_{(1^2,0^{N-2};1^2,0^{N-2})}  \left((A_1)_{11} (A_1)_{22}- (A_1)_{12}(A_1)_{21}\right)$ before averaging over the Weyl group.

It is expected that the baryonic operators \eqref{UVbaryons} and \eqref{chiralopsFT_Y12}  are realized holographically by M5-brane configurations. This identification arises  quite naturally from the holographic EFT that will be derived in the following sections. We will reproduce the microscopic results \eqref{dimensions_Y12_FT} using the holographic EFT and provide more details about the matching of these operators in section \ref{sec:Y12model}.

\subsection{The alternate \texorpdfstring{$Q^{111}$}{Q111} quiver}
\label{sec:Q111quiver}

\begin{figure}[t]
\centering
	\includegraphics[scale=0.6]{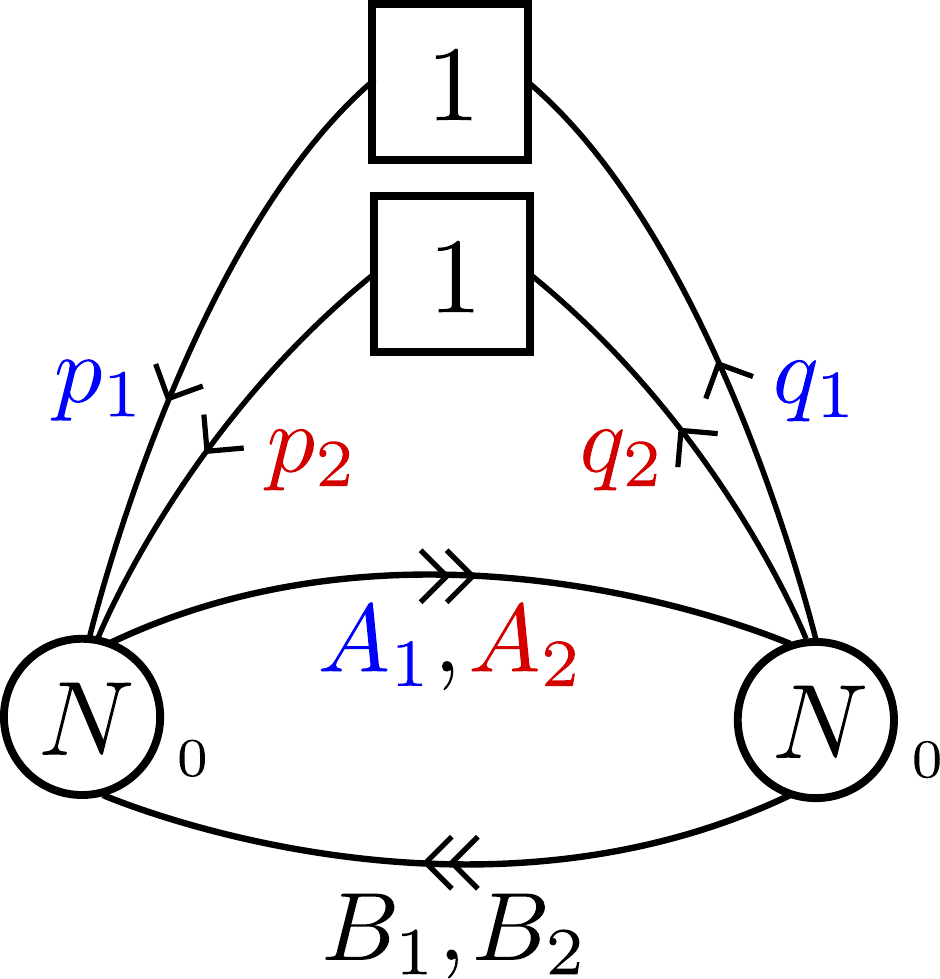}
\caption{Quiver diagram for the worldvolume theory of $N$ M2-branes probing the cone over $Q^{1,1,1}$. Coloured fields are present in extra superpotential terms, here $\Delta W = p_1 A_1 q_1 + p_2 A_2 q_2$.}
	\label{fig:quiverQ111}
\end{figure}

The $Q^{111}$ quiver is partly similar to the  $Y^{12}(\bP^2)$ one. Hence, in the following discussion we will be faster, omitting the details that can be easily adapted from  the previous sections. As for the $Y^{12}(\bP^2)$ quiver, we start with an associated quiver gauge theory, whose quiver diagram is depicted in Fig. \ref{fig:quiverQ111},
 identified in \cite{Jafferis:2009th,BeniniClossetCremonesi2010}. It is again a flavored version of the ABJM theory, now with vanishing bare Chern-Simons levels, which includes two pairs of flavor chiral fields $p_{1,2}, q_{1,2}$ and a superpotential  
 \be
W=\tr(A_1B_1A_2B_2-A_1B_2A_1B_2)+p_1A_1q_1+p_2A_2q_2\ .
\ee 

The theory we are interested in is again obtained by applying the S operation. Not only we apply it to the topological symmetry associated with the $U(1)_{\rm B}$ baryonic gauge symmetry \eqref{baryonicsymm}, as for the $Y^{12}(\bP^2)$ model, by  adding a coupling of the form \eqref{Bc} to a dynamical vector multiplet $\cala^1$. We also perform an S operation for a flavor symmetry $U(1)_{\rm F}$. There are different equivalent choices of $U(1)_{\rm F}$, related by different combinations with the diagonal $U(1)$ subgroup of $U(N)\times U(N)$. We choose  $U(1)_{\rm F}$ so that $p_2,q_2$ are neutral under it, while $p_1$ and $q_1$ have charges $+1$ and $-1$ respectively. The second S operation is obtained by modifying the UV Lagrangian by a minimal coupling of the flavors to a second dynamical vector multiplet $\cala^2$, which then gauges $U(1)_{\rm F}$.  The resulting quiver theory  has an extended gauge group $U(N)\times U(N)\times U(1)\times U(1)$. 

Because of parity anomalies, in addition to \eqref{Bc} one also needs  to add half-integral mixed Chern-Simons coupling to preserve gauge invariance (see also  appendix F of \cite{CremonesiMekareeyaZaffaroni2016}):
\be
\frac{1}{4\pi}\int\d^4\theta\, (2\cala^1-\cala^2)\,\Sigma_{\rm B}
\ee
One can correspondingly define two topologically conserved current multiplets, 
\be\label{J12Q111}
\calj^1=\frac1{2\pi}\Sigma^1\quad,\quad \calj^2=\frac1{2\pi}\Sigma^2\,,
\ee
with $\Sigma^a=\frac\ii4 D\bar D \cala^a$, $a=1,2$. We will denote by $U(1)^a_{\rm T}$, $a=1,2$,  the corresponding topological global symmetries.

As for the $Y^{12}(\bP^2)$ model, we may now dualize $\cala^1$ to a chiral multiplet $\chi_1$, such that $e^{-2\pi\chi_1}$ has charge $+1$ under the global $U(1)_{\rm T}^1$ and  the baryonic $U(1)_{\rm B}$ gauge symmetry.  On the other hand, the dualization of $\cala^2$ to a chiral field $\chi_2$ will be possible only in the low-energy EFT, since in the UV quiver gauge theory $\cala^2$ couples to charged matter.

\subsubsection{Moduli space}

The semiclassical moduli space proceeds as for the $Y^{12}(\bP^2)$ model and is completely analogous. It can again be described in terms of $N$  copies of a fibered conifold \eqref{conifoldfib}, where now
\be 
\xi_{\rm eff}(s_I,\sigma)=-\sigma^1+\frac12\sigma^2-\frac12|s_I|-\frac12|s_I-\sigma^2|
\ee
where $\sigma^a$, $a=1,2$, are the lowest components of the linear multiplets $\Sigma^a=\frac{\ii}{4}D\bar D \cala^a$. After dualizing the low-energy $U(1)^N$ vector multiplets to chiral multiplets $t_I$, one can describe the complete geometric moduli space as a fibration over $\mathbb{R}^2$ parametrized by $(\sigma^1,\sigma^2)$. The fiber is given by $N$ symmetrized copies of the Calabi-Yau four-fold $X$ which is obtained by resolving the cone over $Q^{111}$, with $(\sigma^1,\sigma^2)\in \mathbb{R}^2$ parametrizing the K\"ahler moduli. 

 $X$ admits a GLSM/toric description 
\be\label{Q111_GLSM0}
\begin{array}{ c | c c c c c c | c }
	& Z_1 & Z_2 & Z_3 & Z_4 & Z_5 & Z_6 & \mathrm{FI}\\ \hline
	U(1)_1 & 1 & 1 & 0 & 0 & -1 & -1 & \sigma^1\\
	U(1)_2 & 0 & 0 & 1 & 1 & -1 & -1 & \sigma^2\\
	\hline
	U(1)_{\text{\tiny M}} & 0 & 0 & 0 & -1 & 1 & 0 & s_I-\sigma^2
\end{array}
\ee
where the second line describes the $U(1)_{\text{\tiny M}}$ and the corresponding FI that specifies the connection with the above description in terms of the fibered conifold \eqref{conifoldfib}.

By introducing complex coordinates $z^i$ on $X$, we then expect the SCFT to admit a low-energy EFT for two vector multiplets $\cala^a=(\cala^1,\cala^2)$ and $N$ neutral symmetrized chiral multiplets $z^i_I$. The linear multiplets $\Sigma^a$ can be identified with current multiplets of topological symmetries. Motivated by the M-theory perspective that we will be discussed in the following sections, we will refer to them as {\em Betti} symmetries.  At the EFT level, one may dualize the vector multiplets \eqref{J12Q111} to chiral multiplets $\rho_a$, such that $e^{-2\pi\rho_a}$ carry  charge $+1$ under the $a$-th Betti symmetry.  
The dual M-theory viewpoint  will allow us to  derive the large-$N$ holographic EFT of the SCFT and to more concretely describe such dualization.

\subsubsection{Chiral operators}\label{subsec:chiral_Q111}

The construction of gauge invariant chiral operators in the $Q^{111}$ model, before and after the double S operation, follows the same logic as for the $Y^{1,2}(\bP^2)$ model. 

Let us first focus on the `mesonic' operators, which do not carry any $U(1)_{\cala^a}$ magnetic charges and whose VEVs parametrize the geometric moduli space discussed in the previous subsection, with $\sigma^a=0$. These are ordinary mesons constructed out of the $A, B$ bifundamentals, and dressed monopole operators with equal magnetic charges ${\mathfrak q}=(q_1,\ldots,q_N; q_1,\ldots,q_N)$ under the two $U(N)$ gauge groups. Since the geometric moduli space of the $U(N)\times U(N)$ theory on the worldvolume of $N$ M2-branes is the $N$-th symmetric product of the geometric moduli space of the $U(1)\times U(1)$ theory on a single M2-brane, we will again focus on $N=1$ first  \cite{BeniniClossetCremonesi2010}. As above, we use lower case letters to denote bifundamental fields and monopole operators in this abelian theory. The basic bare monopole operators $t$ and $\tilde t$ with magnetic charges ${\mathfrak q}=(1;1)$ and $(-1;-1)$ respectively both have electric charges $(1;-1)$ under the $U(1)\times U(1)$ gauge group of the quiver, and obey the quantum relation 
\be
t\tilde t= a_1 a_2~.
\ee 
It is useful to trivialise the quantum relation by introducing four independent variables $r, \tilde r$ and $s, \tilde s$ with charges $+1$ and $-1$ under a new $U(1)$ gauge symmetry, such that
\begin{equation}\label{rs}
    t=rs~, \quad \tilde t = \tilde r \tilde s~, \quad a_1 = \tilde rs~, \quad a_2= \tilde s r~.
\end{equation}
This leads to the following identification with the homogeneous coordinates of the toric description \eqref{Q111_GLSM0}, see appendix \ref{appsub:counting_Q111} and \cite{BeniniClossetCremonesi2010} for more details:
\be\label{quivertoricQ111}
Z_1\equiv b_{1}\quad,\quad Z_2\equiv b_{2}\quad,\quad Z_3\equiv r\quad,\quad Z_4\equiv \tilde r\quad,\quad Z_5\equiv s\quad,\quad Z_6\equiv \tilde s\,.
\ee
The introduction of the new variables $r, \tilde r, s, \tilde s$ may seem formal at first sight, but it turns out that they are needed to describe the resolution of $C(Q^{111})$ and to express the monopole operators of the theory obtained after the double S operation. In that case, introducing $\bC^*$ valued GLSM variables $X_a$ of charge $-\delta_a^b$ under the $b$-th $U(1)$ in \eqref{Q111_GLSM0}, one finds the identification 
\be
X_2 r = t_{1;0,1}~, \quad X_2 \tilde r = t_{0;0,1}~, \quad X_{2}^{-1}s = t_{0;0,-1}~,\quad X_{2}^{-1}\tilde s = t_{-1;0,-1}~, \quad X_1 = t_{0;1,0}~,
\ee
where $t_{m;n^1,n^2}$ is the abelian monopole operator of magnetic charge $q=(m;m)$ under the $U(1)^2$ gauge group of the quiver and $(n^1,n^2)$ under the extra $U(1)^2$ gauge group introduced in the double S operation. See appendix \ref{appsub:counting_Q111} for details of how this identification comes about. One can then express a general abelian bare monopole operator as 
\be\label{abel_monopole_Q111}
t_{m;n^1,n^2} = X_1^{n^1} X_2^{n^2} r^{\frac{m}{2}+\frac{|m|}{2}} \tilde r^{-\frac{m-n^2}{2}+\frac{|m-n^2|}{2}} 
 s^{\frac{m-n^2}{2}+\frac{|m-n^2|}{2}}
\tilde s^{-\frac{m}{2}+\frac{|m|}{2}} ~.
\ee

In the $U(N)\times U(N)$ theory, we may again use the bare monopole operators of magnetic charges ${\mathfrak q}=\pm 1_{r} \equiv((\pm 1)^r,0^{N-r}; (\pm 1)^r,0^{N-r})$ for $r=1,\dots,N$, and dress them by bifundamentals of the residual gauge theory to generate all chiral `mesonic' operators. Let us not  dwell on this, and move on instead to discussing  monopole operators for the vector multiplets $\cala^1, \cala^2$, which contain the previous operators as a subcase. We first note that no matter is charged under $U(1)_{\cala^1}$, therefore the corresponding monopole operators factor out and obey the classical relation that a monopole operator of charge $n^1$ is the $n^1$-th power of a monopole operator of charge $1$.
\be\label{monopole_A1_factor}
T_{{\mathfrak q};n^1,n^2}= T_{{\mathfrak q};n^1,0} T_{{\mathfrak q};0,n^2} = (\calt_1)^{n^1} T_{{\mathfrak q};0,n^2}~,
\ee
where $\calt_1\equiv T_{{\mathbf 0};1,0}$ is invertible. In the dual description in terms of the chiral field $\chi$, we can identify $\calt_1$ with $e^{-2\pi \chi_1}$. This factorization does not hold for monopole operators for the vector multiplet $\cala^2$, which has charged matter.

It is not hard to construct  three basic sets of gauge invariant chiral `baryonic' operators, whose elements are labelled by an integer ${\mathfrak m}=0,1,\ldots, N$, with magnetic charges $(n^1,n^2)$ equal to $(1,0)$, $(0,1)$ and $(-1,-1)$ respectively:\footnote{We will see in a later section that these choices of $(n^1,n^2)$ correspond to the generators of the walls between the three K\"ahler chambers of $Q^{111}$.}  
\be\label{UVbaryonsQ111}
  \begin{aligned}
 \calb^{(1)}_{\mathfrak m}&= T_{0;1,0} ({\det B})_{N-{\mathfrak m},{\mathfrak m}}\equiv  \calt_{1} (\det B)_{N-{\mathfrak m},{\mathfrak m}}
 \quad~~~~~&&(0 \le {\mathfrak m}\leq N)\\
 \calb^{(2)}_{{\mathfrak m}}&= T_{+1_{N-\mathfrak m};0,1}  \quad~~~~~&&(0\le {\mathfrak m}\leq N)\\
 \calb^{(3)}_{{\mathfrak m}}&=T_{-1_{N-\mathfrak m};-1,-1}  \equiv \calt_{1}^{-1} T_{-1_{N-\mathfrak m};0,-1}  \quad~~~~~&&(0\le {\mathfrak m}\leq N)~.
  \end{aligned}
  \ee
Each of these sets of operators has the lowest dimension for baryonic operators, 
\begin{equation}
    \Delta(\calb^{(i)}_{{\mathfrak m}}) = \frac{N}{3}~,
\end{equation}
and transforms in the $(N+1)$-dimensional irreducible representation of the $i$-th $SU(2)$ global symmetry of the SCFT, corresponding to the $SU(2)^3$ isometries of $Q^{111}$.

More general baryonic operators of higher dimension can be constructed similarly. Again, we can be more explicit if we assume that the magnetic charges ${\mathfrak q}=(q_1,\ldots,q_N; q_1,\ldots,q_N)$ for $U(N)\times U(N)$ are generic and break it to its maximal torus. The massless modes in this monopole background are described by $N$ copies of an effective abelian ABJM quiver with gauge group $U(1)\times U(1)$ and effective Chern-Simons levels $k^+=k^-=(-1;1)$. 
We label by $t_I$ and $\tilde t_I$ the monopole operators of magnetic charge $(\pm 1;\pm 1)$ of the $I$-th abelian ABJM quiver in the residual gauge theory, which have electric charges $- k^+$ and  $- k^-$ respectively, and by $a_{1,I}$, $a_{2,I}$, $b_{1,I}$, $b_{2,I}$ the bifundamentals in the same abelian quiver. The quantum relations $t_I \tilde t_I = a_{1,I}a_{2,I}$ for all $I$
can be solved by introducing new variables $r_I, \tilde r_I$, $s_I, \tilde s_I$ as in \eqref{rs}.  Using \eqref{abel_monopole_Q111} for each $I$, we can then write the general chiral operators as \be\label{chiralopsFT_Q111}
\begin{split}
\calo_{{\bf \mathfrak q}, n;\,\bf\alpha, \beta}&= T_{\mathfrak{q};n^1,n^2}   \left(\prod_{I} (a_{1,I})^{\alpha^I_1} (a_{2,I})^{\alpha^I_2} (b_{1,I})^{\beta^I_1} (b_{2,I})^{\beta^I_2}\right)_{\rm sym}~\\
&= \prod_{a=1}^2 X_a^{n^a} \bigg(\prod_{I}\bigg[(r_I)^{\frac{q_I}{2}+\frac{|q_I|}{2}+\alpha_2^I} 
(\tilde r_I)^{-\frac{q_I-n^2}{2}+\frac{|q_I-n^2|}{2}+\alpha_1^I}
(s_I)^{\frac{q_I-n^2}{2}+\frac{|q_I-n^2|}{2}+\alpha_1^I}\\
& \qquad \qquad \qquad \times
(\tilde s_I)^{-\frac{q_I}{2}+\frac{|q_I|}{2}+\alpha_2^I} 
(b_{1,I})^{\beta^I_1} (b_{2,I})^{\beta^I_2}\bigg]\bigg)_{\rm sym}~,
\end{split}
\ee 
where the powers $\alpha_i^I$ and $\beta_i^I$ of the bifundamentals are non-negative integers which satisfy the $U(1)\times U(1)$ gauge invariance constraint
\be\label{gauge_inv_Y12}
\alpha^I_1+\alpha^I_2 - \beta^I_1 - \beta^I_2 = \xi_{\rm eff}(q_I,n) \equiv -n^1+ \frac{n^2}{2} -\frac{1}{2}|q_I| - \frac{1}{2}|q_I-n^2|~
\ee
for each $I$. Again, the subscript $\rm sym$ denotes symmetrization over the Weyl group. The operators \eqref{chiralopsFT_Y12} are `mesonic' if $(n^1,n^2)=0$ and `baryonic' if $(n^1,n^2)\neq 0$. If some $q_I$ are equal, the above discussion should be corrected along the same lines as in section \ref{subsec:chiral_Y12}.

We will reproduce the operators \eqref{UVbaryonsQ111} and \eqref{chiralopsFT_Q111} and their quantum numbers from the perspective of the holographic EFT in section \ref{sec:Q111model}.


\section{Holographic M-theory backgrounds}
\label{sec:prelrem}

We now turn to the dual supergravity point of view. As explained in the introduction, our aim is to derive the holographic (i.e.\ large-$N$) EFT and match it with field theory expectations. Even though the models dual to the SCFTs discussed in section \ref{sec:FT} will serve as specific examples, the discussion will have various degrees of generality. We will start by considering a wide class of   M-theory models dual to $\caln=2$  three-dimensional SCFTs which share the same topological properties of toric models. We will then specifically  restrict  to toric models, describing in more detail the structure of their holographic EFT and of the associated semiclassical chiral ring. The models introduced in section \ref{sec:FT} belong to this class and will later be used to concretely apply and check our general results.

The qualitative features of M-theory backgrounds  corresponding to the SCFT vacua along the geometric moduli space have been already described in the introduction, see Figure \ref{fig:MS2}.  In this section we describe in more detail their precise structure (see also \cite{Klebanov:1999tb,Martelli:2008cm,Benishti:2010jn} for previous work) and some of their geometric properties which will be useful in the following sections.   

\subsection{The supergravity solutions}

The M-theory vacua we will focus on have the general structure \cite{Becker:1996gj} 
\be\label{Mback}
\begin{aligned}
\d s^2_{11}&=H^{-\frac23}\d s^2_{\mathbb{R}^{1,2}}+\ell^2_{\text{\tiny P}}\, H^{\frac13}\d s^2_X\ ,\\
F_4&=\d\text{vol}_{\mathbb{R}^{1,2}}\wedge \d H^{-1} ~.
\end{aligned}
\ee
where $\ell_{\text{\tiny P}}$ is the 11-dimensional Planck length and in our conventions the internal line element $\d s^2_X$ is dimensionless.\footnote{In our conventions the M2-brane has tension/charge $2\pi/\ell^3_{\text{\tiny P}}$.}
The AdS vacuum of Figure \ref{fig:MS1} corresponds to the choice $X=C(Y)$,  $\d s^2_{C(Y)}=\d r^2+r^2\d s^2_Y$ and $H_{\rm AdS}=\frac{L^6}{r^6}$, so that $\d s^2_{11,{\rm AdS}}=\d s^2_{\rm AdS_4}+\ell^2_{\text{\tiny P}}\,L^2\d s^2_Y$, where  $L^6=\frac{N}{6\text{vol}(Y)}$  and the $\rm AdS_4$ metric is
\be\label{AdSmetric}
\d s^2_{\rm AdS_4}=\frac{r^4}{L^4}\d s^2_{\mathbb{R}^{1,2}}+\frac{\ell^2_{\text{\tiny P}}L^2}{r^2}\d r^2 ~.
\ee
 For the more general vacua of Figure \ref{fig:MS2}, the internal Calabi-Yau metric  $\d s^2_X$ is a resolution of the conical metric 
 and the warp factor $H$ is the unique solution of 
\be\label{warpeq}
\Delta_X H=*_X\sum^N_{I=1}\delta^8_I
\ee
that vanishes asymptotically for $r\rightarrow \infty$.  In (\ref{warpeq}), $\delta^8_I$ is an 8-form with delta function support at the position of the $I$-th M2-brane. The equation (\ref{warpeq}) is then solved by
\be\label{warp}
H(y)=\sum_I G_X(y;y_I)~,
\ee
where $G_X(y;y')=G_X(y';y)$ is the Green's function of the Laplace operator obtained from the CY metric on $X$, in some real coordinates $y^m$, $m=1,\ldots,6$. 
Note that $G_X(y;y')$ is the unique asymptotically vanishing Green's function, so that \eqref{warp} is the unique solution with asymptotically AdS behaviour: 
\be\label{WasyAdS}
H\simeq \frac{L^6}{r^6}+\ldots\quad~~~~~~~\text{for $r\rightarrow\infty$}~.
\ee  
See \cite{Martelli:2007mk} for a useful summary on the properties of  Green's functions on asymptotically conical  CY spaces. In the following we will just use the existence and uniqueness of these Green's functions, with no need to know their explicit form.

The moduli parametrizing these vacua are given by $4N_{\rm M2}$ complex dimensionless coordinates $z^i_I$, $I=1,\ldots,N_{\rm M2}$, which give the positions of the $N_{\rm M2}=N$ M2-branes  on $X$ in a complex coordinate system $z^i$, 
and $n_{\rm K}=b_2(X)$ K\"ahler moduli $\sigma^a$, $a=1,\ldots, n_{\rm K}$, which measure the resolution of the ambient Calabi-Yau fourfold. The latter combine with the $C_6$ axionic moduli into $b_2(X)$ complex
moduli $\rho_a$, to be defined later on. By supersymmetry, all the complex moduli are bottom components of chiral multiplets.  
Part of the moduli $\rho_a$ may actually be considered non-dynamical: this depends on some freedom in the choice of boundary conditions for the supergravity fields \cite{Breitenlohner:1982jf}, which correspond to different SCFTs \cite{Klebanov:1999tb}. We will initially consider all $\rho_a$ as dynamical complex coordinates on the moduli space $\calm$. We will return to the other possibilities in the following sections.

\subsection{Some geometric generalities}
\label{sec:geom}

In order to derive the holographic EFT we first need to collect some geometric properties of the internal  Calabi-Yau four-fold $X$, following the three-fold case discussed in \cite{Martucci:2016hbu}. 
Recall that we assume that $X$ is a crepant resolution of the Calabi-Yau cone $C(Y)$. It can be proven that in each cohomology class  of  $H^2(X;\mathbb{R})\simeq H^{1,1}(X;\mathbb{R})$ belonging to the K\"ahler cone there exists a unique K\"ahler form $J_X$ defining a Calabi-Yau metric which asymptotically reduces to the conical one \cite{10.2969/jmsj/06431005}.  In  $H^2(X;\mathbb{R})$  one can then expand
\be\label{cohoJexp}
[J_X]=\frac{\ell_{\text{\tiny P}}}{2\pi}\,\sigma^a[\omega_a]~,
\ee
where $\omega_a$, $a=1,\ldots, b_2(X)$, form a basis of harmonic  $(1,1)$-forms,\footnote{\label{foot:prim} As in \cite{Benishti:2010jn,Martucci:2016hbu}, one can argue that the harmonic (1,1)-forms $\omega_a$ are primitive: $J\lrcorner\omega_a\equiv \frac12J^{mn}(\omega_a)_{mn} =0$. On the other hand, a primitive $(1,1)$ closed two-form is harmonic, since it is automatically co-closed.}
and $\sigma^a$ are the real K\"ahler moduli of $X$. The $\frac{1}{2\pi}\,\ell_{\text{\tiny P}}$ factor in (\ref{cohoJexp}) is introduced for later convenience and implies that  $\sigma^a$ have dimension of a mass. 

We choose to normalize $\omega_a$ so that they represent a basis of  the integral cohomology group $H^2(X;\mathbb{Z})$. This is Poincar\'e dual to the relative homology group $H_6(X,Y;\mathbb{Z})$, the homology group of six-chains with boundary on $Y$. 
For definiteness, we require that $X$  shares the same topological properties of toric Calabi-Yau four-folds: $X$ has vanishing odd Betti numbers $b_{2k+1}(X)=0$, $k=0,\ldots,3$, no complex structure moduli, and
\be
b_2(X)=b_6(X)+b_2(Y)~,
\ee 
see for instance \cite{Benishti:2010jn}. 
Correspondingly, we can split the basis of harmonic (1,1)-forms as  
\be\label{omegasplit}
\omega_a=(\hat\omega_\alpha,\tilde\omega_\sigma)~,
\ee
with $\alpha,\beta,\ldots=1,\ldots,b_6(X)$ and $\sigma,\tau,\ldots=1,\ldots,b_2(Y)$. The $\hat\omega_\alpha$'s define a basis of the relative cohomology group $H^2(X,Y;\mathbb{Z})$: these two-forms are  Poincar\'e dual to  compact divisors $\hat D_\alpha$ which provide a basis of $H_6(X;\mathbb{Z})$. On the other hand  $\tilde\omega_\sigma$ are Poincar\'e dual to non-compact divisors $\tilde D_\sigma$, which have a non-trivial boundary on $Y$. For instance,  $b_6(X)=0$ in both the $Y^{12}(\mathbb{P}^2)$ and $Q^{111}$ models, while  they have $b_2(Y)=1$ and $b_2(Y)=2$ respectively -- see sections \ref{sec:Y12model} and \ref{sec:Q111model} below.

The harmonic (1,1)-forms $\hat\omega_\alpha$ are $L_2$-normalizable \cite{Hausel:2002xg}:
\be\label{L2norm}
\int_X \hat\omega_\alpha\wedge *_X \hat\omega_\beta<\infty~.
\ee
On the other hand,  the harmonic (1,1)-forms $\tilde\omega_\sigma$ are only $L^{\rm w}_2$-normalizable, meaning that they are not $L_2$-normalizable but are normalizable using the  warped measure: $\int_X H\,\tilde\omega_\sigma\wedge *_X \tilde\omega_\rho<\infty$. This property crucially derives from the asymptotic AdS behaviour (\ref{WasyAdS}) of the warping -- see for instance \cite{Martelli:2008cm} for a discussion on the analogous Calabi-Yau three-fold case. The singularities of $H$, which locally diverge like the sixth inverse power of the distance from the localised M2-brane, are integrable and do not create problems. Clearly, $L_2$-normalizable forms are also $L^{\rm w}_2$-normalizable, therefore in general 
\be\label{L2wnorm}
\int_X H\,\omega_a\wedge *_X \omega_b<\infty
\ee
for any pair of harmonic (1,1) forms $\omega_a,\omega_b$. We will see below how this warped integral appears as the kinetic matrix of the holographic EFT.

The cohomological expansion \eqref{cohoJexp} translates into the expansion of $J_X$ as a two-form,
\be\label{genJexp}
J_X=J_0+\frac{\ell_{\text{\tiny P}}}{2\pi}\sigma^a\omega_a~,
\ee
 where $J_0$ is an exact form. By the $\del\delbar$-lemma, it can be globally written as 
\be
J_0=\frac{\ell_{\text{\tiny P}}}{2\pi}\,\ii\del\delbar k_0~,
\ee
where $k_0(z,\bar z;\sigma)$ is a globally defined function,  which also depends on the K\"ahler moduli $\sigma^a$. At $\sigma^a=0$ the internal space $X$ reduces to the Calabi-Yau cone and $\frac{1}{2\pi}\,\ell_{\text{\tiny P}}\,k_0$ to its K\"ahler potential, which is $\frac12 r^2$ up to K\"ahler transformations. 

In the following we will also need locally defined `potentials' $\kappa_a(z,\bar z;\sigma)$ such that
\be\label{defkappa}
\omega_a=\ii\del\delbar\kappa_a~.
\ee
The potentials $\kappa_a(z,\bar z;\sigma)$ define metrics $e^{-2\pi\kappa_a}$ on the line bundles $\calo(D_a)$ associated with the divisors $D_a=(\hat D_\alpha,\tilde D_\sigma)$ introduced above. The transition functions of these line bundles translate into $\sigma^a$-independent hol+$\overline{\text{hol}}$ contributions to  $\kappa_a$, which clearly preserve (\ref{defkappa}).

To remove some arbitrariness in $\kappa_a(z,\bar z;\sigma)$ and  $k_0(z,\bar z;\sigma)$, we impose the conditions 
\begin{subequations}\label{bcond}
\begin{align}
&\frac{\del k_0}{\del \sigma^a}=-\sigma^b\frac{\del\kappa_b}{\del \sigma^a}\label{k0der}\\
&\frac{\del\kappa_a}{\del \sigma^b}\rightarrow 0 \quad~~~~\text{as $r\rightarrow\infty$}\label{kappaboundary}\\
& k_0(z,\bar z;\sigma=0)=k_{C(Y)}(z,\bar z)=\frac{\pi}{\ell_{\text{\tiny P}}} r^2~.\label{k0v0}
\end{align}
\end{subequations}
Consistently with (\ref{k0der}), we can also introduce the combination
\be\label{kdec}
k_X=k_0+\sigma^a\kappa_a\,,\quad\quad \kappa_a=\frac{\del k_X}{\del \sigma^a}
\ee
which can be identified with a rescaled K\"ahler potential $k_X$ on $X$, normalized such that 
\be\label{JXk}
J_X=\ii\frac{\ell_{\text{\tiny P}}}{2\pi}\del\delbar k_X\,.
\ee
Notice that with these normalizations $k_X$ and $k_0$ have dimensions of a mass. 

The conditions (\ref{bcond}) completely   fix $k_0$.  On the other hand, even though the potentials $\kappa_a$ are only defined locally, their $\sigma^b$-derivatives $\del\kappa_a/\del \sigma^b$ are globally defined and so are the conditions \eqref{kappaboundary}.  By combining the arguments of \cite{Martucci:2016hbu,Martucci:2016pzt}, one can derive the
identity
\be\label{derkappa}
\frac{\del\kappa_a}{\del \sigma^b}=\frac{\ell_{\text{\tiny P}}}{2\pi}\int_{X,y'}G_X(y;y')(J\wedge J\wedge \omega_a\wedge \omega_b)(y')~,
\ee
where we recall that $G_X(y;y')$ is the Green's function on $X$.


\section{Holographic EFT}
\label{sec:G4=0}

In section \ref{sec:prelrem} we organised the fields entering the  EFT in a set of $n=4N_{\rm M2}+n_{\rm K}$ chiral multiplets $\phi_A=(z^i_I,\rho_a)$. (We use the same symbol for chiral superfields and for their lowest bosonic components.)  In this section we describe the two-derivative holographic EFT for these  chiral multiplets,  which can be derived from M-theory. 

As anticipated in the introduction,  the holographic EFT can be obtained from the rigid/decompactification limit of the three-dimensional supergravity theories of \cite{Martucci:2016pzt},
similarly to what was done in \cite{Martucci:2016hbu} for four-dimensional EFTs with type IIB holographic realizations.  A  new ingredient,  absent in the IIB models of \cite{Martucci:2016hbu}, is the possibility to turn on an internal  flux $F^{\rm int}_4$ -- see for instance \cite{Benishti:2009ky} for a general discussion and references to the previous literature. In this paper we assume  $F^{\rm int}_4=0$, but the incorporation of non-vanishing flux can be achieved without difficulties.  With this restriction the M-theory vacua are completely analogous to the type IIB vacua discussed in \cite{Martucci:2016hbu}. Hence we can adapt step by step the rigid/decompactification limit  discussed in  appendix A therein to obtain a very similar holographic EFT.

\subsection{Mixed formulation with chiral and linear multiplets}
\label{sec:HEFTlinear}

Since $\Im\rho_a$ parametrize the $C_6$ axions, 
the \emph{perturbative}\footnote{Unless explicitly stated, in this paper EFT implicitly indicates the `perturbative' EFT. Non-perturbative effects will be discussed elsewhere.}  holographic EFT  is symmetric under constant shifts of  $\Im\rho_a$. This means that we can dualize the chiral multiplets  $\rho_a$ into a set of linear multiplets $\Sigma^a$ \cite{Lindstrom:1983rt} and use the latter to describe the effective theory, see for instance \cite{deBoer:1997kr,Intriligator:2013lca}. 

The three-dimensional $\caln=2$ linear multiplets $\Sigma^a$  satisfy by definition the constraint 
\be
D^2\Sigma^a=\bar D^2\Sigma^a=0
\ee
which, as reviewed in appendix \ref{app:3dsuperspace}, can be locally solved as follows
\be\label{linearvector}
\Sigma^a=\frac{\ii}{4}D\bar D\cala^a=\sigma^a+\ldots-\frac12\epsilon^{\mu\nu\rho} \theta\gamma_\mu \bar\theta F^a_{\nu\rho}
\ee
in terms of a set of abelian vector multiplets $\cala^a$, whose real scalars  $\sigma^a$ coincide with the K\"ahler moduli on $X$, and whose vectors $A^a$ (with field strengths $F^a=\d A^a$) appear in the decomposition $C_3=\frac{1}{2\pi}\ell^3_{\text{\tiny P}}A^a\wedge \omega_a+\ldots$  of the M-theory three-form $C_3$. The holographic EFT  is completely specified by a `kinetic potential'  $\calf(z,\bar z,\Sigma)$ such that 
\be\label{linHEFT}
\begin{aligned}
S'_{\text{\tiny HEFT}}=&\,\int\d^3 x\d^4\theta\, \calf(z,\bar z,\Sigma)\\
=&\,\frac1{4}\int\calf_{ab} \Big(\d \sigma^a\wedge *\d \sigma^b+F^a\wedge * F^b\Big)-\int\calf^{IJ}_{i\bar\jmath}\,\d z_I^i\wedge*\d\bar z_J^{\bar\jmath}\\
& -\frac{\ii}{2}\int\left(\calf^I_{ai}\d z_I^i-\calf^{I}_{a\bar\imath}\d\bar z_I^{\bar\imath}\right)\wedge F^a+\text{(fermionic terms)}~,
\end{aligned}
\ee
where $\calf_{ab}\equiv\frac{\del^2\calf}{\del \sigma^a\del \sigma^b}$, $\calf^I_{ai}\equiv\frac{\del^2\calf}{\del \sigma^a\del z^i_I}$, etc. Here $\calf(z,\bar z,\Sigma)$ can be obtained by following almost verbatim the rigid/decompactification limit of \cite{Martucci:2016pzt}. The final result is 
\be\label{dualF}
\calf(z,\bar z,\Sigma)=\sum_I k_X(z_I,\bar z_I;\Sigma)~,
\ee
where  $k_X(z,\bar z; \sigma)$ is the K\"ahler potential of the  metric $\d s^2_X$ introduced in \eqref{kdec}. Note that   $k_X(z_I,\bar z_I;\Sigma)$ is only locally defined and can change as follows 
\be\label{kXchange}
k_X(z_I,\bar z_I;{\Sigma})\quad\rightarrow\quad k_X(z_I,\bar z_I;{\Sigma})+\Sigma^a\Re f_a(z_I)~,
\ee
 where $f_a(z_I)$ are holomorphic functions, see section \ref{sec:geom}. Of course, the Lagrangian \eqref{linHEFT} is invariant under the shift \eqref{kXchange}.

We see that the kinetic potential $\calf$ is simply the sum of $N$ contributions obtained by evaluating the K\"ahler potential  \eqref{kdec} at the positions of the $N$ mobile M2-branes and considering the K\"ahler moduli and their superpartners as dynamical. Our holographic derivation implies therefore that the large-$N$ EFT in the formulation with linear multiplets is described by a sort of Hartree approximation, in which the M2-branes move independently in a mean background which only depends on the linear multiplets $\Sigma^a$. 

Note that  the EFT theory  \eqref{linHEFT} is invariant under the symmetric group ${\rm S}_N$ of all the permutations acting on the $N$ indices $I$ of the fields $z^i_I$. Since the $N$ M2-branes are identical, ${\rm S}_N$ must be considered as a {\em gauge} group of the EFT. The presence of this  discrete gauge symmetry will be important in the following. The fixed points of non-trivial subgroups of ${\rm S}_N$ correspond to configurations at which the above EFT is expected to break down because of the appearance of some new degrees of freedom.

The result (\ref{dualF}) is consistent with the intuitive idea that, if we freeze all the K\"ahler moduli  $\sigma^a$, the two-derivative EFT describing a number of non-coincident M2-branes probing  the Calabi-Yau geometry is just the sum of the separate non-linear sigma models describing each M2-brane. However, we stress  that the K\"ahler potential \eqref{kdec}, which appears in \eqref{kXchange}, is {\em not} an arbitray K\"ahler potential associated with the Calabi-Yau metric, since it  must satisfy the conditions  \eqref{bcond}.  It can at most change by a constant or by a shift of the form \eqref{kXchange} while, for instance, one cannot add to  $k_X(z_I,\bar z_I;\sigma)$ any real non-linear function of the K\"ahler moduli.   

 Let us introduce the positive definite symmetric matrix
\be\label{calg}
\calg_{ab}\equiv \frac{\ell_{\text{\tiny P}}}{\pi}\int_X H\,\omega_a\wedge *_X\omega_b = -\frac{\ell_{\text{\tiny P}}}{2\pi}\int_X H\,\omega_a\wedge \omega_b\wedge J\wedge J
\ee
which, recalling \eqref{L2wnorm}, is always finite. The second equality follows from the primitivity of $\omega_b$, see footnote \ref{foot:prim}.
We also introduce 
\be\label{caladef}
\cala^I_{ai}\equiv\frac{\del\kappa_a(z_I,\bar z_I;\sigma)}{\del z^i_I}~,
\ee
where $\kappa_a(z,\bar z;\sigma)$ are the potentials defined in \eqref{defkappa}. $ \cala^I_{ai}$ can be regarded as a connection on the line bundle $\calo(D_a)$ over the Calabi-Yau geometry probed by the $I$-th M2-brane. 
Then, using \eqref{kdec}, \eqref{derkappa} and \eqref{warp}, one can show that the bosonic terms in the holographic EFT  Lagrangian \eqref{linHEFT} take the form
\be\label{linHEFTbos}
\begin{aligned}
\call'_{\text{\tiny HEFT}}=&\,-\frac1{4}\,\calg_{ab}(z,\bar z,\sigma)\Big(\d\sigma^a\wedge *\d \sigma^b + F^a\wedge * F^b\Big)-\frac{2\pi}{\ell_{\text{\tiny P}}}\sum_{I}g_{i\bar\jmath}(z_I,\bar z_I;\sigma)\, \d z^i_I\wedge *\d\bar z^{\bar\jmath}_I\\
&\,-\frac{\ii}{2}\left(\cala^I_{ai}\d z^i_I- \bar\cala^I_{a\bar\imath}\d \bar z^{\bar\imath}_I\right)\wedge F^a+\text{(fermionic terms)}~,
\end{aligned}
\ee
where $g_{i\bar{\jmath}}$ is the K\"ahler metric on the Calabi-Yau manifold $X$. We stress that the K\"ahler moduli $\sigma^a$ can be considered  dynamical precisely because the kinetic matrix \eqref{calg}
  is finite  thanks to the asymptotically AdS behaviour \eqref{WasyAdS} of the warping, see \eqref{L2wnorm}.\footnote{If we were to undo the near-horizon limit and add an arbitrary constant to $H$, 
 then only the $L^2$-normalizable K\"ahler moduli $\hat \sigma^\alpha$  would be dynamical. It is the near-horizon limit that allows us to consider the $L^2$-non-normalizable modes dynamical.}


\subsection{Formulation with chiral multiplets}
\label{sec:HEFTchiral}

So far we have worked with the linear multiplets $\Sigma^a$, but we can dualize them to chiral multiplets $\rho_a$ \cite{Lindstrom:1983rt}, see also \cite{deBoer:1997kr,Intriligator:2013lca}.  The dual formulation depends just on the real part of the chiral multiplets, which is given by
\be\label{defrho}
\Re\rho_a\equiv \frac12\frac{\del\calf}{\del\Sigma^a}=\frac12\sum_I\kappa_a(z_I,\bar z_I;\Sigma)~.
\ee
This relation is obtained by unconstraining $\Sigma^a$ in \eqref{linHEFT}, adding the term $-\int\d^3 x\d^4\theta(\rho_a+\bar\rho_a)\Sigma^a$, and extremizing the resulting action with respect to $\Sigma^a$.
The dual action is then of the form
\be\label{HEFT}
\begin{aligned}
S_{\text{\tiny HEFT}}&=\int\d^3x\d^4\theta\, \calk(\phi,\bar\phi)\\
&=-\int \calk^{A\bar B}(\phi,\bar\phi)\d\phi_A\wedge *_3\d\bar\phi_{\bar B}+\text{(fermionic terms)}~,
\end{aligned}
\ee
where $\calk^{A\bar B}\equiv \frac{\del^2\calk}{\del\phi_A\del\bar\phi_{\bar B}}$ and the K\"ahler potential $\calk$ is given by the Legendre transform 
\be\label{HEFTK}
\begin{aligned}
\calk&=\calf-\Sigma^a \frac{\del\calf}{\del\Sigma^a}\\
&=\sum_I k_0(z_I,\bar z_I;\Sigma)~.
\end{aligned}
\ee
Here $\calk$ must be considered as the function of the chiral superfields $\phi_A\equiv(z^i_I,\rho_a)$ (and their complex conjugates) 
obtained by inverting \eqref{defrho} and expressing  $\Sigma^a$ as functions of $(z^i_I,\bar z^{\bar\imath}_I,\rho_b+\bar\rho_b)$. In particular,  the $\Sigma^a$'s generically depend on all $z^i_I$'s, so the Hartree approximation does not hold in the formulation with chiral multiplets. 

In general it is not possible to obtain an explicit formula for the functions $\Sigma^a$. Nevertheless, by keeping the dependence of $\rho_b+\bar\rho_b$ hidden in $\Sigma^a$,
one can still compute the derivatives of $\calk$.
We then obtain the Lagrangian
\be\label{boschiral}
\call'_{\text{\tiny HEFT}}=-\, \calg^{ab}(z,\bar z,\sigma)\nabla\rho_a\wedge *\nabla\bar\rho_b-
\frac{2\pi}{\ell_{\text{\tiny P}}} \sum_I g_{i\bar\jmath}(z_I,\bar z_I;\sigma)\d z^i_I\wedge *\d\bar z^{\bar\jmath}_I+(\text{fermionic terms})
\ee
where $\calg^{ab}$ is the inverse of \eqref{calg}, $\nabla\rho_a$ is  the covariant derivative \be\label{covderho}
\nabla\rho_a\equiv \d\rho_a-\cala^I_{ai}\d z^i_I~
\ee
associated with the connection \eqref{caladef},
and $\nabla\bar\rho_a\equiv\overline{\nabla\rho_a}$. Again, in these formulas we must view $\sigma^a$ as the functions of 
$(z^i_I,\bar z^{\bar\imath}_I,\rho_b+\bar\rho_b)$ obtained by inverting 
\eqref{defrho}. 

This formulation in terms of chiral multiplets clarifies the complex structure of the moduli space $\calm$: it is a fibered space, whose base is the $N$-th symmetric product of the internal resolved cone $X$ (whose $I$-th copy is parametrized by the four coordinates $z^i_I$), while the fibers are parametrized by chiral fields $e^{-2\pi\rho_a}$, which transform as sections of the product of $N$ copies of the line bundle $\calo_X(-D_a)$.  We stress that this parametrization depends on the choice of $\calf$, which is not unique. 
Indeed, the effective action does not change if we make the transformation
\be\label{red1}
\calf(z,\bar z,\Sigma)\quad\rightarrow \quad\calf'(z,\bar z,\Sigma)=\calf(z,\bar z,\Sigma)+ \Sigma^a [g_a(z)+\bar g_a(\bar z)]~,
\ee
where $g_a(z)$ depends holomorphically on the chiral fields $z^i_I$. This transformation generalises \eqref{kXchange} and corresponds to the holomorphic change of fiber coordinates 
\be\label{red2}
\rho_a\quad\rightarrow \quad \rho_a+g_a(z)~.
\ee

As reviewed  in appendix \ref{app:Kcone}, (spontaneously broken) superconformal symmetry requires the K\"ahler potential \eqref{HEFTK} to have scaling dimension one. Checking that this is the case requires specifying the proper scaling dimensions of the chiral fields. In sections \ref{sec:Y12model} and \ref{sec:Q111model} we will  verify that  this consistency condition is explicitly satisfied by the holographic  EFTs of the  $Y^{12}(\mathbb{P}^2)$ and $Q^{111}$ models.

\subsection{Betti symmetries and non-perturbative effects}
\label{sec:HEFTmixed}

Since the harmonic two-forms split into $L_2$ and $L_2^{\rm w}$ normalizable as in \eqref{omegasplit}, it is natural to do the same for the associated linear and chiral multiplets: $\Sigma^a=(\hat\Sigma^\alpha,\tilde\Sigma^\sigma)$, $\rho_a=(\hat\rho_\alpha,\tilde\rho_\sigma)$.  
We note that while the $\hat\omega_\alpha$'s span a canonical cohomology subgroup $H^2(X,Y;\mathbb{Z})\subset H^2(X;\mathbb{Z})$, this is not true for the $\tilde\omega_\sigma$'s, since they can mix with the $\hat\omega_\alpha$'s as follows:
\be\label{omegashift}
(\hat\omega_\alpha,~\tilde\omega_\sigma)\quad\rightarrow\quad (\hat\omega_\alpha,~\tilde\omega_\sigma+m^\alpha_\sigma\,\hat\omega_\alpha)~,
\ee
where $m^\alpha_\sigma\in \mathbb{Z}$. This freedom is reflected in  the mixed redefinitions of linear and chiral   multiplets 
\begin{subequations}\label{mixedredef}
\begin{align}
(\hat\Sigma^\alpha,~\tilde\Sigma^\sigma) \quad &\to \quad  (\hat\Sigma^\alpha-m^\alpha_\sigma\tilde\Sigma^\sigma,~\tilde\Sigma^\sigma) ~, \\
(\hat\rho_\alpha,~\tilde\rho_\sigma) \quad &\to \quad  (\hat\rho_\alpha,~\tilde\rho_\sigma+m^\alpha_\sigma\,\hat\rho_\alpha) ~.
\end{align}
\end{subequations}
This suggests that the separation into $L_2$ and $L_2^{\rm w}$ normalizable modes could be better described by the mixed linear/chiral multiplets $(\tilde\Sigma^\sigma,\hat\rho_\alpha)$, which do not suffer from the above ambiguity. The corresponding effective theory can be obtained from \eqref{linHEFT} by dualizing only the $\hat\Sigma^\alpha$ linear multiplets.  
This gives the action
\be\label{mixedlag}
\tilde S_{\text{\tiny HEFT}}=\int\d^3 x\d^4\theta\, \calg(z,\bar z,\hat\rho,\overline{\hat\rho},\tilde\Sigma)~,
\ee
where
\be\label{kincalg}
\calg\equiv \sum_I\left[k_0(z_I,\bar z_I; \hat\Sigma,\tilde\Sigma)
+\tilde\Sigma^\sigma\tilde\kappa_\sigma(z_I,\bar z_I; \hat\Sigma,\tilde\Sigma)  \right] ~.
\ee
Here $\hat\Sigma^\alpha$ must be considered as the functions of $(\hat\rho+\overline{\hat\rho},z,\bar z,\tilde\Sigma)$ that are obtained by inverting 
\be
\hat\rho_\alpha+\overline{\hat\rho}_\alpha=\sum_I \hat\kappa_\alpha(z_I,\bar z_I; \hat\sigma, \tilde\Sigma) ~.
\ee
Notice that the mixed redefinition \eqref{mixedredef} induces the shift
\be
\calg\rightarrow \calg+2m^\alpha_\sigma\,\tilde\Sigma^\sigma\Re\hat\rho_\alpha\,,
\ee
which does not change the action \eqref{mixedlag} since  $\tilde\Sigma^\sigma$ are linear and $\hat\rho_\alpha$ chiral superfields.

The `naturalness' of this formulation has a physical interpretation.
As we discussed in section \ref{sec:geom}, $\tilde\Sigma^\sigma$ are counted by $b_2(Y)$. Hence, according to the usual AdS/CFT dictionary, they correspond to Betti multiplets which are dual to {\em exact} $U(1)$ global symmetries of the SCFT. Indeed, we can regard $\tilde\calj^\sigma\equiv \frac1{2\pi}\tilde\Sigma^\sigma$ as the associated conserved current supermultiplet. Notice that by using the description in terms of $\tilde\Sigma^\sigma$ and their associated vector multiplets $\tilde\cala^\sigma$, we make  these $U(1)$ symmetries topological,  in the sense that the currents $\tilde\calj^\sigma$ are conserved off-shell: $D^2\tilde\calj^\sigma=\bar D^2\tilde\calj^\sigma=0$.

On the other hand, the  $b_6(X)$ linear multiplets   $\hat\Sigma^\alpha$ are only associated to {\em perturbative} $U(1)$ global symmetries. These are expected to be broken by non-perturbative effects generated by Euclidean M5-branes wrapping the non-trivial six-cycles in $X$, which  indeed identify a  $b_6(X)$-dimensional lattice. The implications of these non-perturbative effects on the our holographic EFTs will be addressed elsewhere -- see  \cite{Benishti:2010jn} for previous discussions. For the moment suffice it to say that the non-perturbative corrections to the effective theory are weighted by an exponential factor of the form $e^{2\pi n^\alpha\hat\rho_\alpha}$, for some integers $n^\alpha\in\mathbb{Z}$. The  $U(1)$ global symmetry associated with the current multiplet $\hat\calj^{\alpha}=\frac1{2\pi}\hat\Sigma^\alpha$ generates a shift of $\Im\hat\rho_\alpha$. We then see that the holographic EFT action \eqref{mixedlag} is invariant under such shifts, which are instead generically broken by non-perturbative corrections.


\section{Freezing moduli by the \texorpdfstring{$S$}{S} operation}
\label{sec:Soperation}

 In section \ref{sec:G4=0} we saw that the low-energy  spectrum  includes $b_2(Y)$ linear multiplets $\tilde\Sigma^\sigma$ which correspond to exact global Betti $U(1)_\sigma$ symmetries of the dual SCFT, with associated current multiplets $\tilde\calj^\sigma=\frac1{2\pi}\tilde\Sigma^\sigma$. For instance, in the explicit examples of section \ref{sec:FT},  these current multiplets are those in \eqref{Y12Tcur} and \eqref{J12Q111} respectively. We  observe that these global symmetries are spontaneously broken  along the moduli space, since they shift the imaginary part $\Im\tilde\rho_\sigma$ of the dual chiral coordinates. Each $\tilde\rho_\sigma$ is then the supersymmetric  Goldstone boson of a spontaneously broken $U(1)_\sigma$ symmetry. 

According to the AdS/CFT correspondence, a $U(1)_\sigma$ global symmetry in the SCFT corresponds to a `Betti multiplet' in the KK-reduced theory on AdS$_4$. Such a multiplet contains a scalar $\varphi^\sigma$ of mass $m^2=-2/L^{2}$ with the asymptotics $\varphi^\sigma(x,r)\simeq \frac{\varphi^\sigma_0(x)}{r^2}+\ldots$ for $r\rightarrow \infty$. This value of the mass allows for the `alternate' {\em Neumann}  quantisation \cite{Breitenlohner:1982jf,Klebanov:1999tb} in which $\varphi^\sigma$ is dual to a scalar operator  $J^\sigma(x)$ of dimension $\Delta=1$ and VEV  $ \langle J^\sigma(x)\rangle = \varphi^\sigma_0(x)$ \cite{Klebanov:1999tb}.  The operator $ J^\sigma$ is the moment map of the $U(1)_\sigma$ symmetry, the lowest component of an SCFT current multiplet
\be
\calj^\sigma=J^\sigma-\theta\gamma^\mu\bar\theta j^\sigma_\mu+\ldots
\ee
 In our holographic EFT this is represented by $\calj^\sigma_{\text{\tiny HEFT}}= \frac1{2\pi}\tilde\Sigma^\sigma$ and 
then we are naturally led to identify 
\be
\langle J^\sigma(x)\rangle=\frac{1}{2\pi}\tilde\sigma^\sigma(x)
\ee
and $ \varphi^\sigma_0\simeq\tilde\sigma^\sigma$. 
We then see that the holographic EFTs described in section \ref{sec:G4=0} precisely correspond to the Neumann quantization of all the Betti multiplets. 

On the other hand, one may use the standard {\em Dirichlet} quantisation  \cite{Breitenlohner:1982jf,Klebanov:1999tb} in which $\varphi^\sigma$ are dual to operators $\calo_\sigma$ of scaling dimension $\Delta=2$ of a different theory. 
In this case $ \varphi^\sigma_0(x)\sim\tilde\sigma^\sigma(x)$ must be  interpreted as external sources which  deform the dual SCFT by $\int \tilde\sigma^\sigma(x)\calo_\sigma(x)$.  By supersymmetry, the same interpretation extends to the whole linear multiplets $\tilde\Sigma^\sigma$, which should be considered as  non-dynamical background multiplets of spurionic scaling dimension $\Delta_{\tilde\Sigma}=1$. Hence, one should remove these linear multiplets from the holographic EFT.

From the dual SCFT viewpoint, the two quantizations are related by the $S$ operation  \cite{Kapustin:1999ha,Witten:2003ya}  reviewed in section \ref{sec:FT} -- see also \cite{Klebanov:2010tj,Benishti:2010jn} for a discussion in the present context. For instance, in the examples of section \ref{sec:FT} we focused on modified quiver models obtained as a result of an $S$ operation. These models correspond holographically to the Neumann quantization of the Betti multiplets. On the other hand, the quiver models  that we started from before applying the $S$ operation, which have $U(N)\times U(N)$ gauge group, correspond to Dirichlet quantizations of the Betti multiplets.

The holographic EFTs  corresponding to Neumann or Dirichlet quantizations must be related by an $S$ operation as well. Since applying the $S$ operation twice returns the initial theory with reversed sign of the current multiplet, the holographic EFT corresponding to Dirichlet  quantization can be obtained by applying the $S$ operation with sign-reversed current to the holographic EFTs described in section \ref{sec:G4=0}.
More explicitly, the the `Dirichlet' theory can be obtained  by coupling the current multiplets  $-\tilde\calj^\sigma=-\frac1{2\pi}\tilde\Sigma^\sigma$ of the `Neumann' holographic EFT \eqref{mixedlag} to dynamical external vector multiplets $\calb_\sigma$:  
\be\label{theoryA}
S_{\rm D}[\tilde\Sigma,\calb]=\tilde S_{\text{\tiny HEFT}}[\tilde\Sigma]+\frac{1}{2\pi}\int\d^3 x\d^4\theta\, \tilde\Sigma^\sigma\calb_\sigma~.
\ee
We used the mixed description of section \ref{sec:HEFTmixed}, keeping the dependence on the chiral fields $z^i_I,\hat\rho_\alpha$ implicit, since they can be considered as spectators under the $S$ operation. Since $D^2\tilde\Sigma^\sigma=\bar D^2\tilde\Sigma^\sigma=0$ off-shell, $S_{\rm D}$ is invariant under background gauge transformations. Furthermore the second term in \eqref{theoryA} is consistent with superconformal invariance.\footnote{In fact, superconformal symmetry allows the low-energy effective theory to contain  additional CS terms of the form $-\frac{1}{4\pi}\kappa^{\sigma\rho}\int\d^3 x\d^4\theta\, \calb_\sigma\Theta_{\rho}$
where we have introduced the background linear multiplet
$\Theta_\sigma\equiv \frac{\ii}{4}D\bar D \calb_\sigma$. 
We choose the microscopic coupling in such a way that \eqref{theoryA} holds.}   

It is clear that the effect of the $\tilde\Sigma^\sigma\calb_\sigma$ coupling  in \eqref{theoryA} is to dynamically set $\tilde\Sigma^\sigma$ to zero. Hence the `Dirichlet' holographic EFT is obtained simply by setting $\tilde\Sigma^\sigma=0$ in \eqref{mixedlag}, while the other superfields $z^i_I,\hat\rho_\alpha$ remain dynamical. From the M-theory geometrical viewpoint, this corresponds to freezing the `Betti' K\"ahler moduli $\tilde\sigma^\sigma$ to zero.     

More generically, one may couple  the topological currents $\calj^\calb_\sigma=\frac1{2\pi}\Theta_\sigma$, with $\Theta_\sigma\equiv \frac{\ii}{4}D\bar D \calb_\sigma$, to  non-dynamical external vector multiplets  $\cala^\sigma$:
\be\label{theoryAb}
S'_{\rm D}[\tilde\Sigma,\calb;\cala]=\tilde S_{\text{\tiny HEFT}}[\tilde\Sigma]+\frac{1}{2\pi}\int\d^3 x\d^4\theta\, \tilde\Sigma^\sigma\calb_\sigma-\frac{1}{2\pi}\int\d^3 x\d^4\theta\, \Theta_\sigma\cala^\sigma~.
\ee
Now the equations of motion of $\calb_\sigma$ set $\tilde\Sigma^\sigma=\frac{\ii}{4}D\bar D \cala^\sigma$. If for instance we choose $\cala^\sigma=-2\ii\theta\bar\theta\xi^\sigma$, for constant $\xi^\sigma$, then we get the on-shell condition  $\tilde\Sigma^\sigma=\xi^\sigma$. From the CFT viewpoint, $\xi^\sigma$  correspond to  FI parameters or real masses, which explicitly  break the conformal symmetry.  From the geometrical M-theory viewpoint, they correspond to non-vanishing but non-dynamical `Betti' K\"ahler moduli $\tilde\sigma^\sigma=\xi^\sigma$.

Finally, the initial `Neumann' holographic EFT is obtained by performing a further $S$ operation, i.e.\ by promoting the background $\cala^\sigma$ to dynamical vector multiplet. The on-shell relation $\tilde\Sigma^\sigma=\frac{\ii}{4}D\bar D \cala^\sigma$ obtained by  integrating out $\calb_\sigma$  implies that $\cala^\alpha$ is redundant and the EFT reduces to the original $S_{\text{\tiny HEFT}}[\tilde\Sigma]$. One may also apply the $S$ operation to any strict subset of  Betti multiplets, giving a set of mixed `Neumann/Dirichlet' models. However, in the rest of the paper we will focus on purely `Neumann' models.


\section{Holographic EFT of toric models}
\label{sec:toric}

Toric models in M-theory provide a large class of examples which have been extensively studied in the literature, see  \cite{Martelli:2008si,Hanany:2008cd,Ueda:2008hx,Imamura:2008qs,Hanany:2008fj,Franco:2008um, Aganagic:2009zk,Benishti:2009ky, BeniniClossetCremonesi2010,Benishti:2010jn,Benini:2011cma,Closset:2012ep}  for a sample of papers. 
In such models the internal Calabi-Yau space admits a $U(1)^4$ group of isometries. It is then useful to describe the internal metric $\d s^2_X$ appearing in the M-theory background \eqref{Mback} in a manifestly toric way. We refer to \cite{abreu2000kahler}  for a nice introduction to the aspects of the toric K\"ahler geometry that will be used in the following and more mathematical references on the subject.   Note that the following discussion can be readily adapted to other dimensions and then, for instance, applied also to AdS$_5$/CFT$_4$ settings.

\subsection{Symplectic formulation of the K\"ahler structure}
\label{sec:symp}

One  can introduce a set of complex coordinates 
\be
z^i\equiv x^i+\ii\varphi^i\quad~~~~~~\text{with}\quad \varphi^i\simeq \varphi^i+1\,,
\ee 
such that $e^{-2\pi z^i}\in\mathbb{C}^*$ has charge $+1$ under the $i$-th toric $U(1)$.   Then the K\"ahler potential $k_X$ appearing in \eqref{JXk}  can be chosen to depend only on $x^i=\Re z^i$ (and on the K\"ahler moduli $\sigma^a$),  so that the general toric metric $\d s^2_X$  takes the form 
\be\label{toricmetric}
\d s^2_X=\frac{\ell_{\text{\tiny P}}}{2\pi}G_{ij}(x;\sigma)\d z^i\d\bar z^{j}=\frac{\ell_{\text{\tiny P}}}{2\pi}G_{ij}(x;\sigma)\left(\d x^i\d x^j+\d\varphi^i\d\varphi^j\right)\,,
\ee 
where 
\be\label{toricmetric2}
G_{ij}(x;\sigma)=\frac{1}{2}\frac{\del^2 k_X}{\del x^i\del x^j}\,.
\ee
The toric space $X$ is obtained by appropriately completing  the dense open subset $(\mathbb{C}^*)^4\subset X$ parametrized by $e^{-2\pi z^i}$, see below for more details. 

We can describe the toric space as the classical vacuum moduli space of a gauged linear sigma model (GLSM) in terms of a set of  $d\geq 4$ homogeneous coordinates $Z_A$, which are chiral multiplets in the GLSM. They transform under a continuous $U(1)^{d-4}$  gauge group defined by the charges $Q^{aA}$: $Z_A\rightarrow e^{\ii \alpha_a Q^{aA}}Z_A$ 
(no sum over $A$). The toric space inherits a {\em canonical} non Ricci-flat K\"ahler form 
$J^{\rm can}_X=\frac{\ii }{2\pi}\del\delbar k^{\rm can}_X$
 and a canonical K\"ahler potential $k^{\rm can}_X$ by symplectic reduction of the flat K\"ahler form $\frac{\ii}{2\pi}\sum_A \d Z_A\wedge\d \bar Z_A$ on $\mathbb{C}^d$. The associated   D-flatness conditions reads 
\be\label{Dflatness}
\sum_AQ^{aA}|Z_A|^2=\ell_{\text{\tiny P}}\sigma^a\,.
\ee
The $\sigma^a$'s parametrize the K\"ahler moduli and span the K\"ahler cone $\mathbb{K}_{X}$ of $X$.
\footnote{We use a minimal GLSM description of the resolved geometry where the $\sigma^a$'s are linearly independent and the homogeneous coordinates $Z^A$ are in one-to-one correspondence with the toric divisors.} 
 The Ricci-flat  K\"ahler potential can then be written in the form $k_X=k^{\rm can}_X+\Delta k_X$, where $\Delta k_X$ is a globally defined function.
 
 The relation between toric and homogeneous coordinates is given by
\be\label{zZrelation}
e^{-2\pi z^i}=\prod^d_{A=1} Z_A^{v^i_A}\,,
\ee
up to overall constants, where  ${\bf v}_A=(v^1_A,\ldots, v^4_A)\in N_\mathbb{Z}\simeq \mathbb{Z}^4$  are $d$ integral vectors generating the one-dimensional cones of the toric fan $\calt$ associated with the resolved toric variety $X$. In the following $\calt(n)$ will denote the set of $n$-dimensional cones of $\calt$. Hence ${\bf v}_A\in N_{\mathbb{Z}}$ generate the elements of $\calt(1)$. They are such that 
\be\label{qvcon} 
Q^{aA}{\bf v}_A={\bf 0} \quad~~~~~~\forall a=1,\ldots, d-4~.
\ee
By definition, the complete gauge group of the gauged linear sigma model leaves the combination appearing on the r.h.s.\ of \eqref{zZrelation} invariant. Hence
 $e^{-2\pi z^i}$ are good  gauge-invariant coordinates on the toric variety.  

In order to better understand the geometric structure of $X$, it proves useful to go to the symplectic description by introducing the dual variables 
\be\label{lmindef}
l_i=-\frac{1}{2}\frac{\del k_X}{\del x^i}
\ee
which define dual vectors ${\bm l}=(l_1,\ldots,l_4)\in M_{\mathbb{R}}\equiv M_{\mathbb{Z}}\otimes\mathbb{R}$, where  $M_{\mathbb{Z}}$ is the lattice dual to $N_{\mathbb{Z}}$.  
The corresponding dual symplectic potential is given by
\be\label{sympot}
F_X(l;\sigma)=k_X+2l_i\,x^i~,
\ee
where $x^i$ must be considered as the functions of $l_i$ and $\sigma^a$ obtained by inverting \eqref{lmindef}.\footnote{This dualization from $x^i$ to $l_i$ is analogous to the inverse of the dualization from $\Re\rho_a$ to $\Sigma^a$, see \eqref{defrho} and \eqref{HEFTK}.}

In the symplectic  variables $(l_i,\varphi^i)$ the K\"ahler form \eqref{JXk} reads $J_X=\frac{\ell_{\text{\tiny P}}}{2\pi}\,\d l_i\wedge\d\varphi^i$ and the metric becomes 
\be\label{metricdual}
\d s^2_X=\frac{\ell_{\text{\tiny P}}}{2\pi}\left(G^{ij}\d l_i\d l_j +G_{ij}\d\varphi^i\d\varphi^j\right)~,
\ee
where the matrix
\be\label{Gdual}
G^{ij}(l;\sigma)=-\frac12\frac{\del^2 F_X}{\del l_i\del l_j}
\ee
is the inverse of $G_{ij}$.

The real variables $l_i$ take values in a Delzant  polytope \cite{Delzant,guillemin1994,abreu2000kahler}
\be\label{poly}
\calp_\sigma=\{{\bm l}\in M_{\mathbb{R}}:\ s_A(l;\sigma)\equiv \langle {\bm l},{\bf v}_A\rangle+\chi_A\geq 0\} 
\ee
where $ \langle {\bm l},{\bf v}_A\rangle\equiv l_i\,v^i_A$ is the canonical pairing between the dual vector spaces $N_{\mathbb{R}}$ and $M_{\mathbb{R}}$, and the $\chi_A$'s satisfy 
\be\label{chisigma}
Q^{aA}\chi_A=\sigma^a\,.
\ee
In this symplectic description, one can think of $X$ as a $T^4$ fibration over the polytope $\calp_\sigma$, where the `action' coordinates $l_i$ parametrize the base $\calp_\sigma$ and the `angles' $\varphi^i$ parametrize the $T^4$ fibers. On the $A$-th facet $\{s_A(l;\sigma)=0\}$ of $\calp_\sigma$  the $T^4$ fibration reduces to a $T^3$ fibration describing the toric divisor $\cald^A$, since the $S^1$ parametrized by $\varphi^i(t)=t v^i_A$ with $t\in [0,1]$ degenerates.

From the condition \eqref{chisigma} we get the cohomological identity  $\sigma^a[\omega_a]=\chi_A[\cald^A]$, where the cohomology classes $[\cald^A]=Q^{aA}[\omega_a]$ are   Poincar\'e dual to the toric divisors  $\cald^A\equiv\{Z_A=0\}$, and then we can  identify the K\"ahler  class with 
 \be\label{toricJdec}
[J_X]=\frac{\ell_{\text{\tiny P}}}{2\pi}\,\chi_A\,[\cald^A]\ .
\ee
This relation  provides a generically {\em redundant} parametrization of the K\"ahler class in terms of the $d$ variables $\chi_A$.  Indeed, from \eqref{qvcon} we see that $v^i_A [\cald^A]=0$ in cohomology and then  \eqref{toricJdec}  is unaffected by shifts of the form
\be\label{chilshift}
\chi_A\rightarrow \chi_A+\langle {\bm c},{\bf v}_A\rangle\quad,\quad {\bm l}\rightarrow {\bm l}-{\bm c}\,,
\ee
for some ${\bm c}\in M_{\mathbb{R}}$, which leave \eqref{poly} invariant.  One can  use this redundancy to  restrict the $\chi_A$'s to be positive \cite{cox1999mirror,cox2011toric}: 
 \be\label{chipos}
 \chi_A\geq 0\, .
 \ee
In order to guarantee that  \eqref{toricJdec} stays within the closure of the K\"ahler cone $\mathbb{K}_{X}$ one needs to impose that  $\int_{C}J=\frac{\ell_{\text{\tiny P}}}{2\pi}\sigma^a\int_{C}\omega_a\geq 0$ for any effective curve $C\subset X$. By picking the set of effective curves $\{C_\alpha\}$ generating the Mori cone of $X$, the closure of the K\"ahler cone is defined by the set of conditions
\be\label{chiKcone}
n^A_\alpha\chi_A\geq 0\quad,\quad \text{with $n^A_\alpha\equiv C_\alpha\cdot \cald^A$\,.}
\ee 
Notice that since $v^i_A[\cald^A]=0$, this condition is indeed invariant under \eqref{chilshift}.

We can also invert the relation \eqref{chisigma} between the $\chi_A$'s and the K\"ahler moduli $\sigma^a$. Let us choose the integral basis $[\omega_a]\in H^2(X;\mathbb{Z})$ dual to the basis of two-cycles  $C^a\in H_2(X;\mathbb{Z})$ naturally associated to the toric fan (see \cite{Denef:2008wq} for an explicit description), that is such that  $\int_{C^b}\omega_a=\delta_a^b$. 
We can then explicitly solve \eqref{chisigma} by setting
\be\label{chidec}
\chi_A=M_{Aa}\sigma^a\,,
\ee
where $M_{Aa}\in\mathbb{Z}$ are a set of integers such that $[\omega_a]=M_{Aa}[\cald^A]$.  Indeed, by recalling that the charges $Q^{aA}$ are identical to the intersection numbers $C^a\cdot \cald^A$, we immediately conclude that $Q^{aA}M_{Ab}=\delta^a_b$ and then \eqref{chisigma} is satisfied. 
\footnote{\label{foot:Kcones} Since $v^i_A[\cald^A]=0$, $M_{Aa}$ is defined up to shifts $M_{Aa}\rightarrow M_{Aa}+r_{ai}v^i_A$, for arbitrary $r_{ai}\in\mathbb{Z}$. These shifts must be accompanied by the redefinitions  $l_i\rightarrow l_i-r_{ai}\sigma^a$ and are of the form \eqref{chilshift}, with ${\bf c}={\bf r}_a\sigma^a$. 
This corresponds to the K\"ahler transformation  $k_X\rightarrow k_X+(z^i+\bar z^i)r_{ai}\sigma^a$ (while the dual symplectic potential $F_X$ is invariant). }

A toric crepant resolution $X$ of a cone $C(Y)$ corresponds to a toric fan $\calt$ with a maximal set of one-dimensional cones $\Delta\equiv\calt(1)$, generated by the $d$ vectors ${\bf v}_A\in N_{\mathbb{Z}}$. Each resolution $X$ has its own $(d-4)$-dimensional K\"ahler cone $\mathbb{K}_{X}$, which may be regarded as a convex polyedral cone in 
\be\label{extKms}
A_+(\Delta)\equiv\{\chi_A\,[\cald^A] \quad\text{such that}\quad \chi_A\geq 0\}\,.
\ee
By changing the triangulation of the toric fan one gets different crepant resolutions, connected by one or more flops.\footnote{Note that in the parametrization \eqref{chidec} a shift of $M_{Aa}$ of the kind discussed in footnote \ref{foot:Kcones} may be necessary under a flop, since the fan changes.} 
Under these transitions, the K\"ahler cones  glue together and form the {\em extended K\"ahler cone} 
\be\label{extCone}
\mathbb{K}^{\rm ext}_\Delta\subset A_+(\Delta)
\ee

Note that, generically, $\mathbb{K}^{\rm ext}_\Delta$  does {\em  not} fill the entire  $A_+(\Delta)$. 
The excluded region  corresponds to 
{\em  orbifold branches} with $n< d-4$ K\"ahler moduli. Their inclusion allows one to define the `GKZ decomposition' of $A_+(\Delta)$ \cite{oda1991,Gelfand1994discriminants}.

In this paper we assume that the asymptotic Sasaki-Einstein space $Y$ is smooth. Hence, the only allowed  orbifold branches correspond to  toric diagrams with fewer internal points and exist only if $b_6(X)\neq 0$. Conversely, if $b_6(X)= 0$ then the union of all K\"ahler cones fills the entire $A_+(\Delta)$, which can then be identified with the extended K\"ahler cone of the toric variety: $\mathbb{K}^{\rm ext}_{\Delta}\equiv A_+(\Delta)$.

\subsection{Superconformal symplectic structure}
\label{sec:sympEFT}

We would  like to investigate the possible form of the symplectic potential \eqref{sympot} by imposing compatibility with the expected superconformal invariance of the associated holographic EFT. Recall that $F_X$ has been obtained as the Legendre transform   \eqref{sympot} of the Calabi-Yau K\"ahler potential $k_X$. On the other hand, $k_X$ enters the definition of the EFT kinetic function \eqref{dualF}. Hence, in our holographic EFT the change to symplectic coordinates \eqref{lmindef} can be interpreted as moving from a description in terms of $N$ chiral multiplets $z^i_I$, $I=1,\ldots,N$,  to one in terms of $N$ vector multiplets $V^I_i$ with associated linear multiplets $ L^I_i\equiv   \frac{\ii}{4}D\bar D V^I_i= l_i^I+\ldots- \frac12\epsilon^{\mu\nu\rho}\theta\gamma_\mu\bar\theta F^ I_{i\nu\rho}+\ldots$ defined by
 \be
 L^I_i=-\frac{1}{2}\frac{\del\calf}{\del \Re z^i_I}\,,
\ee
where $\calf(z,\bar z,\Sigma)$  is given in \eqref{dualF}. The corresponding effective Lagrangian then takes the form $\int\d^4\theta \tilde\calf(L,\Sigma)$ with
\be\label{Ftilde}
\tilde\calf(L,\Sigma)=\calf+2L_i^I\,\Re z^i_I=\sum_I F_X(L^I;\Sigma)~.
\ee

In terms of the lowest superfield components, we can interpret $\tilde\calf(l,\sigma)$ as the symplectic potential on the entire geometric (perturbative) moduli space $\calm$, in the same sense in   which $F_X(l;\sigma)$ (for fixed $\sigma$) is the symplectic potential of $X$. Indeed, combining the toric and the $b_2(X)$ (perturbative)  $U(1)$ symmetries, $\calm$ can be regarded as a $4N+b_2(X)$-dimensional  toric K\"ahler space. Actually, from its dual SCFT interpretation, we expect $\calm$ to be a K\"ahler {\em cone} of the kind described in appendix \ref{app:Kcone}. 
We can use this observation to constrain the possible form of the function $F_X(l;\sigma)$.

In analogy with our discussion on the linear multiplets $\Sigma^a$, we may be tempted to interpret $\frac1{2\pi}L^I_i$ as current supermultiplets associated with the $4N$ toric symmetries of our EFT. However, one must take into account the discrete gauge group ${\rm S}_N$, which acts by permuting the $I$-indices of the linear multiplets $L^I_i$. Hence, $\frac1{2\pi}L^I_i$ are {\em not} good gauge-invariant current supermultiplets or, in other words, must be identified under the ${\rm S}_N$ gauge group. On the other hand, four physically distinct   current supermultiplets can be identified with the gauge-invariant combinations
\be\label{truetoric}
{\cal J}^{\text{\tiny(toric)}}_i\equiv \frac1{2\pi}\sum^N_{I=1}  L^I_i\,.
\ee
They generate the toric  $U(1)^4$ global symmetry of the SCFT which, at the EFT level, acts  by the simultaneously shifts $\varphi^i_I\to \varphi^i_I+ c^i$ with constant $c^i$.

\subsubsection{Case \texorpdfstring{$N=1$}{N=1}}

The symplectic potential \eqref{Ftilde} describes a conical toric K\"ahler  metric for generic $N$ if and only if this happens for $N=1$. We then first consider $N=1$ and investigate the restrictions imposed by superconformal invariance on  $\tilde\calf_{N=1}(l,\sigma)\equiv F_X(l;\sigma)$, regarded as a symplectic potential for the whole $\calm_{N=1}$.\footnote{The procedure of setting $N= 1$ in \eqref{Ftilde} is only formal, since we expect  \eqref{Ftilde} to be valid only in the large-$N$ limit.} We can address this problem by adapting some arguments of \cite{MartelliSparksYau2006}, which considers the similar issue of understanding the symplectic potential of the  (Calabi-Yau) K\"ahler cones that appear as internal geometries of string/M-theory backgrounds. 

In the symplectic toric description, the K\"ahler cone $\calm_{N=1}$ is parametrised   by the  $4+b_2(X)$ {\em axionic} fibral coordinates $\varphi^i$ and $\vartheta_a\equiv \Im\rho_a$ and $4+b_2(X)$ {\em dual saxionic} base coordinates $l_i,\sigma^a$ (which are dual to the {\em saxions} $\Re z^i_I$ and $\Re\rho_a$ respectively). The latter parametrize a polyhedral cone $\calc_{N=1}$ which is obtained by fibering the polytope \eqref{poly} over the K\"ahler cone $\mathbb{K}_{X}$.  This becomes more manifest by trading the dual saxions $(l_i,\sigma^a)$ for $s_A\in\mathbb{R}^{4+b_2(X)}_+$ as defined in \eqref{poly}, or more explicitly by $s_A\equiv v_A^il_i+M_{Aa}\sigma^a$.   Combining the condition \eqref{poly} with the K\"ahler cone condition \eqref{chiKcone}, we obtain the identification of $\calc_{N=1}$ with the rational polyhedral cone:
\be
\calc^X_{N=1}=\{s_A\in\mathbb{R}^{4+b_2(X)}\ |\ s_A\geq 0\,,\  n^A_\alpha s_A\geq 0\}\,.
\ee
Correspondingly, we can introduce some axionic coordinates $\phi^A\sim \phi^A+1$ such that $\varphi^i=v^i_A\phi^A$ and $\vartheta_a=M_{Aa}\phi^A$, which are paired with the dual saxions $s_A$. 

By enlarging the K\"ahler cone to the extended K\"ahler cone \eqref{extCone}, one obtains  the entire geometric saxionic cone $\calc^{\rm ext}_{N=1}$. It is interesting to observe that, under the assumption that the base $Y$ is smooth and $b_6(X)=0$ (i.e.\ that the toric diagram has no internal points), we can make the identification $\calc^{\rm ext}_{N=1}=\calc^{\Delta}_{N=1}$, where 
\be\label{maxCN=1}
\calc^{\Delta}_{N=1}\equiv \{s_A\in\mathbb{R}^{4+b_2(X)}\ |\ s_A\geq 0\} 
\ee
is the maximally enlarged saxionic cone, which is obtained  by enlarging the extended K\"ahler cone $\mathbb{K}^{\rm ext}_\Delta$ to the entire $A_+(\Delta)$. More generally we have the chain of inclusions $\calc^X_{N=1}\subset \calc^{\rm ext}_{N=1}\subset \calc^{\Delta}_{N=1}$.

In the notation of appendix \ref{app:Kcone},  one can identify the conical radial coordinate  $\tau$  over $\calm_{N=1}$  with the dilaton, and the dilation generator, which is one-half of the Euler vector field $\tau\frac{\del}{\del\tau}$,  with
\be
\cald_{N=1}= s_A\frac{\del}{\del s_A}= l_i\frac{\del}{\del l_i}+\sigma^a\frac{\del}{\del \sigma^a}\,.
\ee
The  toric metric on $\calm_{N=1}$ described by the symplectic potential $\tilde\calf_{N=1}(l,\sigma)$ is a cone if the Hessian of $F_X(l;\sigma)$ (with respect to both $l_i$ and $\sigma^a$) is homogenous of degree $-1$ under constant rescalings of $(l_i,\sigma^a)$.  Then, following \cite{MartelliSparksYau2006}, one can argue that the associated $R$-symmetry generator (to be identified with one-half of the Reeb vector field of the Sasaki base of $\calm_{N=1}$) takes the form 
\be\label{Rsymmetry}
\calr_{N=1}=-\frac{1}{4\pi}P^A\frac{\del}{\del \phi^A}=-\frac{1}{4\pi}\left(b^i\frac{\del}{\del \varphi^i}+p_a\frac{\del}{\del \vartheta_a}\right)\,,
\ee
where $P^A$, $b^i$ and $p_a$ are some constants related by
\be\label{Pbp}
b^i\equiv P^Av_A^i\quad,\quad p_a\equiv P^AM_{Aa}\,.
\ee

The symplectic potential   corresponding to the (non-Ricci flat) canonical K\"ahler potential $k^{\rm can}_X$ introduced just above \eqref{Dflatness} can be written in the form \cite{guillemin1994,abreu2000kahler}
\be\label{FXcan}
F_X^{\rm can}(l;\sigma)=-\frac{1}{2\pi}\sum_As_A(l;\sigma)\log s_A(l;\sigma)\,,
\ee
where $s_A(l;\sigma)$ are defined in \eqref{poly}. The symplectic potential $F_X^{\rm can}(l;\sigma)$  is singular on the boundary of the polytope \eqref{poly} in exactly the correct way to lead to a smooth K\"ahler structure on the resolved space $X$ (for fixed K\"ahler moduli $\sigma^a$).
On the other hand, $F_X^{\rm can}$ can also be identified as a symplectic potential  for the entire moduli space $\calm_{N=1}$:
\be\label{calfcan}
\tilde\calf^{\rm can}_{ N=1}(l,\sigma)\equiv F_X^{\rm can}(l;\sigma)\,. 
\ee
Indeed the Hessian of $\tilde\calf^{\rm can}_{N=1}$   is homogeneous of degree $-1$, as required for the corresponding metric on $\calm_{N=1}$ to be conical. One can also check that the corresponding $R$-symmetry generator takes the form \eqref{Rsymmetry}-\eqref{Pbp} with 
\be
P^A_{\rm can}=1\quad~~~~ \text{for any}\quad A=1,\ldots,d\,,
\ee
so that $b^i_{\rm can}=\sum_A v^i_A$ and $p^{\rm can}_{a}=\sum_A M_{Aa}$. Note that $\tilde\calf^{\rm can}_{N=1}$ can be interpreted as the symplectic potential corresponding to a flat metric on the space $\calm^\Delta_{N=1}\simeq \mathbb{C}^{4+b_2(X)}$ obtained by fibering the axionic coordinates $\phi^A\simeq (\varphi^i,\vartheta_a)$  over the maximally enlarged saxionic cone $\calc^\Delta_{N=1}$ introduced in \eqref{maxCN=1}. If $Y$ is smooth and $b_6(X)=0$, $\calm^\Delta_{N=1}$ has a clear physical interpretation as  extended geometric moduli space: $\calm^{\rm ext}_{N=1}= \calm^\Delta_{N=1}\simeq \mathbb{C}^{4+b_2(Y)}$.

A  symplectic potential  corresponding to a more general $R$-symmetry generator \eqref{Rsymmetry} can be obtained by adding 
\be\label{addsmooth}
\frac{1}{2\pi}\left(s_{P_{\rm can}}\log s_{P_{\rm can}}-s_{P}\log s_{P}\right)
\ee
to  \eqref{FXcan}, where 
\be\label{sP}
\begin{aligned}
s_{P}(l;\sigma)&\equiv P^As_A (l;\sigma)=b^i l_i+p_a\sigma^a
\end{aligned}
\ee
and then $s_{P_{\rm can}}=\sum_A s_A$. Note that the combination \eqref{addsmooth} is regular on the entire maximally extended cone $\calc^\Delta_{N=1}\simeq \mathbb{R}^{4+b_2(X)}_{\geq 0}$ parametrized by the $s_A$'s provided that 
\be\label{Pcond}
P^A>0\,.
\ee
Since \eqref{calfcan} defines a flat metric on $\calm^{\Delta}_{N=1}\simeq \mathbb{C}^{4+b_2(X)}$, the sum of \eqref{calfcan} and \eqref{addsmooth} defines a smooth metric on  $\calm^{\Delta}_{N=1}$.

Consider now two symplectic potentials $\tilde\calf_{ N=1}(l,\sigma)$ and $\tilde\calf'_{ N=1}(l,\sigma)$ having the same $R$-symmetry generator \eqref{Rsymmetry}. Then, as in \cite{MartelliSparksYau2006} one can show that this happens if and only if they differ by a  homogeneous function $f_X(l,\sigma)$  of degree one (up to an irrelevant constant): $f_X(\lambda l,\lambda\sigma)=\lambda f_X(l,\sigma)$. 
From the identification $\tilde\calf_{ N=1}(l,\sigma)\equiv F_X(l;\sigma)$, we can then write the expected most general  $F_X(l;\sigma)$  in the form
\be\label{FXdec}
F_X(l;\sigma)=F_X^{\rm can}+\frac1{2\pi}s_{P_{\rm can}}\log s_{P_{\rm can}}-\frac1{2\pi}s_{P}\log s_{P}+f_X
\ee
As emphasised above, the first three terms on the r.h.s.\ define a smooth metric over the maximally extended space $\calm^{\Delta}_{N=1}$, while $f_X(l,\sigma)$ is  at least  expected to be smooth on the interior of each $\calc^X_{N=1}$ and to extend to $\calc^{\rm ext}_{N=1}\subset \calc^{\Delta}_{N=1}$. Hence $f_X$ should encode  information on the potential  phase transitions at the K\"ahler walls \cite{Witten:1996qb} separating the different K\"ahler chambers of   $\calm_{N=1}\subset\calm^{\rm ext}_{N=1}$ connected by flops.

We stress that the form \eqref{FXdec} of the symplectic potential is dictated by the expected superconformal invariance of the associated holographic EFT.  In fact, the constants $P^A\sim(b^i,p_a)$ and the explicit form of the homogeneous function $f_X$ should be completely fixed by the Calabi-Yau condition of the internal resolved space $X$, once the appropriate boundary conditions \eqref{bcond} on the dual K\"ahler potential $k_X$ are imposed.\footnote{The K\"ahler potential $k_X$ corresponding to \eqref{FXdec} can be easily obtained from \eqref{sympot}:
\be\label{kXdec}
k_X=-\frac{1}{2\pi}\sum_AM_{Aa}\sigma^a\log \left(\frac{s_A}{s_{P_{\rm can}}}\right)-\frac{1}{2\pi}p_a\sigma^a\log s_P+f_X-l_i\frac{\del f_X}{\del l_i}
\ee
Note that $p_a$ encode the violation of homogeneity: $k_X(\lambda l;\lambda\sigma)=\lambda k_X(l;\sigma)-\frac{1}{2\pi}\left(\lambda\log\lambda\right) p_a\sigma^a$.}  
In particular, we will see that the constants $P^A\sim(b^i,p_a)$ are related to the scaling dimensions of the chiral fields of the dual theory.  These aspects will become explicit when we discuss our  examples in sections \ref{sec:Y12model} and \ref{sec:Q111model}.  For the time being, let us just   note that in the conical limit $\sigma^a\rightarrow 0$  $F_X(l;\sigma)$ reduces to the form of conical potentials $F^{\rm cone}_X(l)$ identified in \cite{MartelliSparksYau2006} and $\frac{b^i}{2\pi}\frac{\del}{\del\varphi^i}$  to the associated Reeb vector. In this limit,  $r$ defined by 
\be\label{asympt_r}
\frac12r^2=\frac{\ell_{\text{\tiny P}}}{2\pi}\,b^il_i
\ee 
can be identified with the asymptotic conical  radial coordinate.

\subsubsection{Back to  \texorpdfstring{$N\gg 1$}{N>>1}}
\label{sec:MN>>1}

We can now go back to general $N\gg 1$. By adapting the notation introduced for the $N=1$ case, the dual saxionic variables $(l^I_i,\sigma^a)$ can take values in the chain of increasing dual saxionic cones $\calc^X\subset \calc^{\rm ext}\subset \calc^\Delta$ obtained by fibering $N$ copies of the polytope \eqref{poly} over $\mathbb{K}_X\subset \mathbb{K}^{\rm ext}_\Delta\subset A_+(\Delta)$ respectively. By fibering over $\calc^X\subset \calc^{\rm ext}\subset \calc^\Delta$ the axionic variables $(\varphi^i_I,\vartheta_a)$ one gets the field spaces  $\calm\subset \calm^{\rm ext}\subset \calm^\Delta$, respectively.

 The  EFT is expected to exists at least on each K\"ahler chamber $\calm$ and, up to possible phase transition at the K\"ahler walls, on $\calm^{\rm ext}$ (which coincides with $\calm^\Delta$ if $b_6(X)=0$ and $Y$ is smooth).   At the leading large-$N$ order and up to possible non-perturbative corrections if $b_6(X)\neq 0$,  the holographic EFT is defined by \eqref{Ftilde} with $F_X(l;\sigma)$ of the form \eqref{FXdec}. The corresponding $R$-symmetry generator is  
\be\label{Rsymmetry2}
\calr=-\frac{1}{4\pi}\left(b^i\sum_I\frac{\del}{\del \varphi^i_I}+N p_a\frac{\del}{\del \vartheta_a}\right)\,,
\ee
while the dilation generator is
\be\label{dilDop}
\cald= \sum_I l^I_i\frac{\del}{\del l^I_i}+\sigma^a\frac{\del}{\del \sigma^a}\,.
\ee
Furthermore, the dilaton $\tau$ of the EFT, as defined in appendix \ref{app:Kcone}, is given by 
\be\label{tau2}
\frac12\tau^2=\frac1{2\pi}\left(b^i\sum_I l^I_i+N p_a \sigma^a\right)~,
\ee
which is proportional to the moment map of the $R$-symmetry.

Using \eqref{FXdec}, one can easily check  that the constants $P^A\sim (b^i,p_a)$ completely determine the violation of homogeneity of $\tilde\calf(L,\Sigma)$, in the sense that
\be\label{nonhomcalf}
\tilde\calf(\lambda L,\lambda\Sigma)=\lambda \tilde\calf(L,\Sigma)-\frac{1}{2\pi}(\lambda\log\lambda) \sum_I s_P(L^I,\Sigma)
\ee
Since the violation of homogeneity is due to a linear combination of linear multiplets, it does not affect the effective Lagrangian $\int\d^4\theta \tilde\calf(L,\Sigma)$, which is then scale invariant.

Now we can go back to the chiral formulation with K\"ahler potential:
\be\label{calktoric}
\begin{aligned}
\calk&=\tilde\calf-2L^I_i \Re z^i_I-2\Sigma^a\Re\rho_a=\frac{1}{2\pi}\sum_I s_P(L^I,\Sigma)
\end{aligned}
\ee
where $L^I_i$ and $\Sigma^a$ must be considered as the functions of $\Re z^i_I$ and $ \Re\rho_a$ that are obtained by inverting the relations
\be\label{invchiral}
\begin{aligned}
\Re z^i_I=\frac12\frac{\del \tilde\calf}{\del L_i^I}=&\,-\frac{1}{4\pi} b^i\left[\log s_P(L^I,\Sigma)+1\right]\\
&- \frac{1}{4\pi} \sum_A v^i_A\log\left(\frac{s_A(L^I,\Sigma)}{s_{P_{\rm can}}(L^I,\Sigma)}\right)+\frac12\frac{\del f_X(L^I,\Sigma)}{\del L_i^I}\\
 \Re\rho_a=\frac12\frac{\del \tilde\calf}{\del\Sigma^a}=&-\frac{1}{4\pi} p_a\sum_I\left[\log s_P(L^I,\Sigma)+1\right]\\
 &- \frac{1}{4\pi} \sum_A M_{Aa}\sum_I\log\left(\frac{s_A(L^I,\Sigma)}{s_{P_{\rm can}}(L^I,\Sigma)}\right)+\frac12\sum_I\frac{\del f_X(L^I,\Sigma)}{\del\Sigma^a}
\end{aligned}
\ee
Eqs.~\eqref{tau2} and \eqref{calktoric} are indeed in agreement with the general relation $\calk|_{\theta,\bar\theta=0}=\frac12\tau^2$. 

The K\"ahler potential \eqref{calktoric} is  invariant under constant shifts of the immaginary parts of $z^i_I$ and $\rho_a$. One may regard the $\frac{1}{2\pi}L^I_i$ as the current supermultiplets generating  imaginary shifts of the $z^i_I$'s, associated with the toric structure of the Calabi-Yau space $X$. However, as emphasised around  \eqref{truetoric}, the discrete gauge symmetry group ${\rm S}_N$ reduces the actual physical {\em toric} symmetry group to $U(1)^4$ associated with the current supermultiplets \eqref{truetoric}, which shifts all $z^i_I$ with fixed $i$ and different $I$ by the same constant. We will instead dub the  imaginary shifts of $\rho_a$'s, which correspond to shifts of the M-theory gauge six-form periods, as {\em Betti} symmetries. In fact, only $b_2(Y)\subset b_2(X)=b_2(Y)+b_6(X)$ of the Betti symmetries are  exact, while the remaining $b_6(X)$ are  broken by non-perturbative corrections. Borrowing the terminology used for the baryonic symmetries of the IIB models, we will sometimes call the latter {\em `anomalous'} Betti symmetries and the former {\em exact} or {\em `non-anomalous'} Betti symmetries.

To summarize, in the symplectic/vector multiplet formulation the holographic EFT  is  specified by the constants $P^A\sim(b^i,p_a)$ and the homogeneous function $f_X(l,\sigma)$. The constants $b^i$  can be determined without explicit knowledge of the metric by considering the asymptotic conical structure of the metric and using the extremization principle of \cite{MartelliSparksYau2006}. It would be important to have an analogous procedure to determine $p_a$. The remaining information on the holographic EFT  would then be contained in $f_X(l,\sigma)$. 
If $b_6(X)= 0$, the holographic EFT can be extended to the entire $\calm^\Delta=\calm^{\rm ext}$ and is expected to be exact at leading large-$N$ order. On the other hand, if  $b_6(X)\neq 0$ the holographic EFT apparently makes sense only on $\calm^{\rm ext}$ (which is a strict subset of $\calm^\Delta$) and it is certainly valid only in some perturbative sense, because of the presence of non-perturbative effects  generated by M5-brane instantons. 

Below we will compute these quantities in our explicit examples, both of which have $b_6(X)=0$, while explicit models with $b_6(X)\neq 0$ will be further studied elsewhere.

\subsection{Regime of validity}
\label{sec:regime}
 
The symplectic formulation allows for a more explicit characterization of the regime of validity of the holographic EFT. As usual, this is  valid for energy scales  $E$ which are much smaller then  mass  $m_*$ of the lightest massive `resonance'. However, by conformal invariance, $m_*$ is not a fixed mass scale but should rather be determined by some combination of the VEVs of the scalar operators, which determine the spontaneous breaking of the conformal symmetry. Since $m_*$ should scale with weight one, it must be proportional to $\frac{1}{2}\langle \tau^2\rangle$, which can be identified with the squared  effective dilaton decay constant $f^2_{\rm eff}$. In explicit examples, the proportionality factor may be determined by a direct calculation, but we will not attempt to do that. However, 
 a simple guess of the proportionality factor may be obtained by applying the `one-scale-one-coupling' rule -- see for instance \cite{Panico:2015jxa} -- which gives $m_*\simeq g^2_* f^2_{\rm eff}$, where $g_*$ represents the typical coupling governing the perturbative expansion. In our case we may  set  $g^2_*\sim N^{-r}$ for some  $r>0$.  Under these assumptions, the condition $E/m_*\ll 1$ translates into
\be\label{EFTregime}
\frac{E}{ \frac12\langle\tau^2\rangle g^2_*}\sim \frac{ 2\pi \,E\, N^r}{b^i\sum_I \langle  l^I_i\rangle+N p_a \langle\sigma^a\rangle}\ll 1\,. 
\ee

An estimate of $g_*$ can be obtained by passing through the  AdS$_4$ effective action. In units $L=1$, its prefactor $M_4^2\sim N^\frac{3}{2}$ may be identified with $g_*^{-2}$, which leads to $g_*^{2}\sim N^{-\frac32}$ and $r=\frac32$. It would be important to refine or confirm this estimate, but in the present paper we will not try to do this, leaving $r$ unspecified. When \eqref{EFTregime} is not satisfied higher derivative corrections, suppressed by negative powers of $m_*$, may be relevant.  Instead, in the above simplified scheme, radiative corrections are proportional to positive powers of $g_*^2$ and are then subleading at large-$N$. 

In addition to the condition \eqref{EFTregime}, one also needs to take into account that, if two or more M2-branes coincide at a regular point of the internal Calabi-Yau, the low-energy theory should include a  maximally supersymmetric  SCFT, coupled to the rest of the EFT by irrelevant operators. This happens at the orbifold loci of the moduli space, \emph{i.e.} the fixed points of the discrete gauge group ${\rm S}_N$. Due to the `accidental' enhanced supersymmetry of this localised sector, we do not expect associated corrections to our two-derivative EFT. In principle, higher derivative corrections could become relevant as soon as the energy scale $\sqrt{E}$ becomes of the same order of the distance between two M2-branes on the internal Calabi-Yau. However, the accidental maximal supersymmetry may protect some quantities computed from the two-derivative EFT even if the distance between the M2-branes is not so large. It would be important to make these qualitative arguments more precise.

\section{Effective chiral operators}
\label{sec:chiralop}

In this section we study the chiral operators of the toric models introduced in the previous section. The form of the $R$-symmetry generator \eqref{Rsymmetry2}  implies that the chiral fields $e^{-2\pi z^i_I}$ and $e^{-2\pi\rho_a}$ have $R$-charges $\frac12 b^i$ and $\frac12 N p_a$, respectively. These $R$-charges must equal the corresponding scaling dimensions. Hence
\be\label{zrhoscaling}
\Delta(e^{-2\pi z^i_I})=\frac12 b^i~,\quad \Delta(e^{-2\pi\rho_a})=\frac12 N p_a\,.
\ee
These scaling dimensions  can indeed be  verified more directly by  using \eqref{invchiral}, the degree-one homogeneity of $f_X$  and the fact that the linear multiplets $L^i_I,\Sigma^a$ have scaling dimension 1. 

Note that the scaling dimensions \eqref{zrhoscaling} do not necessarily satisfy the usual positivity bounds for the scaling dimensions of  scalar operators. The point is that, generically,  the chiral fields $e^{-2\pi z^i_I}$ and $e^{-2\pi\rho_a}$ 
are  not globally well defined on the moduli space and then  cannot be considered as low-energy realizations of the SCFT chiral operators. Furthermore, good operators must be gauge invariant under the ${\rm S}_N$ discrete gauge symmetry.    
It is then natural to consider chiral operators of the form:    
\be\label{mnchiral}
\calo_{{\bf m},{\bf n}}\equiv e^{-2\pi\, \langle{\bf m}^{I},  {\bm z}_{I}\rangle}|_{\rm Sym}\,e^{-2\pi \langle {\bf n},{\bm \rho}\rangle }\,.
\ee 
with $\langle{\bf m}^{I} , {\bm z}_{I}\rangle\equiv m^I_iz^i_I$ and $\langle {\bf n},{\bm \rho}\rangle\equiv n^a\rho_a$, where we have introduced the lattice points 
\be 
{\bf m}^I=(m^I_1,\ldots,m^I_4)\in M_{\mathbb{Z}}\simeq \mathbb{Z}^4\quad,\quad  {\bf n}\equiv(n^1,\ldots, n^{b_2(X)})\in H_6(X,Y;\mathbb{Z})\simeq \mathbb{Z}^{b_2(X)}\,.
\ee 
In \eqref{mnchiral}  $|_{\rm Sym}$ indicates the function that is obtained by  symmetrizing under the exchange of the M2-brane positions ${\bm z}_I\in N_{\mathbb{R}}+\ii N_{\mathbb{R}}/N_{\mathbb{Z}}$:
\be
e^{-2\pi\, \langle{\bf m}^{I},  {\bm z}_{I}\rangle}|_{\rm Sym}\equiv \frac{1}{N!}\sum_{g\in {\rm S}_N}e^{-2\pi\, \sum_I\langle{\bf m}^{I},  {\bm z}_{g^{-1}(I)}\rangle}=\frac{1}{N!}\sum_{g\in {\rm S}_N}e^{-2\pi\, \sum_I\langle{\bf m}^{g(I)},  {\bm z}_{I}\rangle}~.
\ee

The operators \eqref{mnchiral} are labelled by the `Betti vector' ${\bf n}\in H_6(X,Y;\mathbb{Z})$ and the symmetrized lattice element
\be\label{Ocharges}
{\bf m}^{\rm S}\equiv({\bf m}^1,\ldots ,{\bf m}^N)_{\rm S}\in {\rm Sym}^N M_{\mathbb{Z}}\equiv M_{\mathbb{Z}}^N/{\rm S}_N
\ee
Note that not all these quantized numbers correspond to conserved charges. First of all, as already discussed, because of the ${\rm S}_N$ discrete gauge symmetry only a $U(1)^4$ toric symmetry is preserved. The corresponding charges of the operators \eqref{mnchiral} are given by 
\be\label{toricharges}
{\bf m}^{\rm toric}=\sum_I{\bf m}^I\in M_{\mathbb{Z}}\,.
\ee
Secondly, only $b_2(Y)\subset b_2(X)$ Betti symmetries are exact, while the others
$b_6(X)$ (the `anomalous' ones) are conserved only at the perturbative level but are broken by non-perturbative effects. In other words, splitting ${\bf n}$ into $\hat{\bf n}\equiv (\hat n^1,\ldots, \hat n^{b_6(X)})$ and $\tilde{\bf n}\equiv (\tilde n^1,\ldots, \tilde n^{b_3(Y)})$, once the non-perturbative corrections are included only $\tilde{\bf n}$ survive as exact conserved Betti charges which can be used to organize chiral fields. To avoid this subtlety, from now on we restrict to `non-anomalous' models, that is we assume that
\be\label{b60cond}
b_6(X)=0\quad~~~~~~~~~~~~~~~~\text{(henceforth)}\,.
\ee
As discussed in previous sections, this condition also ensures that we can identify the extended moduli space $\calm^{\rm ext}$ with the maximal extension $\calm^{\Delta}$ which, only if $b_6(X)=0$, does not contain orbifold phases in the corresponding GKZ decomposition.  In order to lighten the notation, we will keep using the notation valid in the general case, \emph{e.g}.\ writing $\sigma^a$ instead of more explicit $\tilde\sigma^\rho$,  implicitly assuming \eqref{b60cond}. Most of the following discussions can be adapted to the more general $b_6(X)\neq 0$ case, which will be considered elsewhere.

One can understand the behaviour of VEVs of the operators \eqref{mnchiral} along the moduli space by using  \eqref{invchiral}.  Assuming a regular enough $f_X$, $\Re z^i_I$ and $\Re\rho_a$ diverge as  one approaches the  union of the sets $\{s_A=0\}$ of  the extended moduli space $\calm^{\rm ext}$. For a fixed choice of the K\"ahler moduli $\sigma^a$, one may reach this boundary by moving the $I$-th coordinates $l^I_i$ onto the $A$-th facet $s_A(l^I,\sigma)=0$ of the polytope \eqref{poly}. In the limit $s_A(l^I,\sigma)\rightarrow 0$,  \eqref{invchiral} reduces to
\be\label{zrhodiv}
\Re z^i_I\simeq  -\frac{1}{4\pi} v^i_A\log s_A(l^I,\sigma)+\text{(regular)}\quad,\quad  \Re\rho_a\simeq   -\frac{1}{4\pi} M_{Aa}\log s_A(l^I,\sigma)+\text{(regular)}
\ee  
Now, the chiral operators \eqref{mnchiral} are globally well defined on $\calm^{\rm ext}$ only if they do not diverge on $\bigcup_{A,I}\{s_A(l^I,\sigma)=0\}$. From \eqref{zrhodiv} it follows that this condition requires
\be\label{chargecond}
s_A({\bf m}^I;{\bf n})\equiv \langle {\bf m}^I, {\bf v}_A\rangle+ \chi_A({\bf n})\geq 0 \quad~~~~ \forall I=1,\ldots,N\,,
\ee
where $\chi_A({\bf n})\equiv M_{Aa}n^a$. 
Comparing the condition \eqref{chargecond} with \eqref{poly}, it is clear that it  can be interpreted as the requirement that the integral vectors ${\bm l}^I={\bf m}^I \in M_\mathbb{Z}$ belong to the polytope \eqref{poly} corresponding to the quantized K\"ahler moduli $\sigma^a=n^a$. 

Recalling the definition \eqref{Pbp} of the constants $P^A$, the scaling dimensions of the chiral operators \eqref{mnchiral} are given by
\be\label{scalingO}
\begin{aligned}
\Delta(\calo_{{\bf m},{\bf n}})&=\frac12 b^i \sum_I m_i^I +\frac12 N p_a n^a\equiv \frac12 P^A\left(\sum_I m^I_i v^i_A+N M_{Aa}n^a\right)\\
&\equiv \frac12 P^A\sum_I s_A({\bf m}^I;{\bf n}) \equiv \frac{1}{2}\sum_I  s_P({\bf m}^I;{\bf n})\,.
\end{aligned}
\ee
By taking into account \eqref{Pcond}, we see that the condition \eqref{chargecond} ensures the positivity of the scaling dimensions of the chiral operators \eqref{mnchiral}.

The condition \eqref{chargecond} can also be understood from a purely holomorphic perspective. The complex coordinates $z^i$ only cover the open subset $X^{\circ}\equiv X\backslash\bigcup_A\cald^A \simeq (\mathbb{C}^*)^4$, where $\cald^A$ are the toric divisors associated to the fan, and any function $e^{-2\pi \langle{\bf m}, {\bm z}\rangle}$, with ${\bf m}\in M_\mathbb{Z}$, extends to a meromorphic function on $X$, whose zeros and poles are located on the divisor ${\rm div}(e^{-2\pi \langle{\bf m}, {\bm z}\rangle})\equiv\langle{\bf m}, {\bf v}_A\rangle \cald^A$. In particular, it  has a pole at $\cald^A$ if $\langle{\bf m}, {\bf v}_A\rangle<0$, which is possible only if ${\rm div}(e^{-2\pi \langle{\bf m}, {\bm z}\rangle})$ is not effective. Analogously, the combination $e^{-2\pi\langle{\bf m}^I,{\bm z}_I\rangle}|_{\rm Sym}$ appearing in \eqref{mnchiral} extends to a meromorphic function  on $X^N$, with zeros and poles along $\langle {\bf m}^I,{\bf v}_A\rangle \cald^A_I$, where $\cald^A_I\subset X^N$ is the pull-back of $\cald^A\subset X$ under the projection of $X^N$ to the $I$-the copy of $X$.

Suppose first that $n^a=0$. 
In this case the condition \eqref{chargecond}  corresponds to requiring that the divisor $\langle{\bf m}^I, {\bf v}_A\rangle \cald^A$ is effective for any index $I$.  Then $e^{-2\pi\langle{\bf m}^I,{\bm z}_I\rangle}|_{\rm Sym}$ extends to a holomorphic function with no poles on the entire  $X^N$ and zeroes along $\langle {\bf m}^I,{\bf v}_A\rangle \cald^A_I$. Hence it can be regarded as a well defined operator over the entire moduli space. 

Consider next the case  $n^a\neq 0$.  Then $e^{-2\pi \langle {\bf n},{\bm \rho}\rangle}$ takes values in the fiber of  ${\rm Sym}^N\calo_X(-n^aD_a)$, see section \ref{sec:HEFTchiral}. But since we can choose $D_a=M_{Aa}\cald^A$, $e^{-2\pi \langle {\bf n},{\bm \rho}\rangle}$ can be considered as globally defined on the open patch $\calu^\circ={\rm Sym}^N X^\circ$, where  $\calo_X(-n^aD_a)$ is trivial. On the other hand, in a different patch $\calu'$ which includes (part of) the divisor $n^aD_a$, one must use different local coordinates  $e^{-2\pi\rho_a'}$ related to $e^{-2\pi\rho_a}$ by a transition function of ${\rm Sym}^N\calo_X(-D_a)$. We can choose these local trivializations so that  $e^{-2\pi \langle {\bf n},{\bm \rho}\rangle}=\prod_I\zeta_{\bf n}(z_I)e^{-2\pi \langle {\bf n},{\bm \rho}'\rangle}$, where $\zeta_{\bf n}$ is a local trivialization over $\calu'$ of a  meromorphic section of  $\calo_X(n^aD_a)$ such that ${\rm div}(\zeta_{\bf n})=\chi_A({\bf n})\cald^A$.
\footnote{Explicitly, we can set $\zeta_{\bf n}=\prod_{A}(\zeta^A)^{n^aM_{Aa}}$ where $\zeta^A(z)$ is a section of $\calo_X(\cald^A)$, \emph{i.e.}  ${\rm div}(\zeta^A)=\cald^A$.} On the local patch $\calu'$,  the operator \eqref{mnchiral} takes the form
 \be
 \calo_{{\bf m},{\bf n}}= e^{-2\pi\, \langle{\bf m}^{I},  {\bm z}_{I}\rangle}|_{\rm Sym}\prod_I\zeta_{\bf n}(z_I)\,e^{-2\pi \langle {\bf n},{\bm \rho}'\rangle }~.
 \ee
 The prefactor is the symmetrized product of $N$ functions 
 \be 
 f_{{\bf n},{\bf m}^I}(z)\equiv \zeta_{\bf n}(z)e^{-2\pi\, \langle{\bf m}^{I},  {\bm z}\rangle}\,,
 \ee
 which define the divisors ${\rm div}(f_{{\bf n},{\bf m}^I})=s_A({\bf m}^I;{\bf n})\cald^A$ in $X$. We then explicitly see that the condition \eqref{chargecond} is equivalent to requiring that ${\rm div}(f_{{\bf n},{\bf m}^I})$ is effective. Hence each $f_{{\bf n},{\bf m}^I}$ defines a regular local holomorphic function vanishing over $s_A({\bf m}^I;{\bf n})\cald^A$. Since this argument can be repeated for any other local  patch $\calu'$, we see that \eqref{chargecond} ensures that the operators \eqref{mnchiral} are globally well defined over the entire moduli space.

{\if More concretely, $e^{-2\pi \langle {\bf n},{\bm \rho}\rangle}$ may be considered as globally defined on ${\rm Sym}^N X^\circ$, since  $\calo_X(-n^aD_a)$ is trivial on it.  On the other hand,  on a different patch which includes (part of) the divisor $n^aD_a\subset \bigcup_A\cald^A$, one can use a different  $\rho_a'$, defined by   $e^{-2\pi \langle {\bf n},{\bm \rho}\rangle}=\prod_I\zeta(z_I)e^{-2\pi \langle {\bf n},{\bm \rho}'\rangle}$, where $\zeta$ is a meromorphic section of  $\calo_X(n^aD_a)$ defining the divisor $n^aD_a$. The prefactor of $e^{-2\pi \langle {\bf n},{\bm \rho}'\rangle}$ in  \eqref{mnchiral} then becomes the symmetrized $N$-product  of functions of the  form $\zeta(z_I)e^{-2\pi \langle{\bf m}^I, {\bm z}_I\rangle}$ (no sum over $I$), whose zeros and poles define the divisor $\langle{\bf m}^I,{\bf v}_A\rangle\cald^A+n^aD_a\equiv s_A({\bf m}^I;{\bf n})\cald^A$. The condition \eqref{chargecond} precisely implies that this divisor is effective and then that $\zeta(z_I)e^{-2\pi \langle{\bf m}^I, {\bm z}_I\rangle}$ (no sum over $I$) extends to a holomorphic section of $\calo(n^aD_a)$. \fi}

Note also that the condition \eqref{chargecond} implies that the class $n^a[D_a]=\chi_A({\bf n})[\cald^A]= s_A({\bf m}^I; {\bf n})[\cald^A]$ contains the effective toric representative $s_A({\bf m}^I;{\bf n})\cald^A$. Hence, it  requires that $n^a[D_a]$ belongs to $A_+(\Delta)$ as defined in \eqref{extKms}. As we have discussed in section \ref{sec:toric}, in `non-anomalous' models, $A_+(\Delta)\equiv \mathbb{K}_\Delta^{\rm ext}$ is decomposed into the GKP fan of K\"ahler cones. Hence the vector ${\bf n}$ of Betti charges  takes values in the closure of at least one of the K\"ahler cones  associated with the base $Y$. This K\"ahler cone appears so far unrelated to the K\"ahler cone identified by the K\"ahler moduli $\sigma^a$. In section \ref{sec:semi}  we will see that the portion of moduli space corresponding to a certain K\"ahler cone $\mathbb{K}_X$ is in fact naturally probed by the operators $\calo_{{\bf m},{\bf n}}(x)$ with ${\bf n}\in \mathbb{K}_X$. This is a physical manifestation of the Kempf-Ness theorem, which relates holomorphic (GIT) quotients and symplectic quotients \cite{Thomas2005NotesOG}.

One can make the global properties of the operators \eqref{mnchiral} manifest by using the $d$ homogeneous coordinates $Z_A$ of the parent  GLSM, see section \ref{sec:symp}. Let us set $N=1$ for notational simplicity (the generalization to $N>1$ is obvious). We can then `homogenize'  the section of the line bundle $\calo(n^aD_a)$ identified by $e^{-2\pi \langle {\bf m}, {\bm z}\rangle}f_{{\bf m},{\bf n}}(z)$ into 
\be\label{projsec}
\prod^d_{A=1}(Z_A)^{s_A({\bf m},{\bf n})}\,.
\ee
This combination is not gauge invariant, having charges $n^a$ under the GLSM gauge group. However we can add $b_2(Y)$ homogeneous coordinates $X_a\in\mathbb{C}^*$ of vanishing scaling dimension and of charge $-\delta_{ab}$ under the $b$-th $U(1)$ gauge group factor, and construct the gauge invariant operator
\be
\prod^d_{A=1}(Z_A)^{s_A({\bf m},{\bf n})}\prod^{b_2(Y)}_{a=1}(X_a)^{n^a}= \prod^d_{A=1} (Z_A)^{\langle {\bf m}, {\bf v}_A\rangle}\prod^{b_2(Y)}_{a=1}\Big[X_a\prod^d_{A=1} (Z_A)^{M_{Aa}}\Big]^{n^a}\,.
\ee
This operator can be identified with \eqref{mnchiral} for $N=1$, that is $\calo_{{\bf m},{\bf n}}\equiv e^{-2\pi\langle {\bf m}, {\bm z}\rangle}e^{-2\pi\langle {\bf n},{\bm \rho}\rangle}$, 
by using \eqref{zZrelation} and setting
\be
e^{-2\pi\rho_a}\equiv X_a\prod^d_{A=1} (Z_A)^{M_{Aa}}\,.
\ee
Alternatively, since $X_a \in \bC^*$ one may gauge-fix $X_a=1$ and use $Z_A$ to parametrize  $\calm^{\rm ext}_{N=1}\simeq \mathbb{C}^d$.  In this gauge, $\calo_{{\bf m},{\bf n}}$ can be identified with \eqref{projsec}, with no residual gauge symmetry.

Recalling \eqref{scalingO}, this description allows us to  interpret the constants $\frac12 P^A$ as the scaling dimensions associated with  the  $Z_A$'s.  These may be regarded as the elementary $N=1$ Betti operators corresponding to the non-compact toric divisors $\cald^A$. These scaling dimensions become $\frac12 N P^A$ once we go back to $N\gg 1$.

\section{Semiclassics at large charge}
\label{sec:semi}

As in the previous section, we will restrict ourselves to  `non-anomalous' models (that is, with $b_6(X)=0$), whose geometric moduli space is not affected by non-perturbative corrections.  The EFT chiral operators  \eqref{mnchiral}  are then expected to correspond to chiral operators of the microscopic SCFT. 
In turn, by the state-operator correspondence, they should correspond to states in the radial quantization  
of the SCFT. In this section we will construct semiclassical solutions of the  holographic EFT which well describe such states for large scaling dimensions \eqref{scalingO}. These semiclassical states may provide a starting point for a systematic investigation of  the  SCFT structure along the lines of \cite{Hellerman:2017veg,Hellerman:2017sur,Hellerman:2018xpi,Hellerman:2021yqz}.

\subsection{Semiclassical states in \texorpdfstring{$\mathbb{R}^3$}{R3}}
\label{sec:statesR3}

We start by using the holographic EFT described in terms of chiral multiplets. In order to lighten some of the following formulas, we will collectively denote them by\footnote{Since in this section we focus on `non-anomalous' models, there are no $\hat\rho_\alpha$ chiral fields and we will keep using $\rho_a$ (instead of $\tilde\rho_\sigma$) to denote the $b_2(Y)$ chiral fields. Hence, the choice of Greek indices $\alpha,\beta,\ldots$ to label the  collective chiral fields $\phi_\alpha$ should not cause confusion.}
\be
\phi_\alpha\equiv (z^i_I,\rho_a)\ .
\ee
We can then interpret  the holomorphic functions \eqref{mnchiral} as linear combinations of chiral operators of the form 
\be\label{coO}
\calo_{\bf q}(x)\equiv \calo_{{\bf m},{\bf n}}(x)\equiv  e^{-2\pi  \langle {\bf q}, {\bm \phi}(x)\rangle}|_{\rm Sym}\ ,
\ee
where $\langle {\bf q}, {\bm \phi}\rangle\equiv q^\alpha \phi_\alpha$ and $q^\alpha\equiv (m^{I}_i,n^a)$. As in previous sections, $|_{\rm Sym}$ denotes the symmetrization with respect to the discrete ${\rm S}_N$ gauge group acting on the $z^i_I$ fields and makes $\calo_{\bf q}(x)$ gauge invariant. Correspondingly, the operators \eqref{coO} are classified by symmetrized lattice elements ${\bf q}^{\rm S}=({\bf m}^{\rm S},{\bf n})\in {\rm Sym}^N M_{\mathbb{Z}}\times H_6(X,Y;\mathbb{Z})$ (with ${\rm Sym}^N M_{\mathbb{Z}}\equiv M_{\mathbb{Z}}^N/{\rm S}_N $), which define also the  four conserved toric charges $m^{\rm toric}_i$ introduced in \eqref{toricharges} and the  ${b_2(Y)}$ (exact) Betti charges $n^a$. 

We start by selecting a generic vacuum along $\calm^{\rm ext}$,  corresponding to a choice of VEVs $\langle \phi_\alpha\rangle=\phi^{\infty}_\alpha$, up to an ${\rm S}_N$ gauge transformation. Of course, we  expect the VEV of the chiral operators \eqref{coO} to be constant and given by
\be\label{OVEV}
\langle \calo_{\bf q}(x)\rangle=e^{-2\pi  \langle {\bf q}, {\bm \phi}^\infty\rangle}|_{\rm Sym}\ .
\ee
It is nevertheless  instructive to explicitly compute \eqref{OVEV} by a semiclassical saddle-point evaluation of the path integral\footnote{The following discussion may be applied to any superconformal EFT described by a K\"ahler potential $\calk(\phi,\bar\phi)$ which depends only on $\phi+\bar\phi$.} 
\be\label{pathOvev}
\begin{aligned}
\langle \calo_{\bf q}(x_0)\rangle&=\int \cald\mu\, \calo_{\bf q}(x_0)\, e^{\ii S}=\int \cald\mu\, e^{-2\pi  \langle {\bf q}, {\bm \phi}(x_0)\rangle}|_{\rm Sym} \,e^{\ii \int\d^3 x\d^4\theta \,\calk(\phi,\bar\phi)}\\
&=\frac1{N!}\sum_{g\in {\rm S}_N}\int \cald\mu\, e^{-2\pi  \langle {\bf q}^{g}, {\bm \phi}(x_0)\rangle} \,e^{\ii \int\d^3 x\d^4\theta \,\calk(\phi,\bar\phi)}\,,
\end{aligned}
\ee
where ${\bf q}^g$ is the image of ${\bf q}$ under $g\in {\rm S}_N$ and  $\cald\mu$ is the path-integration measure, normalized in such a way that $\langle{\bf 1}\rangle=1$. $\calk$ is the K\"ahler potential of our holographic EFT, which depends on $(\phi_\alpha,\bar\phi_\alpha)$ only through their real parts $\Re\phi_\alpha$. 

Note that in  in the path integral we are implicitly modding out the discrete ${\rm S}_N$ gauge symmetry. Working in the `upstairs' covering field space, one should sum over all possible paths compatible with the ${\rm S}_N$ gauge symmetry, hence including the possible `${\rm S}_N$-twisted' sectors. 
However twisted sectors will not contribute to the following  calculation at the leading semiclassical level. Hence, for our purposes, we can loosely regard the last line of \eqref{pathOvev} as the sum of standard path integrals over the covering field space.

In order to investigate the saddle-point contribution to the path integral, it is convenient to go to the dual formulation \cite{Lindstrom:1983rt,deBoer:1997kr,Intriligator:2013lca} in terms of vector/linear multiplets 
\be
L^\alpha\equiv (L^I_i,\Sigma^a)=  \frac{\ii}{4}D\bar D V^\alpha= l^\alpha+\ldots - \frac12\epsilon^{\mu\nu\rho}\theta\gamma_\mu\bar\theta F^\alpha_{\mu\nu}+\ldots
\ee 
with Lagrangian $\int\d^4\theta \tilde\calf(L)$. The kinetic function $\tilde\calf$ has been introduced in \eqref{Ftilde} and in our condensed notation is given by $\tilde\calf=\calk+2 L^\alpha\Re\phi_\alpha$, with 
\be\label{Lphirel}
L^\alpha\equiv (L^i,\Sigma^a)=-\frac12\frac{\del \calk}{\del \Re\phi_\alpha}\quad\Leftrightarrow\quad \Re\phi_\alpha=\frac12\frac{\del \tilde\calf}{\del L^\alpha}
\ee
For our purposes we must also be careful about boundary terms when we perform the duality. Being interested in the bosonic bottom components of chiral operators, we can just focus on the bosonic terms of the action. Let us then recall how the chiral/vector multiplet duality works at the bosonic level. We start from the bosonic terms in the action   $\int\d^3 x\d^4\theta\,\calk(\phi,\bar\phi)$:
\be\label{boschiral1}
-\frac12\int\calg^{\alpha\beta}\d\phi_\alpha\wedge *\d\bar\phi_\beta\ ,
\ee 
with $\calg^{\alpha\beta}\equiv\frac12\frac{\del^2\calk}{\del\Re\phi_\alpha\del\Re\phi_\beta}$. We can now use the dual saxions $l^\alpha$ instead of $\Re\phi_\alpha$, and dualize the axions $\Im\phi_\alpha\simeq \Im\phi_\alpha+1 $ to vectors. This is done by a standard trick, substituting \eqref{boschiral1} with the Lagrangian 
\be\label{boschiral2}
-\frac12\int \calg_{\alpha\beta}(\d l^\alpha\wedge *\d l^\beta+F^\alpha\wedge *F^\beta)-\int\d\Im\phi_\alpha\wedge F^\alpha\ ,
\ee
where $F^\alpha$ is an arbitrary 2-form, $l^\alpha$ and $\Im\phi_\alpha$ are independent bosonic fields, and 
\be\label{calgalpha}
\calg_{\alpha\beta}=-\frac12\frac{\del^2\tilde\calf}{\del l^\alpha\del l^\beta}
\ee
is the inverse of $\calg^{\alpha\beta}$. The Lagrangian \eqref{boschiral2} is quadratic in $F^\alpha$, which can then be integrated out exactly in the path integral. Plugging their equations of motion
\be\label{dualphi}
\d\Im\phi_\alpha=-\calg_{\alpha\beta}*F^\beta
\ee
back into \eqref{boschiral2} and taking into account (the lowest component of) \eqref{Lphirel}, the original action \eqref{boschiral} is reproduced. Notice that \eqref{dualphi} fixes $\Im\phi_\alpha$ in terms of $F^\alpha$, up to a constant zero mode.

Instead, to obtain  the dual vector multiplet description one has to integrate out $\Im\phi_\alpha$ in the path integral. If we apply  this prescription to the evaluation of \eqref{pathOvev}, it is  convenient  perform a Wick-rotation  to Euclidean $\mathbb{R}^3$. From the last line of \eqref{pathOvev} we see that the insertion of the chiral operator  $\calo_{{\bf q}}(x_0)$ implies  that we must extremise  combinations of the form
\be\label{Sins}
\begin{aligned}
&S^{\rm bos}_{\rm E}-\log e^{-2\pi\langle {\bf q},{\bm \phi}(x_0)\rangle}\\
&= \frac12\int\calg_{\alpha\beta}\left(\d l^\alpha\wedge*\d l^\beta+F^\alpha\wedge *F^\beta\right)+\ii\int\d\Im\phi_\alpha\wedge F^\alpha+2\pi q^\alpha\phi_\alpha(x_0)
\end{aligned}
\ee 
with respect to $\Im\phi_\alpha$, which gives the Bianchi identity
\be\label{monsol}
\d F^\alpha=2\pi q^\alpha\delta_3(x_0)\quad~~~~~~ \Rightarrow\quad~~~~~~ \frac{1}{2\pi}\int_{S^2} F^\alpha= q^\alpha\ ,
\ee
where $S^2$ is any two-sphere surrounding $x_0$. 
Hence, the insertion at $x_0$ of a chiral operator  corresponds, in the dual picture in terms of vector multiplets, to the presence of a monopole operator \cite{Kapustin:1999ha,Borokhov:2002ib,Borokhov:2002cg} of charges $q^\alpha\in\mathbb{Z}$. 

Substituting \eqref{monsol} into \eqref{Sins} and keeping  boundary terms, we are left with a sum of path integrals of combination of the form $e^{-S^{\rm bos}_{\rm E}-2\pi\langle {\bf q},{\bm \phi}(x_0)\rangle}$ with
\be\label{Sins2}
\begin{aligned}
&S^{\rm bos}_{\rm E}+2\pi\langle {\bf q},{\bm \phi}(x_0)\rangle\\
&= \frac12\int\calg_{\alpha\beta}\left(\d l^\alpha\wedge*\d l^\beta+F^\alpha\wedge *F^\beta\right)
+2\pi\ii\, \langle {\bf q},\Im{\bm \phi}^\infty\rangle +2\pi \langle {\bf q}, \Re{\bm\phi}(x_0)\rangle\\
&= \frac12\int\calg_{\alpha\beta}\left(F^\alpha+*\d l^\alpha\right)\wedge*\left(F^\beta+*\d l^\beta\right)+2\pi \langle {\bf q},{\bm \phi}^\infty\rangle~.
\end{aligned}
\ee
Here we have taken into account the boundary conditions $\phi_\alpha(x)\rightarrow \phi^\infty_\alpha$ for $|x|\rightarrow \infty$.

From the last line of \eqref{Sins2}, it is evident that the path integral admits a BPS saddle defined by the equations 
\be\label{BPSmonopole}
F^\alpha=-*\d l^\alpha\ .
\ee
Combined with \eqref{monsol}, this produces the Poisson equations $\nabla^2 l^\alpha=-2\pi\,q^\alpha *\delta_3(x_0)$. Given our boundary conditions, this has unique solution
\be\label{BPSsol}
l^\alpha(x)=l^\alpha_\infty+\frac{q^\alpha}{2|x-x_0|}\ .
\ee 
where $l^\alpha_\infty$ is the dual value of $\Re\phi_\alpha^\infty$. Taking into account 
\eqref{BPSmonopole}, we see that the BPS saddle-point is fully specified by  \eqref{BPSsol}. This describes a BPS monopole of charges $q^\alpha$. 

It is clear that, independently of the detailed form of the EFT, the value of \eqref{Sins2} on the BPS solution is given by $2\pi \langle {\bf q},{\bm \phi}^\infty\rangle$.  Hence,  the saddle-point evaluation of \eqref{pathOvev} eventually reproduces  the expected VEV \eqref{OVEV}. Furthermore, it is clear that the argument can be repeated for insertions of more than one chiral operators at arbitrary points, reproducing again the expected result.

In order to better interpret  these semiclassical configurations in our toric models, let us set $x_0=0$ and rewrite them in our original notation
\begin{subequations}\label{BPSsol2}
\begin{align}
 l^I_i(x)&=l^I_{i\infty}+\frac{m^I_i}{2|x|}\label{BPSsol2a}\\
\sigma^a(x)&=\sigma^a_\infty+\frac{n^a}{2|x|}\label{BPSsol2b}
\end{align}
\end{subequations}
Note that not all values of the monopole charges are allowed. Indeed, from  section \ref{sec:toric} we know that we must impose that ${\bm l}^I(x)\in\calp_{\sigma(x)}$ for any $I=1,\ldots,N$ and at any radius $x\in\mathbb{R}^3$, where $\calp_{\sigma(x)}$ is the polytope \eqref{poly}. Imposing this condition at  $|x|=\infty$ just  requires the vacuum configuration $(l^I_{i\infty},\sigma^a_\infty)$ to be acceptable. The condition then holds at any $x$ provided that the monopole/operator charges $({\bf m}^I,{\bf n})$ satisfy \eqref{chargecond}. This provides an alternative `dynamical' derivation of the conditions \eqref{chargecond}.  
 
Equation \eqref{BPSsol2b} also shows that, whatever $\sigma^a_\infty$ we choose, the semiclassical solution eventually flows to the K\"ahler cone selected by $n^a$. This means that the branch of the moduli space corresponding to a certain K\"ahler cone $\mathbb{K}_{X}$ is naturally `probed' by operators $\calo_{{\bf m},{\bf n}}$ with  ${\bf n}\in \mathbb{K}_{X}$, whose semiclassical configuration is described within the K\"ahler cone $\mathbb{K}_{X}$.  The explicit examples of sections \ref{sec:Y12model} and \ref{sec:Q111model} below will help clarifying the meaning of this observation.    

Regarding the reliability of the above semiclassical arguments, by supersymmetry we do not expect higher-derivative and quantum corrections to the classical EFT, nor the inclusion of the twisted sector, to affect \eqref{OVEV}. However, one may wonder under which conditions we can trust our EFT treatment. At a radius $|x|$, the energy scale of the solution is $E(x)\simeq \frac1{|x|}$. Even though $E(x)\rightarrow \infty$ for $|x|\rightarrow 0$, we have to recall that  our superconformal EFT is valid for energies $E$ much smaller then the highest possible cut-off  scale $m_*$, which  may be roughly identified with   $ N^{-r}\tau^2$, see section  \ref{sec:regime}. In our case, as $|x|\rightarrow 0$ the classical BPS profile of the dilaton diverges too. Indeed, by recalling \eqref{tau2} and \eqref{scalingO}, on the BPS solution we have\footnote{Note that $\Delta(\calo_{{\bf m},{\bf n}})\geq 0$ ensures that $\tau^2(x)|_{\rm BPS}\geq 0$, as should be the case.}  
\be
 \frac12\tau^2(x)= \frac12\tau^2_\infty+\frac{1}{2\pi |x|}\,\Delta(\calo_{{\bf m},{\bf n}})\ .
\ee
By taking $E(x)\simeq \frac1{|x|}$ we then see that  $E(x)/m_*\ll 1$ is satisfied by these solutions for any $x$  as long as
\be\label{largeDelta}
\Delta(\calo_{{\bf m},{\bf n}})N^{-r}\gg  1\,.
\ee
In other words, the above BPS semiclassical analysis can be a priori trusted only if the operator $\calo_{{\bf m},{\bf n}}$ we started from has large enough scaling dimensions, that is, large enough toric and/or Betti charges $({\bf m}^I,{\bf n})$. 

 One should in principle also worry that the above semiclassical solutions do not pass through the ${\rm S}_N$ fixed points on the moduli space, at which interacting  SCFT sectors should arise and higher-derivative terms of the EFT theory can in principle become relevant -- see section \eqref{sec:regime}. One may  judiciously choose the monopole charges so as to `avoid' these points. However, these interacting  SCFT sectors have enhanced supersymmetry which protects the form of the moduli-space metric and is then expected to protect the above semiclassical results as well.

\subsection{Semiclassical states on the cylinder}  
\label{sec:semstates}

It is clear that the EFT condition \eqref{largeDelta} can be satisfied even if $l^I_{i\infty}=\sigma^a_\infty=0$. Hence, the above semiclassical analysis is sensible even when the superconformal symmetry is not spontaneously broken by the vacuum. In this case, the above BPS solutions can be regarded as semiclassical states in the radially quantized SCFT in the  conformal vacuum, which are dual through the usual state/operator correspondence to the chiral operators   $\calo_{{\bf m},{\bf n}}$ (or, better, their microscopic SCFT counterpart). As in \cite{Hellerman:2017veg,Hellerman:2017sur}, we may then use  our  holographic EFT expanded about these classical solutions to study the structure of the microscopic SCFT in the large $R$-charge sector.

The self-consistency of the EFT even for unbroken conformal symmetry becomes evident if we map the above BPS solutions with $l^I_{i\infty}=\sigma^a_\infty=0$ to the cylinder $\bR\times S^2_R$
of radius $R$. The $\mathbb{R}^3$ and cylinder metrics are related by a Weyl rescaling as follows:
\be
\d s^2_{\mathbb{R}^3}=\frac{|x|^2}{R^2}\,\d s^2_{\rm cyl}\equiv e^{\frac{2\tau}{R}} \left(\d\tau^2+R^2\d s^2_{S^2}\right)\,,
\ee 
where $\tau=R \log\frac{|x|}{R}\in (-\infty,\infty)$ and $\d s^2_{S^2}$ is the metric on the two-sphere of unit radius.\footnote{The height coordinate $\tau$ on the cylinder should not be confused with the dilaton.} Under a Weyl rescaling $\d s^2=\Omega(x)^2\d \tilde s^2$, we have $\calo(x)=\Omega(x)^{-2\Delta_\calo}\tilde\calo(x)$ for an operator of scaling dimension $\Delta_\calo$. Going back to the unified notation $l^\alpha=(l^I_{i},\sigma^a)$, since linear multiplets have scaling dimension $\Delta=1$,  the solutions \eqref{BPSmonopole}-\eqref{BPSsol} on $\mathbb{R}^3$ with $l^\alpha_{\infty}=0$ and $x_0=0$, are mapped to constant homogeneous configurations on the cylinder 
\be\label{cylBPS}
l^\alpha=\frac{q^\alpha}{2R}\quad,\quad F^\alpha=\frac12 q^\alpha\d\text{vol}_{S^2}\,.
\ee
These solutions are supersymmetric and can of course be obtained as BPS saddles of the two-derivative EFT on the cylinder, which should be valid up to energy scales $E$ such that $R E \ll \Delta(\calo_{{\bf q}})N^{-r}$ -- see section \ref{sec:regime} -- as long as the condition \eqref{largeDelta} is satisfied.
By expanding the EFT around these classical BPS configurations one may study the large-charge sector of the theory  along the lines of \cite{Hellerman:2017veg,Hellerman:2015nra,Alvarez-Gaume:2016vff,Monin:2016jmo,Loukas:2016ckj}. 

As the simplest example, let us consider the two-point correlation function
\be\label{2point}
\langle \overline\calo_{{\bf q}^{\rm S}_{\rm out}}(\tau_{\rm out},\Omega_{\rm out}) \calo_{{\bf q}^{\rm S}_{\rm in}}(\tau_{\rm in},\Omega_{\rm in})\rangle
\ee
in the asymptotic limit in which $\tau_{\rm out}\rightarrow\infty$ and $\tau_{\rm in}\rightarrow-\infty$, where $\Omega$ denote the angular coordinates on $S^2$. In this limit $\calo_{{\bf q}_{\rm in}}(\tau_{\rm in},\Omega_{\rm in})|0\rangle$ and $\langle 0| \overline\calo_{{\bf m}_{\rm out}}(\tau_{\rm out},\Omega_{\rm out})$ project  onto the lowest energy eigenstate with quantum numbers ${\bf q}^{\rm S}_{\rm in}=({\bf m}^{\rm S}_{\rm in},{\bf n}_{\rm in})$ and ${\bf q}^{\rm S}_{\rm out}=({\bf m}^{\rm S}_{\rm out},{\bf n}_{\rm out})$ respectively, which are the above BPS configurations.  The two-point function then reduces to the computation of 
\be\label{amplqq}
\langle {\bf q}^{\rm S}_{\rm out}|e^{-(\tau_{\rm out}-\tau_{\rm in})H_{\rm cyl}}|{\bf q}^{\rm S}_{\rm in}\rangle
\ee
and, in the large-charge limit, can be computed from a semiclassical path integral evaluation of the EFT.     

Let us go back to the Wick rotation of the bosonic action \eqref{boschiral2} and rewrite it in cylindrical variables. Taking into account that $\calg_{\alpha\beta}(l)$ is homogeneous of degree $-1$ by the superconformal symmetry, the bosonic action on the cylinder takes the form 
\be\label{cylaction}
\begin{aligned}
\frac12\int\calg_{\alpha\beta}\left[\left( \d l^\alpha-\frac1R l^\alpha\d\tau\right)\wedge *\left( \d l^\beta-\frac1R l^\beta\d\tau\right)+\,F^\alpha\wedge*F^\beta\right]+\ii\int \d \Im\phi_\alpha \wedge F^\alpha\,.
\end{aligned}
\ee
We recall that in this action $F^\alpha$ and $\Im\phi_\alpha$ must be considered as independent unconstrained fields. As in \cite{Monin:2016jmo}, we can project on the states of definite charges by adding appropriate boundary terms.
In our EFT, the `in' and `out' states are obtained by ${\rm S}_N$-symmetrizing states of definite monopole charges ${\bf q}_{\rm in}$ and ${\bf q}_{\rm out}$, which can be selected by adding the boundary terms to the action
\be\label{cylbd}
\frac\ii2 q^\alpha_{\rm in}\int_{\tau=\tau_{\rm in}}\Im\phi_\alpha\,\d\text{vol}_{S^2}-\frac\ii2 q^\alpha_{\rm out}\int_{\tau=\tau_{\rm out}}\Im\phi_\alpha\,\d\text{vol}_{S^2}\,.
\ee
Indeed, the sum of \eqref{cylaction} and \eqref{cylbd} is linear in $\Im\phi_\alpha$, which can be integrated out. As a result the path integral localizes on field-strengths satisfying the Bianchi identity $\d F^\alpha=0$ and obeying the boundary conditions $F^\alpha|_{\tau=\tau_{\rm in}}=\frac12q^\alpha_{\rm in}\text{dvol}_{S^2}$ and $F^\alpha|_{\tau=\tau_{\rm out}}=\frac12q^\alpha_{\rm out}\text{dvol}_{S^2}$. From $\d F^\alpha=0$ one also gets the charge conservation condition  
$q^\alpha_{\rm in}=q^\alpha_{\rm out}$, which can be more directly obtained by integrating out the constant zero-mode of $\Im\phi_\alpha$. At the leading large-$N$ semiclassical level, the amplitude \eqref{amplqq} is then non-vanishing only for $q^\alpha_{\rm in}=q^\alpha_{\rm out}\equiv q^\alpha$.\footnote{This is an `accidental' conservation law,  due to the accidental extended $U(1)^{4N}$ toric symmetry of the holographic EFT formulated on the covering moduli space. Hence it  is expected to be violated by $1/N$ and higher derivative corrections, since in the complete theory there is no symmetry which guarantees the separate conservation of the monopole charges $m_i^I$. On the other hand,  the four toric charges $m^{\rm toric}_i$  and the $b_2(Y)$ Betti charges $q^\alpha$ are exactly conserved.}

We then assume that $q^\alpha_{\rm in}=q^\alpha_{\rm out}\equiv q^\alpha$. Integrating out $\Im\phi_\alpha$ by carefully taking into account the boundary terms, the amplitude can be written as a path integral of vector multiplets with fixed boundary conditions provided by the BPS configurations \eqref{cylBPS}. As in subsection \ref{sec:statesR3}, at the leading large-$N$ semiclassical order we can focus on the untwisted sector contribution and work on the `upstairs' covering moduli space:
\be\label{PIamp}
\langle {\bf q}^{\rm S},\text{out}|e^{-(\tau_{\rm out}-\tau_{\rm in})H_{\rm cyl}}|{\bf q}^{\rm S},\text{in}\rangle=\frac1{N!}\sum_{g\in{\rm S}_{N}}\int_{{\bf q}^g,\tau_{\rm in}}^{{\bf q}^g,\tau_{\rm out}}\cald\mu\, e^{-S'_{\rm E}}
\ee
where the bosonic action is
\be
S'_{\rm E}|_{\rm bos}=\frac12\int\calg_{\alpha\beta}\left[\left( \d l^\alpha-\frac1R l^\alpha\d\tau\right)\wedge *\left( \d l^\beta-\frac1R l^\beta\d\tau\right)+\,F^\alpha\wedge*F^\beta\right]\ .
\ee
Still keeping track of total derivative terms, this action can alternatively be  written as 
\be\label{ESampl}
\begin{aligned}
S'_{\rm E}|_{\rm bos}=&\,\frac12\int \calg_{\alpha\beta}\left[\left(F^\alpha+ *\d l^\alpha-\frac1R l^\alpha*\d\tau\right)\wedge *\left(F^\beta+ *\d l^\beta-\frac1R l^\beta*\d\tau\right)\right]\\
& -\int\d\Re\phi_\alpha\wedge F^\alpha+\frac{1}{4\pi R}\,b_\alpha\int\d\tau\wedge F^\alpha
\end{aligned}
\ee
where $b_\alpha= b^i$ or $b_\alpha= Np_a$ if $\alpha$ corresponds to $l_\alpha=l^I_i$ or $l_\alpha=\sigma^a$, respectively.
In deriving \eqref{ESampl} we have used \eqref{calgalpha}, the second relation in \eqref{Lphirel} and the quasi-homogeneity \eqref{nonhomcalf}, which implies 
\be\label{Glhom}
\frac{\del^2\tilde\calf}{\del l^\alpha\del l^\beta}l^\beta=-\frac1{2\pi}b_\alpha\,.
\ee

It is clear that \eqref{cylBPS} extremizes \eqref{ESampl}. Furthermore, only the last term of  \eqref{ESampl} does not vanish on-shell and gives
\be
\frac1{2R}\langle {
\bf q},{\bm b}\rangle (\tau_{\rm out}-\tau_{\rm in})\,,
\ee
with ${\bm b}=(b_1,b_2,\ldots) =(b^{i_1},\ldots,b^{i_N}, Np_a)$.
Hence, the semiclassical saddle-point evaluation of \eqref{PIamp}  gives
\be
 \langle \overline\calo_{{\bf q}}(\tau_{\rm out},\Omega_{\rm out}) \calo_{{\bf q}}(\tau_{\rm in},\Omega_{\rm in})\rangle\ \simeq\ \frac1{N!}\sum_{g\in {\rm S}_N} e^{-\frac{1}{2 R}\langle {
\bf q}^g,{\bm b}\rangle (\tau_{\rm out}-\tau_{\rm in})}=e^{-\frac{1}{2 R}\langle {
\bf q},{\bm b}\rangle (\tau_{\rm out}-\tau_{\rm in})}
 \ee for $\tau_{\rm out}-\tau_{\rm in}\rightarrow\infty$. Note that we have used the identities $\langle {
\bf q}^g,{\bm b}\rangle =\langle {
\bf q},{\bm b}^{g^{-1}}\rangle=\langle {
\bf q},{\bm b}\rangle$  for any $g\in {\rm S}_N$. This reproduces the expected result $e^{-\frac{1}{R}\Delta_{\calo_{{\bf q}}} (\tau_{\rm out}-\tau_{\rm in})}$, provided that we identify $\Delta_{\calo_{{\bf q}}}=\frac1{2}\langle {
\bf q},{\bm b}\rangle $, which indeed coincides with \eqref{scalingO}.

 We stress that these results follow just from general properties of our holographic EFTs for toric models, whose details are typically quite complicated. Furthermore, they give the expected result for a general set of charges $q^\alpha$ although they are really justified only if \eqref{largeDelta} holds. This crucially depends on the fact that we have focused on the BPS states \eqref{cylBPS}, which are expected to hold beyond the large charge and large $N$ regimes, see  the end of  section \ref{sec:statesR3}.  The above discussion provides a self-consistency check of our framework and  a starting point for  studying non-BPS states,  anomalous dimensions and higher-order correlators, as in \cite{Hellerman:2017veg,Hellerman:2017sur,Hellerman:2015nra,Alvarez-Gaume:2016vff,Monin:2016jmo,Loukas:2016ckj}. 
 Of course, in order to perform computations which go beyond the BPS sector, one should in principle restrict to the large-charge sector and use the explicit form of the holographic EFT.  We will provide concrete examples of  holographic EFTs in the following, leaving  their application to a more  in-depth study of the large charge SCFT sector to the future.

\section{M-theory interpretation of the EFT states}
\label{sec:effM5}

Since  the fields of the holographic EFT descend from the M-theory holographic description, the above EFT semiclassical states have a direct M-theory interpretation. As we are going to explain, they can be interpreted  as bound-states  of AdS giant gravitons and Euclidean M5-branes wrapping homologically non-trivial divisors. 
The holographic EFT then provides a direct link between these M-theory brane configurations and the corresponding operators in the dual SCFT.

\subsection{AdS giant gravitons}
\label{sec:AdSgiant}

 Let us first set $n^a=0$ and assume that the vacuum does not spontaneously break the conformal symmetry, \emph{i.e.} $l^I_{i\infty}=\sigma^a_\infty=0$. Since $n^a=0$, \eqref{BPSsol2b} implies the  $\sigma^a\equiv 0$ along the BPS flow. Then the internal Calabi-Yau space $X$ is conical and the solutions sit at the K\"ahler walls of the moduli space. At these points we expect the complete validity of our EFT to be questionable,  because of the appearance of light M2-branes wrapping the vanishing cycles. However, these corrections regard the sector described by the linear multiplets $\Sigma^a$, which are not activated in the solution, and are expected to decouple from the remaining multiplets $L^I_i$ as long as the condition \eqref{largeDelta} is satisfied. 
Hence, in this regime these semiclassical solutions remain sensible.   

 The M-theory interpretation of these states is more easily described in the EFT cylinder  coordinates introduced above.  Then the BPS solution \eqref{cylBPS} is
\be\label{n=0sol}
l^I_i=\frac{1}{2R}m^I_i~,\qquad F^i_I=\frac{1}{2}m^I_i\text{dvol}_{S^2}\,,
\ee
where it is natural to set $R$ equal $L\ell_{\text{\tiny P}}$ (the asymptotic AdS radius), while $\sigma^a=0$ and $F^a=0$. The brane interpretation becomes transparent by dualizing the vector fields $A^i_I$ back to scalars $\varphi^i_I\equiv \Im z^i_I$. By using the general formula \eqref{dualphi} (in Minkowski signature), and recalling   \eqref{calgalpha} and \eqref{Glhom}, the solution  \eqref{n=0sol} can be dualized to a time-dependent solution in symplectic coordinates
\be\label{n=0sol2}
l^I_i=\frac{1}{2R}m^I_i~,\qquad \varphi^i_I=-\frac{1}{4\pi R}b^i t+\alpha^i_{I}\,,
\ee
where $\alpha^i_{I}$ are arbitrary constants. 
 
The solutions \eqref{n=0sol2} describe a bunch of M2-branes  wrapping a static $S^2$ in global AdS$_4$ and 
 spinning along the internal angular coordinates $\varphi^i_I$. These configurations are analogous to those found in \cite{Martelli:2006vh} 
to describe AdS giant gravitons \cite{Grisaru:2000zn,Hashimoto:2000zp,Mandal:2006tk} on general IIB toric AdS$_5\times Y$ backgrounds.  The number of AdS giant gravitons is set by the number of non-vanishing monopole charge vectors ${\bf m}_I=(m^1_I,m^2_I,m^3_I,m^4_I)$ as $I=1,\ldots,N$. Note that in our description in terms of EFT monopole states,  we directly get a discrete spectrum of  AdS giant gravitons and of the corresponding scaling dimensions $\Delta_{\bf m}=\frac12 b^i\sum_I m^I_i $ -- see \eqref{scalingO} --  with no need of any additional geometric quantization of the classical family of probe brane solutions as in \cite{Mandal:2006tk,Martelli:2006vh}. It also automatically implements the orbit average prescription proposed  in \cite{Yang:2021kot} for probe branes. 
Furthermore, we automatically get the expected upper bound $N$ on the number of possible AdS giant gravitons, a bound which must be imposed by hand in a probe treatment. 

Our derivation directly relates these M-theory configurations to the EFT  chiral operators 
\be\label{mesope}
M_{\bf m}(x)\equiv \calo_{{\bf m},{\bf n}= {\bf 0}}(x)=e^{-2\pi \langle {\bf m}^I, {\bm z}_I\rangle}|_{\rm Sym}
\ee
and their microscopic SCFT counterpart.
Borrowing the terminology from the analogous AdS$_5$/CFT$_4$ type IIB models, the operators \eqref{mesope} may be regarded as {\em mesonic} operators.

\subsection{Adding M5-branes}
\label{sec:M5}

Let us now consider operators \eqref{mnchiral} with $n^a\neq 0$. We will refer to these types of operators as {\em Betti} operators, since the are charged under the Betti symmetries.\footnote{In the relative IIB models, they would correspond to baryonic operators.}   
The corresponding semiclassical BPS solutions  in $\mathbb{R}^3$  now contain the non-trivial fields
\be\label{M5sol}
\sigma^a=\frac{n^a}{2|x|}\quad,\quad F^a= \frac12 n^a\d\text{vol}_{S^2}\,.
\ee
 In order to understand their M-theory interpretation, we recall that the effective three-dimensional field-strengths $F^a$ arise by expanding $G_4$ along the internal harmonic 2-forms $\omega_a$, which are Poincar\'e dual to a basis of non-compact divisors $D_a$.\footnote{We remind the reader  that we are restricting ourselves to  `non-anomalous' models, \emph{i.e.} $b_6(X)=0$.}  We can then interpret the solution \eqref{M5sol} as describing a Euclidean M5-brane sitting at the origin of the external $\mathbb{R}^3$ and wrapping an internal non-compact  divisor $D$ homologous to $n^aD_a$.  This can be understood by considering  the Bianchi identity  for $G_4$  modified by the presence of the M5-brane:
\be\label{M5bianchi}
\d G_4=\ell^3_{\text{\tiny P}}\delta_5(D)~.
\ee
By decomposing $G_4=\frac{\ell^3_{\text{\tiny P}}}{2\pi}F^a\wedge\omega_a+\ldots$ and integrating \eqref{M5bianchi} over the internal 2-cycles $C^a$ such $C^a\cdot D_b=\delta^a_b$,  we get the Bianchi identities $\d F^a=2\pi n^a\delta_3(0)$, which are indeed satisfied  by the solution \eqref{M5sol}. 

As explained in section \ref{sec:chiralop},
the charges $({\bf m}^I,{\bf n})$ characterising these semiclassical states and the corresponding chiral operators \eqref{mnchiral} must satisfy the constraint \eqref{chargecond}. This implies that $D\simeq n^aD_a$ is linearly equivalent to an  effective toric divisor, that is, we can choose $D= s_A\,\cald^A$, with $s_A\in\mathbb{Z}_{\ge 0}$ such that $Q^{aA}s_A=n^a$. The most elementary Betti operators correspond to $D=\cald^A$ for some $A=1,\ldots,d$ and  have Betti charges $n^a=Q^{aA}$. The above  corresponding semiclassical state then provides an effective three-dimensional description of the backreaction of an M5-brane wrapping the effective divisor  $\cald^A$ in the Calabi-Yau cone $C(Y)$. The toric divisor $\cald^A$ is itself a cone over a five-cycle $\Pi^A\subset Y$, which can then be `rotated' to a static static world-line in AdS$_4$ times $\Pi^A$, as  in \cite{Mikhailov:2002wx,Beasley:2002xv}. This is the more traditional description of a Betti operators in the AdS/CFT correspondence  \cite{Gubser:1998fp,Gukov:1998kn,Fabbri:1999hw}. 

By extending these comments to the most general effective divisor $D= s_A\,\cald^A$, the corresponding  semiclassical state  describes an M5-branes wrapping the five-cycle $\Pi\equiv \del D\subset Y$. We emphasize that  our holographic EFT provides a description of these M5-branes which goes beyond the probe approximation (and indeed, at the two-derivative level, it is in principle justified only for large charges).  For instance, from the first of \eqref{M5sol} we see that the K\"ahler moduli are forced to flow as we move close to the M5-brane, so that the internal Calabi-Yau is `dynamically' resolved in the direction of the extended K\"ahler moduli space $\mathbb{K}^{\rm ext}$ selected by the Betti charges $n^a$. 

Furthermore, for any given choice of admissible Betti charges $n^a$, there is an infinite number of possible choices of toric charges ${\bf m}^I$ satisfying \eqref{chargecond}. Recalling the previous subsection, we can then regard the most general state of charges $({\bf m}^I,{\bf n})$ as  a bound state of AdS giant gravitons and internal M5-branes.  From an alternative angle, as in \cite{Beasley:2002xv,Butti:2006au}, we may regard the different states corresponding to different toric charges ${\bf m}^I$  as the states obtained by quantizing the moduli space of classical M5-branes wrapping a divisor with fixed Betti charges ${\bf n}$. Indeed, one can easily adapt the discussion for generic toric IIB models of section 3 of \cite{Butti:2006au}. In particular, the states with vanishing Betti charge correspond to M5-branes in AdS$_4\times Y$ wrapping a trivial cycle in $Y$, i.e.\ to `internal' giant gravitons \cite{McGreevy:2000cw}. In combination with the arguments of section \ref{sec:AdSgiant}, our semiclassical description provides a direct identification of internal giant gravitons and AdS giant gravitons, whose counting should then clearly give the same result as in \cite{Mandal:2006tk,Biswas:2006tj} -- see \cite{Butti:2006au} for more comments on this point in the case of toric IIB models.

Notice that this M-theory interpretation holds also if $l^I_{i\infty},\sigma^a_\infty\neq 0$, that is if the vacuum  in $\mathbb{R}^3$ spontaneously breaks the conformal symmetry. The VEV computation of section \ref{sec:statesR3} can then be interpreted as the evaluation  of the  path integral with the insertion of Euclidean M5-branes. This  provides a clear justification  of the prescription proposed in \cite{Klebanov:2007us} to compute the VEV of Betti operators by using probe branes on resolved backgrounds.\footnote{See also \cite{Martelli:2008cm,Benishti:2010jn} for applications to more general IIB and M-theory toric models.} In fact, one may apply this prescription  to compute  the VEV of    operators with low Betti and toric charges, in terms of the EFT chiral fields. This can be done by adapting almost verbatim the steps of section 5 of \cite{Martucci:2016pzt} to the present M-theory setting. This would require the use of probe M5-branes and a subsequent quantization of their moduli space. Our holographic EFT approach provides a  realization of  this procedure that includes the M5-brane backreaction. Furthermore, as for the AdS giant gravitons, in our description in terms of monopole states with quantized monopole charges there is no additional need of quantizing the space of classical M5-brane configurations in order to get the appropriate discrete spectrum of states.

We also observe that  \eqref{scalingO} implies that the elementary Betti operators associated with the choice $D=\cald^A$ has scaling dimension $\Delta=\frac12 NP^A$. According to the above interpretation in terms of M5-branes wrapping $\Pi^A\subset Y$, this $\Delta$ should coincide with  $\frac{\pi N \text{vol}(\Pi^A)}{6\text{vol}(Y)}$, where the  volumes are computed using the asymptotic Sasaki-Einstein metric  \cite{Fabbri:1999hw}. Hence, by physical consistency  we get the following interesting identity:
\be\label{Pvolrel}
P^A=\frac{\pi \text{vol}(\Pi^A)}{3\text{vol}(Y)}\,.
\ee
We do not have a general mathematical justification of this identity, but we will explicitly check it in the examples discussed below.


\subsection{Charged EFT particles}

It is interesting to observe that the M-theory derivation naturally suggests the presence of half-BPS massive particles in the holographic EFT. Indeed, one can obtain a three-dimensional particle  by wrapping an M2-brane on  an effective curve $C\subset X$, defining a non-trivial compact two-cycle of the  resolved internal Calabi-Yau space. A simple dimensional reduction of the M2-brane effective action over \eqref{Mback} shows that its  contribution  to the three-dimensional EFT includes the terms
\be\label{partaction}
-\int_\gamma   {\mathfrak{m}}_{\bf e}(\sigma)\,\d{\rm vol}_\gamma +e_a\int_\gamma A^a\quad~~~~\text{with}\quad e_a\equiv  C\cdot D^a\in\mathbb{Z}\,,
\ee
where $\mathfrak{m}_{\bf e}(\sigma)\equiv\langle {\bm \sigma},{\bf e}\rangle\equiv \sigma^a e_a$ is the   mass of the particle,  $\gamma$ is the particle world-line and $\d{\rm vol}_\gamma$ the corresponding line element.  The mass $\mathfrak{m}_{\bf e}(\sigma)$ is positive precisely because $C$ belongs to the Mori cone, which is dual to the K\"ahler cone $\mathbb{K}_X$ parametrized by $\sigma^a$. One can of course consider the corresponding anti-particles with opposite charges $-{\bf e}$ and masses $\mathfrak{m}_{-\bf e}(\sigma)\equiv \langle {\bm \sigma},{\bf e}\rangle>0$. The spectrum of these BPS particles changes  across the K\"ahler walls inside $\mathbb{K}^{\rm ext}$.   

The contributions \eqref{partaction} are  compatible with the proposal made in \cite{Cuomo:2017vzg} to include the spin into the EFT description of  large charge  operators. One important difference is the presence of the first term in \eqref{partaction}, which is possible because  the mass $\mathfrak{m}_{\bf e}(\sigma)$ is moduli-dependent and then can have the correct scaling dimension $\Delta=1$ to be compatible with
conformal invariance. Note that, from the dual chiral formulation viewpoint, these particles are characterized by the axionic monodromy  $\rho_a\rightarrow \rho_a-\ii e_a$ around them. They are analogous to axionic strings in four dimensions that can be obtained from wrapped D3-branes in type IIB models, which are hence related  to the effective strings studied in \cite{Cuomo:2019ejv}.

We leave to the future an EFT  study of the  large charge SCFT  including these localised objects, along the lines of \cite{Cuomo:2017vzg,Cuomo:2019ejv}, and limit ourselves to a few general remarks here. The most basic one is that one cannot neglect the backreaction of these particles in three dimensions. 
Consider a static particle  and parametrise the two transverse directions by a complex coordinates $w$. In the dual chiral formulation of the bulk moduli, its backreaction can be described by the half-BPS flow solutions 
\be\label{rhoflow}
{\bm \rho}={\bm \rho}^{0}-\frac{{\bf e}}{2\pi}\log\frac{w}{w_0}\,,
\ee
with fixed $z^{i}_I\equiv z^{0,i}_I$, and arbitrary integration constant ${\bm\rho}^0$ at $w=w_0$. These kinds of solutions have been recently discussed in \cite{Lanza:2021qsu} in the completely analogous case of half-BPS strings in $\caln=1$ four-dimensional theories. As discussed therein, once rephrased in terms of a floating EFT UV cut-off, the flow solution \eqref{rhoflow} can be interpreted as a classical RG-flow  of the EFT Lagrangian, and in particular of the  mass appearing in \eqref{partaction}. On the other hand, the total IR physical mass (including the backreaction contribution) is given by $\mathfrak{m}^{\rm IR}_{\bf e}\equiv\lim_{|w|\rightarrow\infty}\mathfrak{m}_{\bf e}(\sigma(w))$ and measures the total volume of the holomorphic curve $\calc\subset\calm^{\rm ext}$ described by the embedding \eqref{rhoflow}.  In the present setting, conformal invariance implies that $\calm^{\rm ext}$ is  conical and then  $\calc$, being holomorphic, is necessarily non-compact. Hence these static BPS configurations are affected by IR divergences, since $\mathfrak{m}^{\rm IR}_{\bf e}=\infty$, as expected for  charged particles in three dimensions. However, one may construct non-static finite energy  configurations of  vanishing total charge as in \cite{Cuomo:2017vzg}.   

 These arguments can be immediately generalised to BPS particles carrying axionic charges associated with the toric $U(1)^4$ symmetry. Indeed, by using the formulation in terms of the linear multiplets $L^I_i$ for the mobile M2-brane sector, one may generalize \eqref{partaction} to 
\be\label{genpart}
-\int_\gamma \mathfrak{m}_{{\bf d},{\bf e}}\, {\rm dvol}_\gamma +d^i\sum_I\int_\gamma  A_i^I+e_a\int_\gamma A^a \quad~~~\text{with}\quad  \mathfrak{m}_{{\bf d},{\bf e}}\equiv\sum_I\langle {\bm l^I}, {\bf d}\rangle+\langle{\bm \sigma},{\bf e}\rangle\,,
\ee 
where ${\bf d}=(d^1,\ldots,d^4)\in N_\mathbb{Z}$.
These couplings are compatible with the ${\rm S}_N$ discrete gauge symmetry and generate the axion monodromy $z^i_I\rightarrow z^i_I-\ii d^i$ around the particle. 
 The static BPS backreaction now also includes the flow ${\bf z}_I={\bf z}_I^0-\frac{1}{2\pi}{\bf d}_I\log\frac{w}{w_0}$ and the curve $\calc\subset\calm^{\rm ext}$ projects to $N$ copies of a holomorphic curve inside  $X$. The charges $({\bf d},{\bf e})$ must be such that the mass $\mathfrak{m}_{{\bf d},{\bf e}}$ appearing in \eqref{genpart} is positive.  Since the moduli $(l^I_i,\sigma^a)$ move inside the cone defined by $s_A(l^I,\sigma)\geq 0$ for all $I=1,\ldots,N$, we can guarantee that  $\mathfrak{m}_{{\bf d},{\bf e}}$ remains positive over the entire $\calm^{\rm ext}$ by picking $d^i=v_A^i k^A$ and $e_a=N M_{Aa}k^A$ with $k^A\in\mathbb{Z}_{\geq 0}$.
 Again, in order to get finite energy configurations one must consider non-static configurations of vanishing total charge.

\section{The \texorpdfstring{$Y^{12}(\mathbb{P}^2)$}{Y12(P2)} model}
\label{sec:Y12model}

We now illustrate the above general results with a first concrete example. In this  model the UV quiver theory describes the dynamics of $N$ M2-branes at the tip of the cone over the Sasaki-Einstein space $Y^{12}(\mathbb{P}^2)$.  The UV quiver theory was discussed in section \ref{sec:Y12quiver}. 
This theory has a geometric branch, with a corresponding holographic EFT and a class of BPS semiclassical states, which will be described in detail. We will then match this holographic description with what is expected from the dual quiver point of view.

 According to our general discussion, in order to derive the  holographic EFT we must consider the crepant resolutions of the Calabi-Yau cone over $Y^{12}(\mathbb{P}^2)$.  
As already mentioned in section \ref{sec:Y12quiver}, these can be described in terms of a $U(1)$ gauged linear sigma model with five dimensionless complex homogeneous coordinates $Z_A$, $A=1,\ldots , 5$ (so that $d=5$ in the formulas of section \ref{sec:toric})  of charges 
  \be\label{Y12glsm}
\begin{array}{c|ccccc|r} 
 & Z_1 & Z_2 & Z_3  & Z_4 & Z_5  &{\rm FI} \\
\hline
Q^A& 1 & 1 & 1 & -2 &  -1   & \sigma
\end{array}
\ee
where $\sigma$ is the unique K\"ahler modulus.

\begin{figure}[t]
\centering
	\includegraphics[scale=0.6]{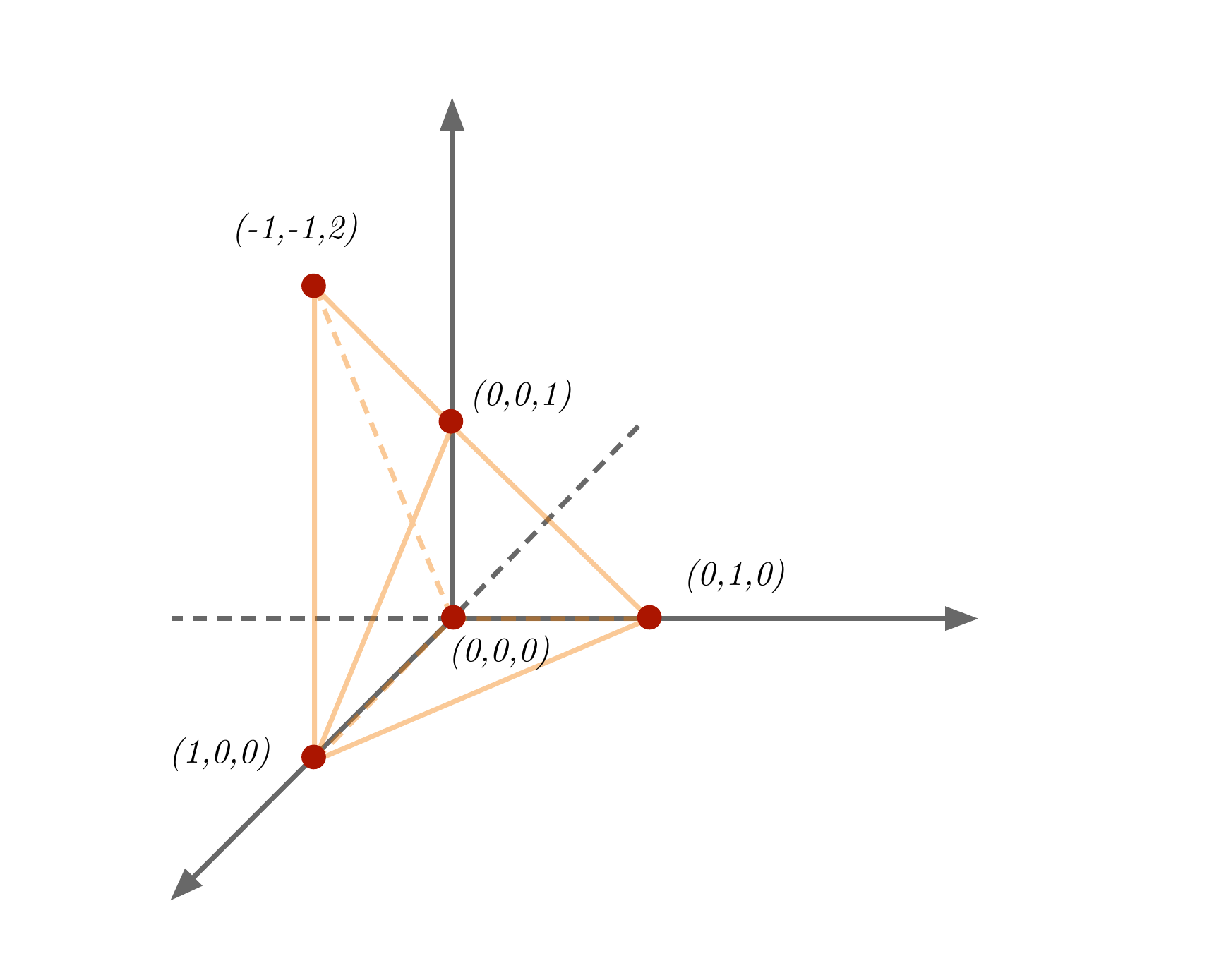}
\caption{Toric diagram of $Y^{1,2}(\bP^2)$}
	\label{fig:Fan-Y12}
\end{figure}

It admits different resolutions, depending on the sign of $\sigma$ and corresponding to the two possible triangulations of the toric diagram, see Fig.~\ref{fig:Fan-Y12}. For later convenience, we choose  
the following set of toric fan generators in $\mathbb{Z}^4$:
\be\label{Y12vec}
\begin{aligned}
&{\bf v}_1=(0,0,1,0)\, ,\ {\bf v}_2=(0,0,0,1)\, ,\ {\bf v}_3=(2,1,-1,-1)\,,\\
&{\bf v}_4=(1,0,0,0)\, ,\ {\bf v}_5=(0,1,0,0)\, . 
\end{aligned}
\ee
Indeed, one can easily check that  $Q^A{\bf v}_A={\bf 0}$. By an appropriate $SL(4,\mathbb{Z})$ transformation, one can map ${\bf v}_A$ to vectors of the form $({\bf w}_A,1)$, where ${\bf w}_A \in \mathbb{Z}^3$ identify the vertices of the toric diagram of Fig.~\ref{fig:Fan-Y12}.\footnote{More explicitly, $M {\bf v}_A^T=({\bf w}_A,1)^T$, where
$$
M = \begin{pmatrix} 0 & 0 & 1 & 0 \\ 0 & 0 & 0 & 1 \\ 1 & 0 & 0 & 0 \\ 1 & 1 & 1 & 1 \end{pmatrix}\quad\Rightarrow\quad \begin{array}{ll}
     & \mathbf{w}_1=(1,0,0)\,, \ \mathbf{w}_2=(0,1,0)\,, \,\mathbf{w}_3=(-1,-1,2)\,, \\
     & \mathbf{w}_4=(0,0,1)\,, \quad \mathbf{w}_5=(0,0,0)\,.
\end{array}
$$

}

A detailed description of the geometrical properties of these resolutions can be found in \cite{Benishti:2009ky}. The fan corresponding to K\"ahler cone $\sigma< 0$  is simplicial and there is a leftover terminal 
$\bZ_2$ orbifold singularity. As a complex space, the  Calabi-Yau corresponding to $\sigma<0$ can be regarded as the total space of the vector bundle $\calo(-1)^{\oplus 3}$ over the weighted projected space $\mathbb{WP}^1_{[1,2]}$. To our knowledge, the Calabi-Yau metric on this space is not known. For this reason, we will restrict our attention to the branch  $\sigma\geq 0$.

\subsection{The resolved geometry}

In the K\"ahler cone $\sigma\geq 0$, the resolution  replaces  the tip of the cone with a $\mathbb{P}^2$. The resolved geometry $X$ is the total space of the line bundle $\calo_{\mathbb{P}^2}(-2)\oplus \calo_{\mathbb{P}^2}(-1)$ and we  can  identify $\sigma$ with the volume of any $\mathbb{P}^1\subset \mathbb{P}^2$, so that the conical geometry corresponds to value  $\sigma=0$. By using \eqref{Y12vec} in \eqref{zZrelation} we get  the following definition of the toric coordinates $e^{-2\pi z^i}$ in terms of homogeneous coordinates:
\be\label{Y12toricz}
e^{-2\pi z^1}=Z_4Z^2_3\equiv \zeta_1 \,,\quad e^{-2\pi z^2}=Z_5Z_3\equiv\zeta_2 \,,\quad e^{-2\pi z^3}=\frac{Z_1}{Z_3}\equiv\lambda_1 \,,\quad e^{-2\pi z^4}=\frac{Z_2}{Z_3}\equiv\lambda_2 \,.
\ee
We have also introduced local coordinates $(\zeta_1,\zeta_2,\lambda_1,\lambda_2)$ which make manifest its bundle structure of $X$: $(\lambda_1,\lambda_2)$  are local coordinates on $\mathbb{P}^2$, $\zeta_1$ is the fibral coordinate of $\calo_{\mathbb{P}^2}(-2)$ and $\zeta_2$ is the fibral coordinate of $\calo_{\mathbb{P}^2}(-1)$. 

In order to compute the holographic EFT, we need the explicit form of  the Ricci-flat K\"ahler potential  $k_X$ of the internal Calabi-Yau metric, see \eqref{JXk}, which enters the mixed formulation of section \ref{sec:HEFTlinear}. From this one can then go to the  formulations in terms of  either  just  chiral multiplets or  just  linear multiplets -- see sections \ref{sec:HEFTchiral} and \ref{sec:sympEFT}. 
The Calabi-Yau metric has been computed in \cite{Martelli:2007pv} for a fixed choice of the K\"ahler modulus $\sigma$ and in a particular  set of real coordinates. One can find the appropriate change of coordinates and, by a simple scaling argument,  the metric for general  $\sigma$. From this we can extract the value of K\"ahler potential $k_X$ which decomposes as in \eqref{kdec} satisfying the appropriate conditions \eqref{bcond}.  Here we just present the relevant results, relegating the details to appendix \ref{app:Y12}. 

The K\"ahler potential $k_X$ takes the form
\be\label{Y12k}
k_X(z,\bar z;\sigma)=U(t_1,t_2;\sigma)+\sigma k_{\mathbb{P}^2}
\ee
where we have introduced the $\mathbb{P}^2$ K\"ahler potential\footnote{$k_{\mathbb{P}^2}$ is normalized  so that $j_{\mathbb{P}^2}\equiv  \ii\del\delbar k_{\mathbb{P}^2}=  \frac1{6\pi}\calr_{\mathbb{P}^2}$ and $\int_{\mathbb{P}^1}j_{\mathbb{P}^2}=1$. It is only locally defined: \eqref{kP2} and then \eqref{Y12k} are valid in the local patch $X\backslash \cald^3$, where $\cald^A=\{Z_A=0\}$ denote the toric divisors.}
\be\label{kP2}
k_{\mathbb{P}^2}=\frac{1}{2\pi}\log\left(1+|\lambda_1|^2+|\lambda_2|^2 \right)
\ee
and the global real `radial' coordinates along the fibers 
\be\label{deft1t2}
\begin{aligned}
t_1&\equiv  |\zeta_1|^2 e^{4\pi k_{\mathbb{P}^2}}=(|Z_4 Z^2_1|+|Z_4 Z_2^2|+|Z_4 Z_3^2|)^2\,,\\ 
t_2&\equiv |\zeta_2|^2 e^{2\pi k_{\mathbb{P}^2}}=|Z_5 Z_1|^2+|Z_5 Z_2|^2+|Z_5 Z_3|^2 \,.
\end{aligned}
\ee

Unfortunately, $U(t_1,t_2;\sigma)$ does not seem to admit a reasonable explicit formula. (As we will presently see, the symplectic potential $F_X$, which enters the formulation with vector multiplets of section \ref{sec:sympEFT}, can be made more explicit.)    Rather, $U(t_1,t_2;\sigma)$ will be defined  in terms of two new real variables 
\be\label{xyt}
x=x(t_1,t_2;\sigma)\,,\quad y=y(t_1,t_2;\sigma)\,,
\ee
 which appear in the solution of \cite{Martelli:2007pv}. These can be in principle obtained as functions of  $(t_1,t_2;\sigma)$ by inverting 
\be\label{Y12t12v}
\begin{aligned}
\left(t_1\right)^{\frac{3 u_+}{8}} &= \sigma   \Big[(x-u_-) (y-u_-)\Big]^{\frac{u_-^2}{|u_--u_*|^2}} \,
\Big| \left[(x-u_*) (y-u_*)\right]^{\frac{-\ii\,u_*^2}{(u_*-u_-) \Im u_*}}\Big|\,,   \\
\left(t_2\right)^{\frac{3 u_-}{4}} &= \sigma   \Big[(x-u_+) (u_+-y)\Big]^{\frac{u_+^2}{|u_+-u_*|^2}} \,
\Big| \left[(x-u_*) (y-u_*)\right]^{\frac{-\ii\,u_*^2}{(u_*-u_+) \Im u_*}}\Big|\,,
\end{aligned}
\ee
where $u_-,u_+$ and $u_*,\overline{u_*}$ are the two real (with $u_-<u_+$) and the two complex zeros, respectively, of the polynomial 
\be\label{Gmathfrak}
{\mathfrak G}(u)\equiv u^4-\frac{4}{3}u^3+\frac83\nu\,.
\ee 
Consistency requires that \cite{Martelli:2007pv}
\be\label{xycons}
x\in[u_+,\infty)\quad,\quad y\in [u_-,u_+]
\ee
and furthermore that $\nu$ is fixed by the conditions
\be\label{s+-fix}
\frac{u_-(u_+-1)}{u_+(u_--1)}=-\frac12\,.
\ee
In order to find the explicit values   of $\nu$, $u_\pm$ and $u_*$, let us set 
\be\label{nu_of_g}
\nu=\frac{(\eta^2-1)^3}{54 \eta^2}
\ee
with $\eta>0$, which allows us to write \be\label{sigma+-_of_g}
\begin{aligned}
u_\pm  = \frac{1}{3}(\eta+1)\left(1\pm\sqrt{\frac{2}{\eta}-1}\,\right) \quad,\quad
u_*=-\frac13(\eta-1)\left(1+\ii\sqrt{\frac{2}{\eta}+1}\,\right)\,.
\end{aligned}
\ee
The condition (\ref{s+-fix}) fixes the value of $\eta$ to be
\be\label{g}
\eta=\frac{1}{6}\left[2+2^{1/3}(58+3\sqrt{78})^{1/3}+2^{1/3}(58-3\sqrt{78})^{1/3} \right]\simeq 1.91799~, 
\ee
which translates into
\be
\nu\simeq 0.0968\,,\quad u_-\simeq 0.771\,,\quad u_+\simeq 1.174\,,\quad u_*\simeq -0.306 -\ii\,0.437~.
\ee

We are now in a position to write down the (implicit) function $U(t_1,t_2;\sigma)$ appearing in \eqref{Y12k}:
\be\label{GY12}
\begin{aligned}
U(t_1,t_2;\sigma)=&\,\frac{2\sigma}{3\pi u_- u_+}\Big[x+y + \frac{1}{\Im{u_*}}\Im\Big(u_*^2 \log[(x-u_*) (y-u_*)]\Big)+2\Re u_*\log \sigma\Big]
\end{aligned}
\ee
where  the logarithmic term appearing  in the second line is necessary in order to satisfy the asymptotic conditions \eqref{bcond}. We stress that in the above formulas $x,y$ must be considered as functions of $t_1,t_2$, and then of $z^i$ or $\lambda_1,\lambda_2,\zeta_1,\zeta_2$, and $\sigma$, as in \eqref{xyt}. More details on the corresponding K\"ahler metric can be found in appendix \ref{app:Y12}. Notice also that all formulas are manifestly invariant under the toric $U(1)^4$ symmetry $\Im z^i\rightarrow \Im z^i+\text{constant}$, which is reflected into the `mixed' holographic EFT of section \ref{sec:HEFTlinear}, which is fully specified by $k_X$.

We can now move to the symplectic description of the resolved metric, which enters the  dual formulation with vector multiplets of section \ref{sec:sympEFT}. This is obtained by going to the   dual coordinates $l_i$   and  the symplectic potential \eqref{sympot}.  By applying \eqref{lmindef} to \eqref{Y12k} we get:
\be\label{Y12li}
\begin{aligned}
l_1&=\frac{(x-u_-)(y-u_-)}{2 u_-(u_+-u_-)}\sigma\\
l_2&=\frac{(x-u_+)(u_+-y)}{u_+(u_+-u_-)}\sigma\\
l_3&=\frac{|\lambda_1|^2}{1+|\lambda_1|^2+|\lambda_2|^2}(\sigma+2l_1+l_2)\\
l_4&=\frac{|\lambda_2|^2}{1+|\lambda_1|^2+|\lambda_2|^2}(\sigma+2l_1+l_2)~.
\end{aligned}
\ee
One can check that they span the polytope 
\be\label{Y12poly}
\calp_\sigma=\{l_1,l_2,l_3,l_4\geq 0\}\cap\{2l_1+l_2+\sigma\geq l_3+l_4\}
\ee
This matches \eqref{poly}-\eqref{chipos} provided that  $M_A=\delta_{A3}$, so that $\chi_A=\delta_{A3}\sigma$. 

One can then compute $F_X$ from \eqref{sympot}. This can be entirely expressed in terms of $l_i$ and $\sigma$ by inverting \eqref{Y12li}. Alternatively, by the linear change of coordinates introduced in \eqref{poly}, we can use  the coordinates
\be\label{Y12sA}
s_1=l_3\,,\quad s_2=l_4\,,\quad s_3=2l_1+l_2-l_3-l_4+\sigma\,,\quad s_4=l_1\,,\quad s_5=l_2\,.
\ee
The symplectic potential $F_X$ can then be rewritten in the form
\eqref{FXdec} with 
\begin{subequations}\label{sPY12}
\begin{align}
s_{P_{\rm can}} &\equiv \sum_A s_A= 3 l_1+2 l_2 + \sigma\label{sPY12a}\\
s_{P} &= \frac{8}{3\,u_+}l_1+\frac{4}{3\,u_-}l_2-\frac{8\,\Re u_*}{3\,u_-u_+}\sigma\label{sPY12b}
\end{align}
\end{subequations}
and\footnote{\label{foot:fX} In computing $f_X$ from \eqref{sympot} we have omitted terms linear in $l_i$ and $\sigma$, that do not affect the metric. This corresponds at most to shifting the dual coordinates $\Re z^i$ by constants.} 
\be\label{fXY12}
\begin{aligned}
f_X(l,\sigma)=&\, \frac{1}{2\pi} s_{P_{\rm can}} \log \frac{2l_1+l_2+\sigma}{s_{P_{\rm can}}}+ \frac{1}{2\pi} s_{P} \log \frac{s_P}{\sigma}
-\frac{1}{2\pi} l_1 \log \frac{2l_1+l_2+\sigma}{l_1}\\
&\,-\frac{1}{2\pi} l_2 \log \frac{2l_1+l_2+\sigma}{l_2}- \frac{4 u_-^2}{3 \pi u_+ |u_--u_*|^2} l_1 \log \frac{l_1}{\sigma}-\frac{2 u_+^2}{3 \pi u_- |u_+-u_*|^2} l_2 \log \frac{l_2}{\sigma}
\\
&\quad\,+\frac{2}{3 \pi u_- u_+\,\Im u_*} \Im\left[\frac{u_*^2\, \alpha}{(u_--u_*)(u_+-u_*)}\log \frac\alpha\sigma\right]\,,
\end{aligned}
\ee
where
\be\label{Y12alpha}
\alpha(l,\sigma) \equiv 2  u_- (u_+-u_*)l_1+  u_+(u_--u_*)l_2 +(u_--u_*)(u_+-u_*)\sigma\,.
\ee
From \eqref{sPY12b} and \eqref{Y12sA} we can then extract the 
constants that characterise the solution:
  \be\label{Y12PA}
  \begin{aligned}
  &P^1=P^2=P^3=-\frac{8\,\Re u_*}{3\,u_-u_+} = \frac{4\eta}{(1+\eta)^2}\simeq 0.901\\ 
  &P^4=\frac{8(4-3u_+)}{9\,u_-u_+}\simeq 0.470\quad,\quad P^5=\frac{4(4-3u_-)}{9\, u_-u_+}\simeq 0.827
  \end{aligned}
  \ee
  or alternatively
\be\begin{aligned}\label{Y12bp}
&b^1=\frac{8}{3\,u_+}\simeq 2.272 \quad,\quad b^2=\frac{4}{3\,u_-}\simeq 1.728\quad,\quad b^3=b^4=0\,,\\
& p=-\frac{8\,\Re u_*}{3\,u_-u_+}= \frac{4\eta}{(1+\eta)^2}\simeq 0.901\ .
\end{aligned}
\ee
According to \eqref{Pbp}, these two sets of constants are related by $b^i=P^Av^i_A$ and $p=P^AM_A=P_3$. 

It is interesting to observe  that  the constants \eqref{Y12PA} indeed satisfy \eqref{Pvolrel} in a rather non-trivial way. This can be shown by observing that the quantities on the two sides of the equalities \eqref{Pvolrel} are the only real roots of the same cubic polynomials with integer coefficients. We refer to appendix \ref{subsubsec:Rvolumes} for the details of this match.

We observe that the terms proportional to $\log\sigma$ obtained by expanding the logarithms in \eqref{fXY12} actually cancel each other.
We have nevertheless chosen to write $f_X$ in the form \eqref{fXY12} in order to manifestly show that it is homogeneous of degree-one in $(l_i,\sigma)$, so that $F_X$ has precisely the structure predicted in section \ref{sec:sympEFT} from holographic arguments. We also remind the reader that the above formulas are sensible only for $\sigma\geq  0$. According to the arguments section  \ref{sec:sympEFT}, for $\sigma<0$ it should be given by a symplectic potential $F_X$ specified by the same $P^A$'s but by a different homogeneous function $f_X$.

\subsection{Holographic EFT}\label{subsec:HEFT_Y12}

In the previous subsection we have seen that the symplectic structure of the resolved cone over $Y^{12}(\mathbb{P}^2)$ can be made quite explicit (at least for $\sigma\geq 0$). This can then be taken as  starting point for describing the associated holographic EFT for vector multiplets, providing a concrete illustration of the general discussion of section \ref{sec:sympEFT}. The dual  formulations (partially) involving chiral multiplets and considered in sections \ref{sec:HEFTlinear} and \ref{sec:HEFTchiral} can then be obtained by subsequent Legendre transformations.  

The EFT action can  be written as 
\be\label{Y12eL}
\int\d^3 x\d^4\theta\,\tilde\calf(L,\Sigma)
\ee
where $\tilde\calf(L,\Sigma)$ is a function of  $4N+1$ linear  multiplets $\Sigma=\sigma+\ldots- \frac12\epsilon^{\mu\nu\rho}\theta\gamma_\mu\bar\theta F_{\nu\rho}$ and $L^I_i=l^I_i+\ldots- \frac12\epsilon^{\mu\nu\rho}\theta\gamma_\mu\bar\theta (F^I_i)_{\nu\rho}$, of the form \eqref{Ftilde}. In the present case, it is explicitly given by
\begingroup
\allowdisplaybreaks
\begin{align}
\tilde\calf(L,\Sigma)=&-\frac1{2\pi}\sum_I\Big[L^I_3\log L^I_3+L^I_4\log L^I_4-\big(2L_1^I+L^I_2+\Sigma\big)\log \big(2L_1^I+L^I_2+\Sigma\big)\Big]\nonumber\\
\begin{split}\label{Y12tildeF}
    &-\frac1{2\pi}\sum_I\left(2L^I_1+L^I_2-L^I_3-L^I_4+\Sigma\right)\log\left(2L^I_1+L^I_2-L^I_3-L^I_4+\Sigma\right) \\
&- \frac{4 u_-^2}{3 \pi u_+ |u_--u_*|^2} \sum_IL^I_1 \log L^I_1-\frac{2 u_+^2}{3 \pi u_- |u_+-u_*|^2} \sum_IL^I_2 \log L^I_2
\end{split}
\\
&+\frac{2}{3 \pi u_- u_+\,\Im u_*} \sum_I\Im\left[\frac{u_*^2}{(u_--u_*)(u_+-u_*)}\,\alpha(L^I,\Sigma)\log \alpha(L^I,\Sigma)\right]\nonumber
\end{align}
\endgroup
where $\alpha(L^I,\Sigma)$ is the superfield obtained from \eqref{Y12alpha}  by substituting $l_i\rightarrow L^I_i$ and $\sigma\rightarrow \Sigma$.

Since the linear multiplets have fixed scaling dimensions $\Delta(L^I_i)=\Delta(\Sigma)=1$, we can explicitly check that $\tilde\calf( L, \Sigma)$ is quasi-homogeneous of degree-one under scaling transformations, as in \eqref{nonhomcalf}, and then \eqref{Y12eL} is  superconformal. Furthermore, the above EFT is well defined on the branch of the moduli space in which $\sigma\geq 0$ and all the vectors ${\bm l}^I\equiv (l^I_1,l^I_2,l^I_3,l^I_4)$ belong to the polytope \eqref{Y12poly}.

It is interesting to observe how the kinetic potential \eqref{Y12tildeF} reduces to a very simple form $\sum_LL\log L+$constant, where the $L$'s are linear combinations of  $L^I_i,\Sigma$. Hence the formulation of the EFT in terms of vector multiplets is particularly simple. 
One can then go to the mixed formulation of section \ref{sec:HEFTlinear} by taking a partial Legendre transform of $\calf(z,\Sigma)=\tilde\calf(L,\Sigma)-2L^I_i\Re z^i_I$,
where $z^i_I$ are the chiral multiplets obtained by dualization of the linear multiplets $L^I_i$, such that $\Re z^i_I=\frac12\frac{\del\tilde F(L,\Sigma)}{\del L^I_i}$. By a further Legendre transform one can then obtain the K\"ahler potential $\calk(z,\rho)=\calf(L,\Sigma)-2\Sigma\Re\rho$ in which $\Sigma$ is traded for the dual chiral mutiplet $\rho$, such that $\Re\rho=\frac12\frac{\del \calf(z,\Sigma)}{\del\Sigma}=\frac12\frac{\del \tilde\calf(L,\Sigma)}{\del\Sigma}$, obtaining the chiral formulation of section \ref{sec:HEFTchiral}. From   \eqref{Y12tildeF} we can easily get the explicit relation between the chiral multiplets and the linear multiplets:  
\begingroup
\allowdisplaybreaks
\begin{align}
\Re z^1_I=&\,-\frac1{2\pi}\log\left(1+\frac{L_3^I+L_4^I}{2L_1^I+L_2^I+\Sigma}\right)-\frac{2u_-^2}{3\pi u_+|u_--u_*|^2}\log L_1 \nonumber\\
&+\frac{2}{3\pi u_+\Im u_*}\Im\Big(\frac{u^2_*}{u_--u_*}\log\alpha\Big)-\frac{2}{3\pi u_+}\nonumber\\ 
\Re z^2_I=&\,-\frac1{4\pi}\log\left(1+\frac{L_3^I+L_4^I}{2L_1^I+L_2^I+\Sigma}\right)-\frac{u_+^2}{3\pi u_-|u_+-u_*|^2}\log L_2\nonumber\\
\begin{split}\label{Y12chilin}
  &+\frac{1}{3\pi u_-\Im u_*}\Im\Big(\frac{u^2_*}{u_+-u_*}\log\alpha\Big)-\frac{1}{3\pi u_-}\\ 
\Re z^3_I=&\,-\frac1{4\pi}\log\Big(\frac{L_3^I}{2L_1^I+L_2^I-L_3^I-L_4^I+\Sigma}\Big)\\
\Re z^4_I=&\,-\frac1{4\pi}\log\Big(\frac{L_4^I}{2L_1^I+L_2^I-L_3^I-L_4^I+\Sigma}\Big)  \end{split}
\\
\Re \rho=&\,\frac1{4\pi}\sum_I\log\Big(\frac{2L^I_1+L^I_2+\Sigma}{2L_1^I+L_2^I-L_3^I-L_4^I+\Sigma}\Big)\nonumber\\
&+\frac{1}{3\pi u_-u_+\Im u_*}\sum_I\Im\log\left[u_*^2\log\alpha(L^I,\Sigma)\right]+\frac{2N\Re u_*}{3\pi u_-u_+}~.\nonumber
\end{align}
\endgroup
These relations can in principle be inverted to get the linear multiplets as functions of (the real part of) the chiral multiplets, from which one can extract the explicit form of the EFT K\"ahler potential -- see \eqref{calktoric} -- 
\be
\calk=\frac2{3\pi}\sum_I\left(\frac{2}{\,u_+}L^I_1+\frac{1}{u_-}L^I_2\right) -\frac{4N\,\Re u_*}{3\pi(u_-u_+)}\,\Sigma
\ee
as a function of the chiral fields. 

From \eqref{Y12chilin} one can extract the scaling dimensions of the chiral multiplets $e^{-2\pi z^i_I}$ and $e^{-2\pi\rho}$, by taking into account that their phases do not scale and that $\Delta(L^I_i)=\Delta(\Sigma)=1$:
\be\label{Y12Deltazrho}
\begin{aligned}
&\Delta(e^{-2\pi z^1_I})=\frac{4}{3\, u_+}\simeq 1.136\ ,\quad\Delta(e^{-2\pi z^2_I})=\frac{2}{3\, u_-}\simeq 0.864 \ ,\\
&\Delta(e^{-2\pi z^3_I})=\Delta(e^{-2\pi z^4_I})=0\ ,\\ &\Delta(e^{-2\pi\rho})=-\frac{4N\,\Re u_*}{3\,u_-u_+}=\frac{2N\eta}{(1+\eta)^2}\simeq N\times 0.4505
\end{aligned}
\ee 
Of course, these are in agreement with the general formula \eqref{zrhoscaling} -- see \eqref{Y12bp}.

We stress that, as emphasized in section \ref{sec:chiralop}, the fields $e^{-2\pi\rho}$ and $e^{-2\pi z_I^i}$ are only locally defined on the moduli space and cannot be interpreted as low-energy 
representations of  chiral operators of the microscopic SCFT. Hence, the fact that $e^{-2\pi z_I^3}$ and $e^{-2\pi z_I^4}$ have vanishing scaling dimension does not constitute any violation of the usual conformal  unitarity bounds. We will come back to this point in the next subsection. 

Note that, even though the above description of the holographic EFT is a priori valid only for $\sigma=\Sigma|_{\theta=\bar\theta=0}> 0$, the formulas \eqref{Y12tildeF} and \eqref{Y12chilin} make sense even for $\sigma=0$, as long as the $l^I_i$'s belong to the interior of the polyedral cone $\calp_{\sigma=0}$, see  \eqref{Y12poly}. Indeed, by setting $\Sigma\equiv 0$ we just get the effective theory for $N$ probe M2-branes moving in the Calabi-Yau cone over $Y^{12}(\mathbb{P}^2)$.  
On the other hand, for $\sigma<0$ the EFT is expected to change, due to the change of the internal resolved Calabi-Yau cone. While we expect the holomorphic  parametrization of the moduli space provided by $e^{-2\pi z^i_I}$ and $e^{-2\pi\rho}$ to smoothly connect the two branches, nothing prevents the complete EFT to exhibit some kind of phase transition on the K\"ahler wall $\sigma=0$. We will not have much more to say about this issue for the   $Y^{12}(\mathbb{P}^2)$ model, since we do not know the explicit Calabi-Yau metric (and then the holographic EFT) for $\sigma<0$. Instead we will come back to this point in section \ref{sec:Q111model}, since for the  $Q^{111}$ model we will be able to describe the holographic EFT on the entire extended K\"ahler moduli space.

\subsection{EFT chiral operators and semiclassical states}

Following the general discussed of section \ref{sec:chiralop}, in the $Y^{12}(\mathbb{P}^2)$ model we can construct chiral operators
\be\label{Y12chiral}
\calo_{{\bf m},n}(x)\equiv e^{-2\pi \langle {\bf m}^I, {\bm z}_I(x)\rangle}|_{\rm Sym}\,e^{-2\pi n\rho(x)}
\ee
where  the charges $m^I_i\in \mathbb{Z}$ are constrained by the lattice polytope conditions \eqref{chargecond}, which in the present case read
\be\label{Y12charges}
 m^I_1,m^I_2,m^I_3,m^I_4\geq 0 \quad,\quad 2m^I_1+m^I_2+n\geq m^I_3+m^I_4~.
\ee
From the general formula \eqref{scalingO}, or more directly from \eqref{Y12Deltazrho},  one can immediately obtain the corresponding scaling dimensions:
\be\label{Y12scaling}
\begin{aligned}
\Delta(\calo_{{\bf m},n})&=\frac{4}{3 u_+}\sum_I m_1^I+\frac{2}{3 u_-}\sum_I m_2^I+\frac{2n\eta}{(1+\eta)^2}\,N\\
&=\frac12 P^A \sum_I s_A({\bf m}^I,n)\,,
\end{aligned}
\ee
where $s_A({\bf m}^I,n)\geq 0$ are the  integral numbers obtained by setting $ l_i=m_i^I$ and $\sigma=n$ in \eqref{Y12sA}.
Note that,  given \eqref{Y12charges}  and the values \eqref{Y12PA} of the $P^A$'s, the unitarity bound $\Delta(\calo_{{\bf m},n})\geq \frac12$ is always satisfied for $N\gg 1$.

From \eqref{Y12scaling} and \eqref{Y12charges} we see that the spectrum of scaling dimensions is degenerate, in the sense that it is determined by the charges $m_1^I,m_2^I,n$, while the charges $m_3^I,m_4^I$ label the `harmonics' along the base $\mathbb{P}^2$ which enter the operator. 

Take for instance the mesonic operators, which have vanishing Betti charge $n=0$. From \eqref{Y12scaling} and \eqref{Y12charges}, it is clear that for any pair of toric charges $m_1^I,m_2^I\geq 0$, the finite number of possible charges $m_3^I,m_4^I\geq 0$ such that $m_3^I+m_4^I\leq 2m_1^I+m_2^I$ label the set of degenerate mesonic operators with the same dimension $\Delta=\sum_I\left(\frac{4}{3 u_+} m_1^I+\frac{2}{3 u_-}m_2^I\right)$. These operators may be written in terms of non-negative powers of the homogeneous coordinates as follows:
\be\label{Y12mesonic}
M_{{\bf m}}=\left(\prod_{I}\left[(Z^I_1)^{m_3^I}(Z^I_2)^{m_4^I}(Z^I_3)^{2m^I_1+m^I_2-m_3^I-m_4^I}(Z^I_4)^{m^I_1}(Z^I_5)^{m^I_2}\right]\right)_{\rm sym}~.
\ee

In the first K\"ahler cone $\sigma>0$, the exceptional locus is given by $ \cald^4\cap\cald^5\simeq \mathbb{P}^2$. It is then clear that the mesonic operators \eqref{Y12mesonic} identically vanish as soon as one M2-brane touches $\cald^4\cap\cald^5$ and are then `blind' to the resolved $\mathbb{P}^2$. A similar conclusion holds for the other K\"ahler cone $\sigma<0$, whose exceptional locus is given by $\cald^1\cap\cald^2\cap\cald^3\simeq \mathbb{WP}^1_{[1,2]}$.

The exceptional locus of the resolved geometry is instead detected by Betti operators, which have  $n\neq 0$. Consider the  `lightest' ones  with  Betti charge $n=1$: they have toric charges  $m^I_1=m^I_2=0$ but possibly non-vanishing $m^I_3,m^I_4$ such that $0\leq m^I_3+m^I_4\leq 1$. Because of the symmetrization in \eqref{Y12chiral}, these operators can be labelled by two integers 
\be
{\mathfrak m}_3=\sum^N_{I=1} m_3^I\quad,\quad {\mathfrak m}_4=\sum^N_{I=1} m_4^I\,,
\ee
such that $0\leq {\mathfrak m}_3+\mathfrak{m}_4\leq N$. We can then denote the corresponding operators as follows\footnote{We emphasize that this effective description of the chiral operators is valid when the EFT fields $z^i_I$ move in the local patch $X\backslash \cald^3$, 
see \eqref{Y12toricz}. In order to include the locus $\cald^3$ in the description one should change local patch by appropriately changing $e^{-2\pi\rho}$, which transforms  as a local section of the line bundle ${\rm Sym}^N\calo_X(-\cald^3)$ -- see discussion below \eqref{chargecond}.
Under such transformation the operators $\calb_{{\mathfrak m}_3,\mathfrak{m}_4}$ preserve the same form but are generically mixed among themselves.} 
\be\label{calBY12}
\begin{aligned}
\calb_{{\mathfrak m}_3,\mathfrak{m}_4}\equiv e^{-2\pi (m^I_3 z^3_I+ m^I_4 z^4_I)}|_{\rm Sym}\,e^{-2\pi \rho}\,,\quad 0\leq m^I_3+m^I_4\leq 1
\end{aligned}\,.
\ee

According to the general discussion of subsection \ref{sec:M5},
these operators can be associated to M5-branes wrapping a divisor belonging to the $\mathbb{P}^2$-family of divisors  $D_{[c_1:c_2:c_3]}=\{c^1Z_1+c^2Z_2+c^3Z_3=0\}$. Indeed, from \eqref{Y12scaling} we see that 
\be
\Delta(\calb_{{\mathfrak m}_3,\mathfrak{m}_4})=\frac12NP^3=\frac{\pi N{\rm vol}(\Pi_{[c_1:c_2:c_3]})}{6{\rm vol}(Y)} \simeq N \times 0.4505 \,,
\ee
with $\Pi_{[c_1:c_2:c_3]}=\del D_{[c_1:c_2:c_3]}$ -- see the comment after \eqref{Y12bp}. 

As mentioned in subsection \ref{sec:M5}, one may consider a probe M5-brane wrapping $D_{[c_1,c_2,c_3]}$ in the resolved M-theory geometry and compute the expectation value of the corresponding operator according   to the prescription proposed in \cite{Klebanov:2007us}. By quantizing the dependence on the projective coordinates $[c^1:c^2:c^3]\in\mathbb{P}^2$ one gets exactly  \eqref{calBY12}. We omit the details here, since they are identical to the IIB case discussed in  \cite{Martucci:2016hbu}. However, as argued in subsection \ref{sec:M5}, our  formulation makes more direct the connection with the  M5-brane  states: in the holographic EFT in terms of vector multiplets, the M5-branes states correspond to  BPS monopoles   which can be conformally mapped to the  chiral operators \eqref{Y12chiral}. In particular, since the monopoles generate a three-dimensional radial flow $\sigma=\frac{n}{2|x|}$, they can be described within the EFT considered in section \ref{subsec:HEFT_Y12} only for $n\geq 0$. This is related to the fact that the operators \eqref{Y12chiral} with $n\geq 0$ are the natural observables to parametrize the moduli space corresponding with $\sigma\geq 0$. 

Consider instead the `lightest' Betti operators of negative Betti charge $n=-1$ and $n=-2$ respectively:
\be\label{calBtildeY12}
\tilde\calb_{1}= e^{-2\pi \sum_Iz_I^2}e^{2\pi\rho}\quad,\quad \tilde\calb_{2}= e^{-2\pi \sum_Iz_I^1}e^{4\pi\rho}
\ee
These can be associated  with Euclidean M5-branes wrapping the toric divisors $\cald^5$ and $\cald^4$ respectively. They have the correct Betti charges since, in relative homology, we have $[\cald^5]= -[\cald^3]$ and $[\cald^5]\simeq -2[\cald^3]$. Furthermore,  they have the expected scaling dimensions
\be
\begin{split}
\Delta(\tilde\calb_{1})&=\frac12NP^5=\frac{\pi N{\rm vol}(\Pi^5)}{6{\rm vol}(Y)}\simeq N \times 0.4136 \\ 
\Delta(\tilde\calb_{2})&=\frac12N P^4=\frac{\pi N{\rm vol}(\Pi^4)}{6{\rm vol}(Y)}\simeq N \times 0.2349\,, 
\end{split}
\ee
where $\Pi^4=\del \cald^5$ and $\Pi^5=\del \cald^5$. 
It is clear that as we move one or more M2-branes onto the resolved $\mathbb{P}^2$ (for $\sigma>0$), corresponding to the locus $e^{-2\pi z^1}=e^{-2\pi z^2}=0$, the VEVs of the operators $\tilde\calb_{1}$ and  $\tilde\calb_{2}$ vanish. Loosely  speaking, in the branch  $\sigma\geq 0$, $\tilde\calb_{1}$ and  $\tilde\calb_{2}$ are blind to the resolution of the Calabi-Yau geometry, which is instead detected by the operators \eqref{calBY12}. In other words, in the branch $\sigma\geq 0$ the moduli space can be probed by the combination of mesonic operators $  M_{{\bf m}}$ and the Betti operators  $\calb_{{\mathfrak m}_3,\mathfrak{m}_4}$,  while the VEVs of  $\tilde\calb_{1}$ and $\tilde\calb_{2}$  are fixed by the chiral ring relations  
\be
\begin{aligned}
\calb_{{\mathfrak m}_3,\mathfrak{m}_4}\tilde\calb_{1}=M_{{\bf m}}|_{m^1_2=\ldots =m^N_2=1}\,,\\
\calb_{{\mathfrak m}_3,\mathfrak{m}_4}\tilde\calb_{2}=M_{{\bf m}}|_{m^1_1=\ldots =m^N_1=1}\,.
\end{aligned}
\ee

We conclude this section by matching chiral operators between the holographic EFT and the microscopic Chern-Simons quiver gauge theory.  The operators \eqref{Y12mesonic} are  holographic EFT  realization of mesonic operators in the quiver gauge theory. Using the correspondence \eqref{quivertoric0} for each $I$, we can match the holographic effective field theory operator \eqref{Y12mesonic} with the `mesonic' chiral operator $\calo_{{\bf \mathfrak q}, n=0;\,\bf\alpha, \beta}$ of equation \eqref{chiralopsFT_Y12} in the microscopic quiver gauge theory, under the following identification of quantum numbers:
\be\label{EFT-FT_identification_Y12}
m_1^I = -\frac{q_I}{2}+ \frac{|q_I|}{2} +\alpha_1^I~, \qquad m_2^I=\alpha_2^I~, \qquad  m_3^I=\beta_1^I~, \qquad m_4^I=\beta_2^I~. 
\ee

Similarly, the Betti operators \eqref{calBY12} and \eqref{calBtildeY12} are the holographic EFT realizations of the baryonic operators \eqref{UVbaryons} of the microscopic quiver gauge theory. In particular, the holographic EFT chiral field $\rho$ and  the UV chiral field $\chi$ are related as follows: 
\be
\rho\simeq \chi-\frac1{2\pi}\log T^{(N)}\,.
\ee

The identifications between operators are to be understood up to a potential mixing with products of lighter operators, which we are not concerned about here, see footnote \ref{sym_polys}.
We have explicitly reproduced the scaling dimensions \eqref{dimensions_Y12_FT} of these operators. The other toric charges can similarly be reproduced. 

More generally, the quiver gauge theory counterparts of the operators \eqref{Y12chiral} are given by the baryonic operators $\calo_{{\bf \mathfrak q}, n;\,\bf\alpha, \beta}$ of equation \eqref{chiralopsFT_Y12}, with the identification \eqref{EFT-FT_identification_Y12}.

\section{The \texorpdfstring{$Q^{111}$}{Q111} model}
\label{sec:Q111model}

We now consider the toric model engineered starting from the Calabi-Yau over the so-called $Q^{111}$ Sasaki-Einstein space \cite{Benishti:2010jn}.\footnote{Aspects of the $Q^{111}$ holographic EFT have been studied in \cite{BilloThesis}.}
Differently from the $Y^{12}(\mathbb{P}^2)$ model,  $Q^{111}$ is a regular Sasaki-Einstein space and will be characterised by rational scaling dimensions. On the other hand, this model has other properties which allow us to concretely test other aspects of our holographic EFTs. 

\begin{figure}[t]
\centering
	\includegraphics[scale=0.6]{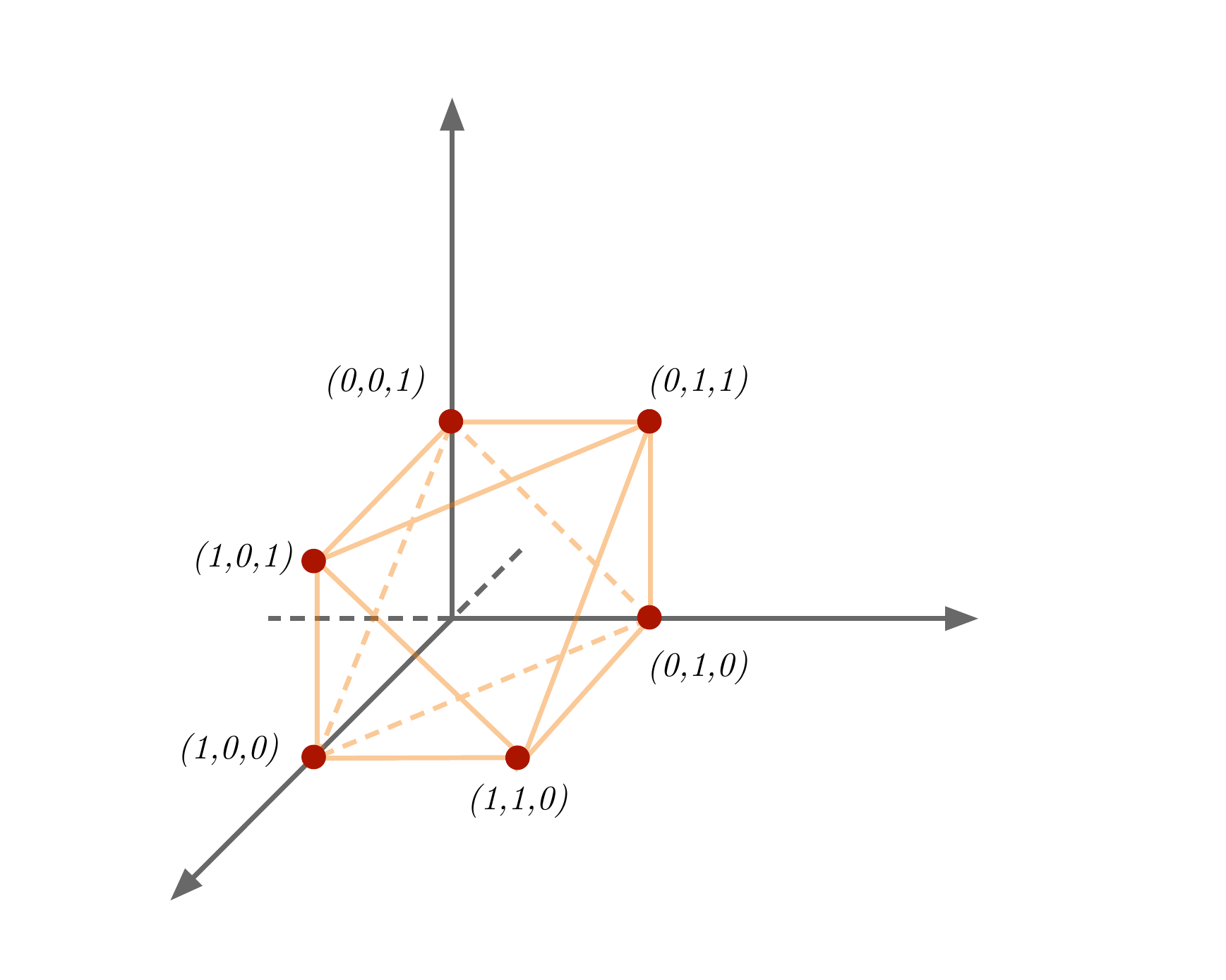}
\caption{The toric fan of $Q^{111}$ in the ${\bf w}$-space.}
	\label{fig:Fan-Q111}
\end{figure}

The resolution of the K\"ahler cone over $Q^{111}$ can be regarded as a toric space with six homogeneous coordinates $Z_A$, $A=1,\ldots,6$, charged under the gauge group $U(1)^2$. The charges of the fields can be chosen as follow 
\be\label{Q111_GLSM}
\begin{array}{ c | c c c c c c | c }
	& Z_1 & Z_2 & Z_3 & Z_4 & Z_5 & Z_6 & \mathrm{FI}\\\hline
	Q^{1A} & 1 & 1 & 0 & 0 & -1 & -1 & \sigma^1\\
	Q^{2A} & 0 & 0 & 1 & 1 & -1 & -1 & \sigma^2\\
\end{array}
\ee
where $\sigma^a$, $a=1,2$, are the K\"ahler moduli. We can choose the following generators of the toric fan edges:
\be
\label{Q111_TFan}
\begin{aligned}
	&\mathbf{v}_1=(1,1,-1,0)\,, \ \mathbf{v}_2=(0,0,1,0)\,, \,\mathbf{v}_3=(1,1,0,-1)\,,
	\\
	&\mathbf{v}_4=(0,0,0,1)\,, \quad \mathbf{v}_5=(1,0,0,0)\,,  \ \mathbf{v}_6=(0,1,0,0)\,.
\end{aligned}
\ee
which indeed satisfy $Q^{1A}\mathbf{v}_A=Q^{2A}\mathbf{v}_A=0$. By an appropriate $SL(4,\mathbb{Z})$-transformation, they identify the toric diagram depicted in Fig.~\ref{fig:Fan-Q111}.\footnote{The vertices ${\bf w}_A \in \mathbb{Z}^3$ of the toric diagram  correspond to  the  vectors $({\bf w}_A,1)^T=M{{\bf v}_A}^T\in \mathbb{Z}^4$ with 
$$
M = \begin{pmatrix} 1 & 0 & 0 & 1 \\ 1 & 0 & 1 & 0 \\ 0 & 1 & 1 & 1 \\ 1 & 1 & 1 & 1 \end{pmatrix}\quad \Rightarrow\quad \begin{array}{ll}
     &  \mathbf{w}_1=(1,0,0)\,, \ \mathbf{w}_2=(0,1,1)\,, \,\mathbf{w}_3=(0,1,0)\,,\\
     & \mathbf{w}_4=(1,0,1)\,, \quad \mathbf{w}_5=(1,1,0)\,,  \ \mathbf{w}_6=(0,0,1)\,.
\end{array}
$$
}

The extended K\"ahler moduli space parametrized by $(\sigma^1,\sigma^2)$ is given by the entire $\mathbb{R}^2$ and is divided in the three K\"ahler cones
\be\label{KcQ111}
\begin{aligned}
	\calc_{\rm I}&= \{\sigma^1\geq 0,\sigma^2\geq 0\}\\
	\calc_{\rm II}&=\{\sigma^1\leq 0,\sigma^2\geq \sigma^1\}\\
	\calc_{\rm III}&=\{\sigma^2\leq 0,\sigma^1\geq \sigma^2\}
\end{aligned}
\ee
It is easy to see that the three K\"ahler cones are isomorphic and can be related by taking different linear combinations of the charges. In each K\"ahler cone, resolving the geometry amounts to substituting the apex of the cone with two $\mathbb{P}^1$s. The resolved geometry $X$ can be identified with the total space of the bundle $\calo_B(-1,-1)\oplus\calo_B(-1,-1)$ over the base $B \simeq \mathbb{P}^1_{(1)} \times \mathbb{P}^1_{(2)}$ and the singular cone is recovered in the limit $\sigma^1,\sigma^2 \to 0$. The different K\"ahler cones are connected by flop transitions, in which one $\mathbb{P}^1$ blows-down and a new one blows-up.

Since the three K\"ahler cones are isomorphic, we  will first describe the explicit K\"ahler and symplectic structures in chamber $\calc_{\rm I}$, and then show how they are glued to same structures on the other K\"ahler cones. In this way, we will be able to extract the holographic EFT on the entire extended (geometrical) moduli space of the model.

\subsection{The resolved geometry}

In this subsection we restrict to the first K\"ahler cone, that is, we assume that $(\sigma^1,\sigma^2)\in \calc_{\rm I}$.
We can use \eqref{Q111_TFan} in \eqref{zZrelation} to get  the  definition of the local toric coordinates $e^{2\pi z^i}$ in terms of homogeneous coordinates:
\be\label{Q111_coord}
\begin{aligned}
&e^{-2\pi z^1}=Z_1Z_3Z_5\equiv \zeta_1 \,,\quad e^{-2\pi z^2}=Z_1Z_3Z_6\equiv\zeta_2 \,,\\
&e^{-2\pi z^3}=\frac{Z_2}{Z_1}\equiv\lambda_1 \,,\quad e^{-2\pi z^4}=\frac{Z_4}{Z_3}\equiv\lambda_2 \,.
\end{aligned}
\ee
The new local coordinates $(\zeta_1,\zeta_2,\lambda_1,\lambda_2)$ are useful to make manifest its bundle structure of $X$: $\lambda_1$ and $\lambda_2$ are respective local coordinates over the $\mathbb{P}^1_{(1)}$ and $\mathbb{P}^1_{(2)}$ in the base $B$, while $\zeta_1$ and $\zeta_2$ are the coordinates over the two $\calo_B(-1,-1)$ fibers. These local coordinates $e^{-2\pi z^i}$ are good on $X\backslash (\cald^1\cup\cald^3)$ where, as above, $\cald^A\equiv \{Z_A=0\}$.

The Calabi-Yau metric on the resolved geometry has been already computed in the literature \cite{Cvetic:2000db,Cvetic:2001ma,Benishti:2010jn}.
We are interested in its K\"ahler structure and here we present just the final result, giving more details in appendix \ref{app:Q111}. 
In the local patch introduced above, the K\"ahler potential can be written as
\be
\label{Q111_k}
k_X(z,\bar z; \sigma) = U(t; \sigma) + \sigma^1 k_{(1)}+ \sigma^2 k_{(2)}\,,
\ee
where $k_{(1)}$ and $k_{(2)}$ are the canonical K\"ahler potentials on $\mathbb{P}^1_{(1)}$ and $\mathbb{P}^1_{(2)}$ respectively, 
\be
\label{Q111_kP}
k_{(1)}=\frac{1}{2\pi} \log(1+|\lambda_1|^2) \,,\quad k_{(2)}=\frac{1}{2\pi} \log(1+|\lambda_2|^2)
\ee
and $U(t; \sigma)$, besides carrying a dependence on the K\"ahler moduli $\sigma^a$, solely depends on the globally defined radial coordinate
\be
\label{Q111t}
\begin{aligned}
t &\equiv (|\zeta_1|^2+|\zeta_2|^2) e^{2\pi k_{(1)}+2\pi k_{(2)}}\\
&=(|Z_1|^2+|Z_2|^2)(|Z_3|^2+|Z_4|^2)(|Z_5|^2+|Z_6|^2)\,.
\end{aligned}
\ee
By imposing the Calabi-Yau condition as well as consistency with \eqref{kdec} and \eqref{bcond}, one obtains $U(t;\sigma)= U_{\rm I}(t;\sigma)$, with
\be\label{Q111Udef}
U_{\rm I}(t;\sigma)\equiv-\frac1{\pi}\lim_{\Lambda\rightarrow\infty}\left[\int_{\sqrt{t}}^\Lambda \frac{\d \tau}{\tau} \, {\mathfrak G}_0(\tau;\sigma)-2^{\frac{1}{4}}\Lambda+\frac13(\sigma^1+\sigma^2)\log\Lambda\right]
\ee
where ${\mathfrak G}_0(\tau;\sigma)$ is the real function defined by the equation \be\label{Q111algeq2}
\frac14{\mathfrak G}_0^4+\frac13(\sigma^1+\sigma^2){\mathfrak G}_0^3+\frac12\sigma^1\sigma^2\,{\mathfrak G}_0^2=\frac12\tau^4\,. 
\ee 
We observe that ${\mathfrak G}_0(\tau;\sigma)$ monotonically increases from $0$ to $\infty$ for $\tau\in[0,\infty)$. One can then trade $\tau$ for ${\mathfrak G}_0$ as integration variable in the integral appearing in \eqref{Q111Udef}.  Omitting irrelevant terms linear in $(\sigma^1+\sigma^2)$,  we can write \eqref{Q111Udef} in the less implicit form:
\be
\label{Q111Uexpl}
\begin{aligned}
U_{\rm I}(t;\sigma)=&\,\frac1{\pi}{\mathfrak G}_0(\sqrt{t},\sigma)+\frac{1}{3\pi}(\sigma^1+\sigma^2)\left[{\mathfrak G}_0(\sqrt{t},\sigma)-\log\left(|\zeta_1|^2+|\zeta_2|^2\right)\right]\\
&-\frac1{3\pi}(\sigma^1+\sigma^2)\log(1+|\lambda_1|^2)-\frac1{3\pi}(\sigma^1+\sigma^2)\log(1+|\lambda_2|^2)\\
&+\frac1{6\pi}\sqrt{A(\sigma) }\log\left(\frac{3{\mathfrak G}_0(\sqrt{t},\sigma)+2(\sigma^1+\sigma^2)-2\sqrt{A(\sigma) }}{3{\mathfrak G}_0(\sqrt{t},\sigma)+2(\sigma^1+\sigma^2)+2\sqrt{A(\sigma) }}\right)
\end{aligned}
\ee
where
\be
A(\sigma) \equiv (\sigma^1)^2+(\sigma^2)^2-\frac52\sigma^1\sigma^2 ~.
\ee 
 Note that the logarithm appearing in \eqref{Q111Uexpl} is real also for $A<0$, provided we choose the same branch for all $\sqrt{A}$'s (say, $\sqrt{-|A|}=\ii\sqrt{|A|}$) and $\log$ such that $\log z=\log |z|+\ii \theta$ where $z=|z|e^{\ii\theta}$ with $- \pi < \theta < \pi$ (that is, the $\log$ branch cut along the negative horizontal semi-axis.). Indeed, in this way the last term of  (\ref{Q111Uexpl}) is holomorphic in $A(\sigma)$ at $A(\sigma)=0$.

We observe that the function \eqref{Q111Uexpl} depends on ${\mathfrak G}_0$, which is only implicitly determined by \eqref{Q111algeq2}. Hence, our description of K\"ahler structure of the resolved geometry is somewhat indirect. On the other hand, as we will presently see (and as we have already experienced in the $Y^{12}(\mathbb{P}^2)$ model), it turns out that the symplectic description of the resolved geometry allows for a  completely explicit description. 

The symplectic description  is obtained by going to the dual variables \eqref{lmindef}. In the present case we obtain
\be
\label{Q111_li}
\begin{aligned}
	& l_1 =  \frac{|\zeta_1|^2}{|\zeta_1|^2+|\zeta_2|^2}\,{\mathfrak G}_0\big(\sqrt{t};\sigma\big)\,,\quad l_2 =  \frac{|\zeta_2|^2}{|\zeta_1|^2+|\zeta_2|^2}\,{\mathfrak G}_0\big(\sqrt{t};\sigma\big)
	\\
	& l_3 =\frac{|\lambda_1|^2 }{1+|\lambda_1|^2}\left[ \sigma^1 + {\mathfrak G}_0\big(\sqrt{t};\sigma\big)  \right] \,,\quad l_4= \frac{|\lambda_2|^2 }{1+|\lambda_2|^2}\left[ \sigma^2 + {\mathfrak G}_0\big(\sqrt{t};\sigma\big) \right] 
\end{aligned}
\ee
By direct inspection one can easily realise that they span the polytope
\be
\label{Q111_Polyt}
\calp_{\sigma} = \{l_1,l_2,l_3,l_4\geq 0\} \cap \{l_1+l_2-l_3\geq-\sigma^1 \} \cap \{l_1+l_2-l_4\geq -\sigma^2\}
\ee
This is compatible with the one directly obtained from \eqref{poly} for the toric fan \eqref{Q111_TFan}, if we take $M_{A1} = \delta_{A1} $, $M_{A2} =\delta_{A3}$, and, accordingly,  $\chi_{A} =  \delta_{A1} \sigma^1+ \delta_{A3} \sigma^2$.

We can now compute $F_X$ from \eqref{sympot}. By inverting \eqref{Q111_li}, it can be explicitly expressed in terms of $l_i$ and $\sigma$, or alternatively the coordinates introduced in \eqref{poly}:
\be\label{Q111sA}
\begin{aligned}
&s_1 = l_1+l_2-l_3+\sigma^1 \,,\quad s_2 = l_3\,,\quad s_3 = l_1+l_2-l_4+\sigma^2\,,\\	
&s_4 = l_4\,,\quad s_5 = l_1\,,\quad s_6 = l_2\,.
\end{aligned}
\ee
In particular, the symplectic potential $F_X$ can then be rewritten in the form
\eqref{FXdec} with 
\begin{subequations}\label{sPQ111}
\begin{align}
s_{P_{\rm can}} &= 3 l_1+3 l_2 + \sigma^1+\sigma^2\label{sPQ111a}\,,\\
s_{P} &= 2 l_1+2 l_2 + \frac23\left(\sigma^1+\sigma^2\right)\label{sPQ111b}\,,
\end{align}
\end{subequations}
and $f_X(l,\sigma)=f^{\rm I}_X(l,\sigma)$ with\footnote{As for the $Y^{12}(\mathbb{P}^2)$ model, in  $f^{\rm I}_X$  we have omitted terms linear in $l_i$ and $\sigma$ -- see footnote \ref{foot:fX}.} 
\be\label{fXI}
\begin{split}
	f^{\rm I}_X(l,\sigma) =&\,  \frac{1}{4 \pi} s_P \log \frac{\calh(l,\sigma)}{s_P}+\frac1{6\pi}\sqrt{A(\sigma) }\log\left(\frac{3 (l_1+l_2)+2(\sigma^1+\sigma^2)-2\sqrt{A(\sigma) }}{3(l_1+l_2)+2(\sigma^1+\sigma^2)+2\sqrt{A(\sigma) }}\right)
	\\
	&+\frac{1}{2\pi} (l_1+l_2+\sigma^1) \log\frac{l_1+l_2+\sigma^1}{\calh(l,\sigma)}+\frac{1}{2\pi} (l_1+l_2+\sigma^2) \log \frac{l_1+l_2+\sigma^2}{\calh(l,\sigma)}\,,
\end{split}
\ee
where
\be
\calh(l,\sigma)\equiv \sqrt{3 (l_1+l_2)^2 + 4(\sigma^1+\sigma^2)(l_1+l_2) + 6\sigma^1\sigma^2}\,.
\ee
Comparing \eqref{sPQ111b} and \eqref{sP} we obtain
\be
P^A=\frac23\quad~~~\text{for}\quad A=1,\ldots,6
\ee
or, equivalently, 
\be
b^1= b^2=2\quad,\quad b^3=b^4=0\quad,\quad p_1=p_2= \frac23\,.
\ee
One can also check that relation \eqref{Pvolrel} is  satisfied.

\subsection{Extended description of the resolved geometry}

The descriptions of the resolved Calabi-Yau spaces in the other K\"ahler cones can be obtained via a reshuffling of the projective coordinates and of the charges. For instance, the resolution on the K\"ahler cone II can be obtained by swapping $(Z_1,Z_2)$ and $(Z_5,Z_6)$ and considering as new sets of charges $-Q^{1A}$ and $Q^{2A}-Q^{1A}$, corresponding to new K\"ahler moduli $-\sigma^1$ and $\sigma^2-\sigma^1$. (Notice that the radial variable \eqref{Q111t} is invariant under such redefinitions.) 
In this way we can get a unified description of the resolved geometry and select a K\"ahler potential $k_X$ which has  the form \eqref{Q111_k} at any point $(\sigma^1,\sigma^2)\in\mathbb{R}^2$. The function  $U(t;\sigma)$ can  be written in the form \eqref{Q111Udef}:
\be\label{Q111Udef2}
U(t;\sigma)=-\frac1{\pi}\lim_{\Lambda\rightarrow\infty}\left[\int_{\sqrt{t}}^\Lambda \frac{\d \tau}{\tau} \, {\mathfrak G}(\tau;\sigma)-2^{\frac{1}{4}}\Lambda+\frac13(\sigma^1+\sigma^2)\log\Lambda\right]
\ee
where now
\be
{\mathfrak G}(\tau,\sigma)=\left\{\begin{array}{ll}
{\mathfrak G}_{\rm I}(\tau,\sigma^1,\sigma^2)={\mathfrak G}_0(\tau,\sigma^1,\sigma^2)&\quad\text{for $(\sigma^1,\sigma^2)\in\calc_{\rm I}$}\\
{\mathfrak G}_{\rm II}(\tau,\sigma^1,\sigma^2)={\mathfrak G}_0(\tau,-\sigma^1,\sigma^2-\sigma^1)-\sigma^1&\quad\text{for $(\sigma^1,\sigma^2)\in\calc_{\rm II}$}\\
{\mathfrak G}_{\rm III}(\tau,\sigma^1,\sigma^2)={\mathfrak G}_0(\tau,\sigma^1-\sigma^2,-\sigma^2)-\sigma^2&\quad\text{for $(\sigma^1,\sigma^2)\in\calc_{\rm III}$}
\end{array}\right.
\ee

It is important to observe that ${\mathfrak G}(\tau,\sigma)$, solves the differential equation
\be\label{GCYcond}
{\mathfrak G}(\tau,\sigma)\left[{\mathfrak G}(\tau,\sigma)+\sigma^1\right]\left[{\mathfrak G}(\tau,\sigma)+\sigma^2\right]\frac{\del {\mathfrak G}(\tau,\sigma)}{\del \tau}=2\tau^3
\ee
 for any $(\sigma^1,\sigma^2)\in \mathbb{R}^2$. Indeed, \eqref{GCYcond} is equivalent to the Calabi-Yau condition for the K\"ahler potential \eqref{Q111_k}. The different descriptions of ${\mathfrak G}(\tau,\sigma)$ on the different K\"ahler cones can then be associated with different values of the integration constant  that one gets by integrating \eqref{GCYcond} and imposing appropriate regularity conditions. Indeed, \eqref{GCYcond} can be integrated into
  \be\label{Q111algeq2gen}
\frac14{\mathfrak G}^4+\frac13(\sigma^1+\sigma^2){\mathfrak G}^3+\frac12\sigma^1\sigma^2\,{\mathfrak G}^2=\frac12\tau^4 +h(\sigma)
\ee
where $h|_{\calc_{\rm I}}=h_{\rm I}$, $h|_{\calc_{\rm II}}=h_{\rm II}$ and $h|_{\calc_{\rm III}}=h_{\rm III}$, with
\be
h_{\rm I}\equiv 0\quad~~~~~ h_{\rm II}\equiv -\frac1{12}(\sigma^1)^3(\sigma^1-2\sigma^2)\quad~~~~~ h_{\rm III}\equiv -\frac1{12}(\sigma^2)^3(\sigma^2-2\sigma^1)\,.
\ee
Note that $h_{\rm II}-h_{\rm I}$ is proportional to $(\sigma^1)^2$. Hence ${\mathfrak G}$ and its first and second derivatives in $\sigma^1$ are continuous across the K\"ahler wall $\sigma^1=0$ which separates $\calc_{\rm I}$ and $\calc_{\rm II}$. Analogous observations hold for the behaviour of ${\mathfrak G}$ across the other   K\"ahler walls.   

A more explicit form of $U_{\rm II}\equiv U|_{\calc_{\rm II}} $ and $U_{\rm III}\equiv U|_{\calc_{\rm III}} $ can be obtained from that of $U_{\rm I}\equiv U|_{\calc_{\rm I}} $, see  \eqref{Q111Uexpl}, and the relations
\be\label{UIUIIUIII}
\begin{aligned}
U_{\rm II}(t; \sigma^1,\sigma^2) &= U_{\rm I}(t; -\sigma^1,\sigma^2-\sigma^1) -\frac{1}{2\pi}\sigma^1\log t\,,\\
U_{\rm III}(t; \sigma^1,\sigma^2) &= U_{\rm I}(t; \sigma^1-\sigma^2,-\sigma^2) -\frac{1}{2\pi}\sigma^2\log t\,. 
\end{aligned}
\ee 

Let us now pass to the symplectic formulation of section \ref{sec:symp} in terms of the variables \eqref{lmindef}. We have seen that in $\calc_{\rm I}$ they are related to the complex coordinates as in \eqref{Q111_li}. On the other K\"ahler cones $\calc_{\rm II}$ and $\calc_{\rm III}$  one has just to modify \eqref{Q111_li} by substituting ${\mathfrak G}_0\equiv {\mathfrak G}_{\rm I}$ with ${\mathfrak G}_{\rm II}$ and ${\mathfrak G}_{\rm III}$, respectively.  
 By inverting these relations, one can then compute $F_X$ from \eqref{sympot}. By taking into account \eqref{UIUIIUIII}, one finds that  $F^{\rm II}_X\equiv F_X|_{\calc_{\rm II}}$ can be obtained by making the following substitutions in $F^{\rm I}_X\equiv F_X|_{\calc_{\rm I}}$:
\be\label{symItoII}
\begin{array}{l}
\sigma^1\rightarrow -\sigma^1\ ,\quad \sigma^2\rightarrow \sigma^2-\sigma^1\\
l_1\rightarrow l_1+l_2-l_3+\sigma^1 \\
l_2\rightarrow l_3\ ,\quad l_3\rightarrow l_2\quad l_4\rightarrow l_4
\end{array}\quad~~~ \text{(From $\calc_{\rm I}$ to $\calc_{\rm II}$)}\,.
\ee
Analogously, we can obtain $F^{\rm III}_X\equiv F_X|_{\calc_{\rm III}}$ from $F^{\rm I}_X$ by making the substitutions
\be\label{symItoIII}
\begin{array}{l}
\sigma^1\rightarrow \sigma^1-\sigma^2\ ,\quad \sigma^2\rightarrow -\sigma^2\,,\\
l_1\rightarrow l_1+l_2-l_4+\sigma^2 \\
l_2\rightarrow l_4\ ,\quad l_3\rightarrow l_3\ ,\quad l_4\rightarrow l_2\,.
\end{array}\quad~~~ \text{(From $\calc_{\rm I}$ to $\calc_{\rm III}$)}\,.
\ee

Note that the polytope $\calp_\sigma$ introduced in \eqref{Q111_Polyt} is invariant under \eqref{symItoII} and \eqref{symItoIII}, so that  $l_i$'s  parametrize $\calp_\sigma$ for any $(\sigma^1,\sigma^2)\in\mathbb{R}^2$. Furthermore, one can easily check that the changes \eqref{symItoII} and \eqref{symItoIII} correspond to permutations of the variables $s_A$'s introduced in \eqref{Q111sA} and leave $s_{P_{\rm can}}$ and $s_P$ invariant. Hence, as expected, $F^{\rm II}_X$ and $F^{\rm III}_X$ can be written in the form  \eqref{FXdec} with the same $s_{P_{\rm can}}$ and $s_P$ \eqref{sPQ111} found for $F^{\rm I}_X$, but with different homogeneous terms $f^{\rm II}_X$ and $f^{\rm III}_X$ which can be obtained from \eqref{fXI} by making the substitutions \eqref{symItoII} and \eqref{symItoIII}, respectively. $f^{\rm I}_X$, $f^{\rm II}_X$ and $f^{\rm III}_X$ can be glued into a single   $f_X$ defined on the entire extended K\"ahler moduli space $(\sigma^1,\sigma^2)\in\mathbb{R}^2$. 
From the above comments on the behaviour of ${\mathfrak G}$ across  the K\"ahler wall, we can  conclude that $f_X$ (and then $F_X$) can have at most discontinuities in $(\sigma^1,\sigma^2)$ of the `third order' (that is, regarding its third derivatives in $(\sigma^1,\sigma^2))$ on the K\"ahler walls.

\subsection{Holographic EFT}

We can now extract the information regarding the holographic EFT. We may straightforwardly repeat, mutatis mutandis, most of the observations made for the $Y^{12}(\mathbb{P}^2)$ model. We will then be brief, leaving some details to the reader.  As for the $Y^{12}(\mathbb{P}^2)$ model, let us start with the formulation in terms of the vector multiplets and the corresponding   $4N+2$ linear multiplets $L_i^I=l_i^I+\ldots- \frac12\epsilon^{\mu\nu\rho}\theta\gamma_\mu\bar\theta (F_i^I)_{\nu\rho}$ and  $\Sigma^a=\sigma^a+\ldots- \frac12\epsilon^{\mu\nu\rho}\theta\gamma_\mu\bar\theta F^a_{\nu\rho}$, $a=1,2$. The EFT action can  be written as in \eqref{Y12eL} with $\tilde F(L,\Sigma)$ obtained by using $F_X$ computed above in \eqref{Ftilde}, that is (omitting a term linear in $3L_1^I+3L^I_2+\Sigma^1+\Sigma^2$) 
\be\label{Q111tildeF}
\begin{aligned}
\tilde\calf(L,\Sigma)=&\,\sum_I\left[-\frac1{2\pi}\sum_A s_A(L^I,\Sigma)\log s_A(L^I,\Sigma)\right]\\
&+\sum_I\left[\frac1{4\pi}s_P(L^I,\Sigma)\log s_P(L^I,\Sigma)+ f_X(L^I,\Sigma)\right]\\
\end{aligned}
\ee
where $s_A(l,\sigma)$ and $s_{P}(l,\sigma)$ are defined in \eqref{Q111sA} and \eqref{sPQ111b}, and $f_X$ is given by \eqref{fXI} in the first K\"ahler cone and by $f^{\rm II}_X$ and $f^{\rm III}_X$, described in the previous subsection, in the other two K\"ahler cones. 

One could now repeat almost verbatim the discussion of section \ref{subsec:HEFT_Y12}. In particular, one can explicitly see  that $\tilde F(L,\Sigma)$ is almost-homogeneous in the sense of \eqref{nonhomcalf} under rescalings of $L^I_i$ and $\Sigma^a$ (which have scaling dimension 1) as required by the superconformal invariance of the EFT. One can also go to the dual formulations involving chiral multiplets. In particular, from \eqref{Q111tildeF} one can compute the explicit form of $\Re z^i_I=\frac12\frac{\del\tilde F}{\del L^I_i}$ and $\Re\rho_a=\frac12\frac{\del\tilde F}{\del \Sigma^a}$. These can be  in principle inverted  to extract the explicit form of the EFT K\"ahler potential, see \eqref{calktoric}, 
\be
\calk=\frac{1}{\pi}\sum_I\left(L_1^I+L_2^I\right)+\frac{N}{3\pi}(\Sigma^1+\Sigma^2)\,.
\ee
One can also verify the following scaling dimensions of the dual chiral fields
\be
\label{Q111_scalzrho}
\begin{aligned}
&\Delta(e^{-2\pi z^1_I})=\Delta(e^{-2\pi z^2_I})=1\,,\\
&\Delta(e^{-2\pi z^3_I})=\Delta(e^{-2\pi z^4_I})=0\,,\\
&\Delta(e^{-2\pi \rho_1})= \Delta(e^{-2\pi \rho_2})=\frac{1}3N
\end{aligned}
\ee
As for the $Y^{12}(\mathbb{P}^2)$ model, there is no unitarity problem with  \eqref{Q111_scalzrho}, since the chiral fields $e^{-2\pi z^i_I}$ and $e^{-2\pi\rho_a}$ are not part of the EFT ring of scalar chiral operators, which will be discussed in the following section.     

\subsection{EFT chiral operators and semiclassical states}

According to the general discussion of section \ref{sec:chiralop}, the chiral operators of the $Q^{111}$ model can be classified by the toric and Betti charges $m^I_i$ and $n^1,n^2$ and   admit the following EFT description
\be\label{Q111chiral}
\calo_{{\bf m},{\bf n}}(x)\equiv    e^{-2\pi \langle {\bf m}^I, {\bm z}_I(x)\rangle}|_{\rm Sym}\,e^{-2\pi n^1\rho_1(x)-2\pi n^2\rho_2(x)} 
\ee
in the holographic EFT. The charges must satisfy the constraints  \eqref{chargecond}, which in the present case become
\be\label{Q111charges}
\begin{aligned}
 m^I_1,m^I_2,m^I_3,m^I_4\geq 0  \quad,\quad m^I_1+m^I_2-m^I_3\geq -n^1 \quad,\quad m^I_1+m^I_2-m^I_4\geq -n^2\,.
\end{aligned}
\ee
From the general formula \eqref{scalingO}, or more directly from \eqref{Q111_scalzrho}, we get the corresponding scaling dimensions:
\be\label{Q111scaling}
\Delta(\calo_{{\bf m},{\bf n}})=\sum_I \big(m_1^I+m_2^I\big)+\frac{1}{3}N(n^1+n^2)
\ee
From \eqref{Q111charges} it is clear that the unitarity bound $\Delta(\calo_{{\bf m},{\bf n}})>\frac12$ is satisfied for any set of charges $m^I_i,n^a$ (since we are assuming large-$N$, and then $N>1$). As discussed in section \ref{sec:semstates},  for large charges these chiral operators can be conformally mapped to semiclassical states of the holographic EFT.

The mesonic operators, which have vanishing Betti charges ($n^1=n^2=0$), can be written as non-negative powers of the homogeneous coordinates 
\be\label{Q111_mesonic}
M_{\bf m}=\Big(\prod_I(Z^I_1)^{m^I_1+m^I_2-m^I_3}(Z^I_2)^{m^I_3}(Z^I_3)^{m^I_1+m^I_2-m^I_4}(Z^I_4)^{m^I_4}(Z^I_5)^{m^I_1}(Z^I_6)^{m^I_2}\Big)_{\rm sym}
\ee
and vanish if one of the M2-branes touches the exceptional $ \mathbb{P}^1\times \mathbb{P}^1$ of the resolved geometry.  
These exceptional loci can instead be probed by Betti operators. The  basic Betti operators have Betti charges $(n^1,n^2)=(1,0)$, $(n^1,n^2)=(0,1)$  and $(n^1,n^2)=(-1,-1)$ and lowest dimension. These are labeled by an integer ${\mathfrak m}=0,1\ldots,N$ and take the form
\be\label{Q111B1}
\begin{array}{ll}
\calb^{(1)}_{{\mathfrak m}}=e^{-2\pi m^I_3 z^3_I}|_{\rm Sym}\,e^{-2\pi \rho_1}\,,&\quad{\mathfrak m}=\sum^N_{I=1} m_3^I\,,\quad 0\leq m^I_3\leq 1\\
\calb^{(2)}_{{\mathfrak m}}=e^{-2\pi m^I_4 z^4_I}|_{\rm Sym}\,e^{-2\pi \rho_2}\,,&\quad{\mathfrak m}=\sum^N_{I=1} m_4^I\,,\quad 0\leq m^I_4\leq 1\\
\calb^{(3)}_{{\mathfrak m}}= e^{-2\pi m^I_1 z^1_I-2\pi m^I_2 z^2_I}|_{\rm Sym}\,e^{2\pi \rho_1+2\pi \rho_2}\,,&\quad{\mathfrak m}=\sum^N_{I=1} m_1^I\,,\quad m^I_1+m^I_2=1
\end{array}
\ee
These operators are dual to semiclassical monopoles of the holographic EFT which represent the states of  a Euclidean M5-brane wrapping a member of the  $\mathbb{P}^1$-families of divisors $\{c^1Z_1+c^2Z_2=0\}$, $\{c^3Z_3+c^4Z_4=0\}$ and $\{c^5Z_5+c^6Z_6=0\}$,  respectively. 
One can indeed check that their scaling dimension $\Delta=\frac13N$ coincides with the usual geometric formula $\frac{\pi N \text{vol}(\Pi)}{6\text{vol}(Q^{111})}$, where $\Pi=\del\cald^1$, $\Pi=\del\cald^3$ and $\Pi=\del\cald^5$ respectively (where the volumes are computed by using the asymptotic Sasaki-Einstein geometry). This also follows from \eqref{Pvolrel} which, as already mentioned, indeed holds for the preset model. 

The semiclassical monopoles associated with the operators \eqref{Q111B1} describe flows along the K\"ahler walls of the moduli space. In order to probe the interior of the K\"ahler cones, one must then monomials in pairs of \eqref{Q111B1}. For instance,  $\calb^{(1)}_{{\mathfrak m}_1}\calb^{(2)}_{{\mathfrak m}_2}$ on $\calc_{\rm I}$, $\calb^{(2)}_{{\mathfrak m}_1}\calb^{(3)}_{{\mathfrak m}_2}$ on $\calc_{\rm II}$ and $\calb^{(1)}_{{\mathfrak m}_1}\calb^{(3)}_{{\mathfrak m}_2}$ on $\calc_{\rm III}$.

We conclude this section by matching chiral operators between the holographic EFT and the microscopic Chern-Simons quiver gauge theory. The operators \eqref{Q111_mesonic} are  the holographic EFT  realization of `mesonic' chiral operators in the microscopic quiver gauge theory. Indeed, using the correspondence \eqref{quivertoricQ111} for each $I$, we can match  \eqref{Q111_mesonic}  with the  $\calo_{{\bf \mathfrak q}, \mathbf{n}=0;\,\bf\alpha, \beta}$ of equation \eqref{chiralopsFT_Q111}, if we identify quantum numbers as follows:
\be\label{EFT-FT_identification_Q111}
\begin{split}
    m_1^I &= \frac{q_I-n^2}{2}+ \frac{|q_I-n^2|}{2} +\alpha_1^I~, \qquad m_2^I=-\frac{q_I}{2}+\frac{|q_I|}{2} +\alpha_2^I~,\\
  &\qquad m_3^I=\beta_2^I~, \qquad m_4^I=-\frac{q_I-n^2}{2}+ \frac{|q_I-n^2|}{2} +\alpha_1^I~. 
\end{split}
\ee
More generally, the quiver  counterparts of the  operators \eqref{Q111chiral} are given by the  operators $\calo_{{\bf \mathfrak q}, \mathbf{n};\,\bf\alpha, \beta}$ of equation \eqref{chiralopsFT_Q111}, with the identification \eqref{EFT-FT_identification_Q111}. In particular, the Betti operators \eqref{Q111B1} are the holographic EFT realizations of the baryonic operators \eqref{UVbaryonsQ111} of the microscopic quiver gauge theory.
\footnote{The usual caveat about potential mixing with products of lighter operators applies, see footnote \ref{sym_polys}.} Note that this matching implies the following microscopic interpretation of the holographic EFT chiral superfields $\rho_a$:
\be
e^{-2\pi\rho_1}\simeq T_{0;1,0}\det B_1= e^{-2\pi\chi_1}\det B_1\quad,\quad e^{-2\pi\rho_2}\simeq T_{1_N;0,1}\,.
\ee

In all these cases, the quantum numbers of operators computed in the holographic EFT and in the microscopic quiver gauge theory agree.



\vspace{1cm}

\centerline{\large\em Acknowledgments}

\vspace{0.5cm}

\noindent We thank I.~Garc\'ia-Etxebarria for discussions and collaboration in a related project, and D.~Cassani, S.~Giusto, A.~Monin, D.~Orlando and A.~Zaffaroni for useful discussions.  SC is supported in part by STFC through grant ST/T000708/1, and thanks GGI and the workshop ``Supersymmetric Quantum Field Theories in the Non-perturbative Regime'' for hospitality and support during the preliminary stages of this project. SL is supported by the Dutch Research Council (NWO) via a Start-Up grant. 
SC thanks his collaborators for their patience during the very long gestation of this paper, and dedicates the paper to everyone who is fighting to make UK universities the safe, fair and sustainable place that workers and students deserve.

\vspace{2cm}

\newpage

\centerline{\LARGE \bf Appendix}
\vspace{0.5cm}

\begin{appendix}


\section{Summary on three-dimensional superspace}
\label{app:3dsuperspace}

In this appendix we collect our conventions and some useful computations on three-dimensional $\mathcal{N}=2$  superspace that we have used throughout the paper. Recall that the three-dimensional $\mathcal{N}=2$ superspace is described by three spacetime coordinates $x^\mu$, $\mu = 0,1,2$, and a set of four real Grassmann-odd variables, which we collect in the complex spinor $\theta^\alpha$. As in  \cite{Intriligator:2013lca}, these quantities can be obtained readily by reducing those of \cite{Wess:1992cp} for four-dimensional $\caln=1$ superspace down to three spacetime dimensions as follows. First, we `forget' the third space-time coordinate $x^2$; secondly, since in three spacetime dimensions the irreducible spinorial representation of the Lorentz group is a real spinor $\psi^\alpha$, $\alpha = 1,2$, we delete the `dots' over the anti-chiral spinor indices. In particular, spinor indices are raised and lowered by  the two-dimensional Levi-Civita symbol $\varepsilon^{\alpha\beta}$ as
\begin{equation}
	\label{app:3ds_psia}
	\psi_\alpha = \varepsilon_{\alpha\beta} \psi^\beta\,,\qquad \psi^\alpha = \varepsilon^{\alpha\beta} \psi_\beta
\end{equation}
with $\varepsilon^{01}= - \varepsilon_{01} = + 1$. The contractions of spinor indices are always understood to be NW-SE; for instance:
\begin{equation}
	\label{app:3ds_psicontr}
	\psi \chi \equiv \psi^\alpha \chi_\alpha\,, \qquad \psi\bar\chi\equiv \psi^\alpha\bar\chi_\alpha\,, \qquad {\rm etc.}
\end{equation}
Furthermore, the hermitian conjugate is defined as $\psi^\dagger_\alpha = \bar{\psi}_\alpha$, that implies, for instance, $(\psi \bar \chi)^\dagger = - \chi \bar \psi$. The three-dimensional gamma matrices 
are defined from the Pauli matrices as 
\begin{equation}
	\label{app:3ds_gamma}
	(\gamma^{\mu})_{\alpha}{}^\beta=({\rm i}\sigma_2,\sigma_3,-\sigma_1)\,
\end{equation}
and obey the identity
\begin{equation}
	\label{app:3ds_gammaid}
	(\gamma^\mu)_{\alpha}{}^{\delta} (\gamma^\nu)_{\delta}{}^{\beta} = \eta^{\mu\nu} \delta_\alpha^\beta + \varepsilon^{\mu\nu\rho} (\gamma_\rho)_{\alpha}{}^{\beta}\,,
\end{equation}
with $\eta^{\mu\nu} = {\rm diag}(-1,1,1)$ and the three-dimensional Levi-Civita symbol $\varepsilon^{\mu\nu\rho}$ normalized so that $\epsilon^{012}=1$. 

The supercovariant derivatives are
\begin{equation}
    \label{app:3ds_D}
    \begin{aligned}
    D_\alpha = \frac{\partial}{\partial \theta^\alpha} - \ii (\gamma^\mu \bar\theta)_\alpha \partial_\mu\,, \qquad \bar{D}_\alpha = -\frac{\partial}{\partial \bar\theta^\alpha} + \ii (\gamma^\mu \theta)_\alpha \partial_\mu\,, 
    \end{aligned}
\end{equation}
which satisfy
\begin{equation}
    \label{app:3ds_Drel}
    \begin{aligned}
    \{D_\alpha,D_\beta\} = \{\bar D_\alpha,\bar D_\beta\} = 0\,,\qquad \{D_\alpha,\bar D_\beta\} = - 2 \ii (\gamma^\mu)_{\alpha\beta} \partial_\mu\,.
    \end{aligned}
\end{equation}

As in four-dimensional $\mathcal{N}=1$ supersymmetry, three-dimensional $\mathcal{N}=2$ supersymmetric multiplets can be organized in a superfield expansion in the Grassmann variables $\theta^\alpha$, $\bar{\theta}^\alpha$. Supersymmetric actions are then obtained by appropriately integrating superfields, or combinations thereof, over the Grassmann variables. Specifically, in three-dimensional $\mathcal{N}=2$ superspace, integrals are performed by exploiting $\int {\rm d}^2\theta\,1=0$ and
\begin{equation}
	\label{app:3ds_int}
	\int {\rm d}^2\theta\,\theta^2=-\int {\rm d}^2\bar\theta\,\bar\theta^2=1\,, \qquad \int {\rm d}^4\theta\, \theta^2 \bar\theta^2 \equiv \int {\rm d}^2\theta \int {\rm d}^2\bar\theta\, \theta^2 \bar\theta^2  =  -1  \,.
\end{equation}

We now list the supersymmetric multiplets, as well as their superfield expansions, that are used throughout the work. A chiral superfield $Z$ obeys the relation $\bar D_{\alpha}Z = 0$. Its expansion in bosonic components is
\begin{equation}
	\label{app:3ds_Z}
	Z= z - \ii \theta \gamma^\mu \bar \theta \partial_\mu z + \theta^2 f - \frac14 \theta^2 {\bar\theta}^2 \Box z+\text{(fermions)}\,,
\end{equation}
with $z$, $f$ complex scalar fields. An anti-chiral multiplet $\bar Z$ obeys the relation $D_\alpha \bar Z = 0$ and its component expansion can be obtained from \eqref{app:3ds_Z} by complex conjugation.

A vector multiplet is represented by a real superfield whose components are reduced by enforcing gauge symmetry. A $U(1)$ vector multiplet $V$, in the Wess-Zumino gauge, has the following expansion in bosonic components
\begin{equation}
	\label{app:3ds_V}
	V=-2 {\rm i} \theta\bar\theta \sigma -2\theta\gamma^\mu\bar\theta A_\mu- \theta^2\bar\theta^2 D+\text{(fermions)}
\end{equation}
where $\sigma$ and $D$ are real scalar fields and $A_\mu$ a real vector field. A $U(1)$ gauge transformation is realized by shifting $V\rightarrow V-{\rm i}(\Lambda-\bar\Lambda)$, with $\Lambda$ a chiral superfield. A chiral field of charge $q$ transforms as $\Phi\rightarrow e^{{\rm i} q\Lambda}\Phi$ under a $U(1)$ gauge transformation. The linearized coupling of a current multiplet  ${\cal J}=\ldots-  \theta\gamma^\mu\bar\theta\, j_\mu+\ldots$  to a $U(1)$ vector multiplet $V$ is 
\begin{equation}
-\int {\rm d}^4\theta\, V{\cal J} \supset   j^\mu A_\mu. 
\end{equation}
At the non-linear level, ${\cal J}$ can be obtained  by computing $\delta S=-\int {\rm d}^3x {\rm d}^4\theta \mathcal{J} \delta V $ under  a general variation $\delta V$.

Given a real vector multiplet $V$, we can construct the associated real linear multiplet
\begin{equation}
	\label{app:3ds_sigma}
	\Sigma\equiv \frac{\rm i}4 D\bar D V= \frac{\rm i}4 \epsilon^{\alpha\beta} D_\beta \bar D_\alpha V=-\frac{\rm i}4 \epsilon^{\alpha\beta}\bar D_\alpha D_\beta V\,,
\end{equation}
which obeys $D^2 \Sigma = {\bar D}^2 \Sigma = 0$. Focusing on bosonic components only, its superfield expansion is
\begin{equation}
	\label{app:3ds_sigmaexp}
	\Sigma =\sigma-\frac12 \epsilon^{\mu\nu\rho}\theta\gamma_\mu\bar\theta F_{\nu\rho}+{\rm i} \theta\bar\theta D + \frac14 \theta^2 {\bar\theta}^2 \Box \sigma+\text{(fermions)}
\end{equation}
with $F_{\mu\nu} = 2 \partial_{[\mu} A_{\nu]}$. 

A general $\mathcal{N}=2$ action describing the interaction among neutral chiral multiplets $Z^i$ and linear multiplets $\Sigma^a$ can be obtained from a superspace integral as follows
\be
\label{app:3d_actions}
\begin{aligned}
S=&\,\int\d^3 x\,\d^4\theta\, \calf(Z,\bar Z,\Sigma)\,,
\end{aligned}
\ee
where $\calf(Z,\bar Z,\Sigma)$ is a  real function of the superfields $Z^i$, ${\bar Z}^{\bar\imath}$ and $\Sigma^a$. Substituting the superfield expansions \eqref{app:3ds_Z} and \eqref{app:3ds_sigmaexp}, \eqref{app:3d_actions} produces the following bosonic action
\be
\label{app:3d_actions_2}
\begin{aligned}
S=&\,\frac1{4}\int\calf_{ab} \Big(\d \sigma^a\wedge *\d \sigma^b+F^a\wedge * F^b\Big)-\int\calf_{i\bar\jmath}\,\d z^i\wedge*\d\bar z^{\bar\jmath}\\
& -\frac{\ii}{2}\int\left(\calf_{ai}\d z^i-\calf_{a\bar\imath}\d\bar z^{\bar\imath}\right)\wedge F^a + \int \left(\calf_{i\bar\jmath}\, f^i {\bar f}^{\bar\jmath} - \frac14 \calf_{ab} D^a D^b \right) *1\,,
\end{aligned}
\ee
where, for instance, $\calf_{ai}\equiv\frac{\del\calf}{\del \sigma^a\del z^i}$.

Charged matter can be coupled as follows. For simplicity, we specialize to the case where there is a single $U(1)$ gauge group, with gauge field $A_\mu$ contained in a vector multiplet as in \eqref{app:3ds_V}, and with the linear multiplet $\Sigma$ in \eqref{app:3ds_sigma} providing its super-field-strength. Then, the $\mathcal{N}=2$ supersymmetric Lagrangian containing the gauge kinetic terms, Chern-Simons terms and Fayet-Iliopoulos term is 
\begin{equation}
	\label{app:3ds_LagGauge}
	\begin{aligned}
		\mathcal{L}_{\rm gauge} &= \int {\rm d}^4\theta\, \left(- \frac{1}{e^2} \Sigma^2 - \frac{k}{4\pi} \Sigma V - \frac{\zeta}{2\pi} V\right)\,.
	\end{aligned}
\end{equation}
Matter chiral superfields $Q_i$, with lowest bosonic components $q_i$ and auxiliary fields $f_i$, with charge $n_i$ under the $U(1)$ gauge field can then be coupled in a $\mathcal{N}=2$ supersymmetric fashion via the following Lagrangian:
\begin{equation}
	\label{app:3ds_LagChiral}
	\begin{aligned}
		\mathcal{L}_{\rm matter} &=  \sum\limits_{i=1}^N \int {\rm d}^4\theta\, \bar{Q}_i e^{n_i V + 2\ii m_i \theta \bar\theta} Q_i 
	\end{aligned}
\end{equation}
The bosonic components of the Lagrangians \eqref{app:3ds_LagGauge} and \eqref{app:3ds_LagChiral} are 
\begin{equation}
	\label{app:3ds_LagLinChiral_comp}
	\small{\begin{aligned}
		\mathcal{L}_{\rm B} &= \left(\mathcal{L}_{\rm gauge} + \mathcal{L}_{\rm matter}\right)_{\rm bosonic}
		\\
		&= - \frac{1}{4 e^2} F^{\mu\nu} F_{\mu\nu}  - \frac{1}{2 e^2} \partial^\mu \sigma \partial_\mu \sigma + \frac{1}{2 e^2} D^2 +\frac{k}{4\pi} \varepsilon^{\mu\nu\rho} A_\mu \partial_\nu A_\rho  - \frac{k}{2\pi} \sigma D - \frac{\zeta}{2\pi} D\,
		\\
		&\quad\,+\sum\limits_{i=1}^N \Big\{ - \partial^\mu \bar{q}_i \partial_\mu q_i + \bar{f}_i f_i  +(\ii \partial_\mu \bar{q}_i A^\mu q_i + {\rm c.c.} )  + \bar{q}_i \left[D - A^\mu A_\mu -  (\sigma -m_i)^2 \right] q_i\Big\}\,.
	\end{aligned}}
\end{equation}
The auxiliary fields $f_i$, $D$ can be integrated out by using their equations of motion,
\begin{equation}
\label{app:3ds_LagLinChiral_D}
    f_i = 0\,, \qquad D= - e^2 \left( \sum\limits_{i=1}^N n_i \bar{q}_i q_i - \frac{k}{2\pi} \sigma - \frac{\zeta}{2\pi}\right)\,,
\end{equation}
leading to the on-shell Lagrangian
\begin{equation}
	\label{app:3ds_LagLinChiral_compb}
	\small{\begin{aligned}
		\mathcal{L} &=  - \frac{1}{4 e^2} F^{\mu\nu} F_{\mu\nu}  - \frac{1}{2 e^2} \partial^\mu \sigma \partial_\mu \sigma +\frac{k}{4\pi} \varepsilon^{\mu\nu\rho} A_\mu \partial_\nu A_\rho  - \frac{e^2}{2} \left( \sum\limits_{i=1}^N n_i \bar{q}_i q_i - \frac{k}{2\pi} \sigma - \frac{\zeta}{2\pi}\right)^2 \,
		\\
		&\quad\,+\sum\limits_{i=1}^N \Big\{ - \partial^\mu \bar{q}_i \partial_\mu q_i + \bar{f}_i f_i  +(\ii \partial_\mu \bar{q}_i A^\mu q_i + {\rm c.c.} )  - \bar{q}_i \left[ A^\mu A_\mu +  (\sigma -m_i)^2 \right] q_i\Big\} .
	\end{aligned}}
\end{equation}

\section{K\"ahler cone structure and the dilaton}
\label{app:Kcone}

Let us consider a three-dimensional $\caln=2$ SCFT with moduli space $\calm$, parametrized by some chiral coordinates $X_A$ with scaling dimensions $\Delta_A$. By superconformal invariance,  $\calm$ must be a K\"ahler cone and admit a globally defined homogeneous K\"ahler potential $\calk(X,\bar X)$, with scaling dimension $\Delta_{\calk}=1$.

Let us pick any chiral combination  $\cals$ of dimension $\Delta_\cals=\frac12$ and set $X_A\equiv  \cals^{2\Delta_A}e^{f_A(\zeta)}$, with $\zeta^i$ a set of (locally defined) chiral fields of vanishing scaling dimensions. We can then set
\be
\calk=|\cals|^2e^{2\hat\calk(\zeta,\bar\zeta)}
\ee

A different choice of $\cals$ corresponds to a holomorphic redefinition $\cals\rightarrow \cals e^{-2g(\zeta)}$, which in turn corresponds to a transformation 
\be
\hat\calk(\zeta,\bar\zeta)\rightarrow\hat\calk(\zeta,\bar\zeta)+g(\zeta)+\overline{g(\zeta)}
\ee  
The corresponding K\"ahler metric on the moduli space is 
\be\label{modulimetric}
\d s^2_\calm =2\calk\left[(\d\log\cals+2\hat\calk_{i}\,\d \zeta^{i})(\d \log\bar\cals+2\hat\calk_{\bar\imath}\,\d \bar\zeta^{\bar\imath})
+2\hat \calk_{i\bar\jmath}\,\d \zeta^i\d\bar \zeta^{\bar \jmath}\right]\,.
\ee

One can make manifest the conical structure by introducing the {\em dilaton}, namely the `radial' coordinate  $\tau$ defined by \be\label{defr}\tau^2\equiv 2\calk=2|\cals|^2 e^{2\hat \calk(\zeta,\bar\zeta)}\,.
\ee
Notice that $\tau$ has scaling dimension $\Delta_\tau=\frac12$.
In order to write the metric in conical form, we   set
\be
\cals=\frac{1}{\sqrt{2}}e^{\ii\chi}\tau e^{-\hat {\calk}(\zeta,\bar\zeta)}\,.
\ee
 The metric (\ref{modulimetric}) can then be written as
\be\label{conemetric}
\d s^2_\calm=\d \tau^2 +\tau^2\d s^2_{\cal Y}\,,
\ee
where
\be
\d s^2_{\cal Y}\equiv(\d\chi+\calq)^2+2\hat {\calk}_{i\bar\jmath}\,\d\zeta^i\d\bar\zeta^{\bar\jmath}
\ee
and
\be
\calq\equiv -\ii \left(\hat\calk_{i}\,\d\zeta^{i}-\hat\calk_{\bar\imath}\,\d \bar\zeta^{\bar\imath}\right)\,.
\ee

The  form (\ref{conemetric}) of the metric shows that the moduli space $\calm$ is a cone based on a Sasaki space ${\cal Y}$.
The  $U(1)_R$ $R$-symmetry  acts by shifting the $\chi$ coordinate.  In the language of Sasakian geometry, we identify the so-called contact one-form
$\eta\equiv \d\chi+\calq$
so that we can write the K\"ahler form on $\calm$  as 
\be
J_\calm=\tau\d\tau\wedge\eta +\ii \tau^2\hat {\calk}_{i\bar\jmath}\,\d \zeta^i\wedge \d\bar \zeta^{\bar\jmath}\,.
\ee

\section{The resolved cone over \texorpdfstring{$Y^{12}(\mathbb{P}^2)$}{Y12(P2)}}
\label{app:Y12}

In this section we provide more details of the geometry of the resolved cone over $Y^{12}(\mathbb{P}^2)$, which was chosen as M-theory background in section \ref{sec:Y12model}. We compute explicitly the metric of the resolved cone, also making contact between our notation and that of \cite{Martelli:2007pv}.

The resolved cone $X$ over the Sasaki-Einstein manifold $Y^{1,2}(\mathbb{P}^2)$ may be described as the moduli space of a gauged linear sigma model. Its five complex coordinates $Z_A$ are charged under a single $U(1)$ as
\be\label{Y12glsm2}
\begin{array}{c|ccccc|r} 
	& Z_1 & Z_2 & Z_3  & Z_4 & Z_5  &{\rm FI} \\
	\hline
	Q^A& 1 & 1 & 1 & -2 &  -1   & \sigma
\end{array}
\ee
where $\sigma$ denotes the FI parameter and we restrict to the phase $\sigma \geq 0$. The resolved cone $X$ is then described by the D-flatness condition
\be 
|Z_1|^2+|Z_2|^2+|Z_3|^2-2|Z_4|^2-|Z_5|^2= \ell_{\text{\tiny P}}\sigma
\ee
further modded out by the gauge symmetry
\be
Z_A \rightarrow e^{\ii Q^A \alpha} Z_A\,.
\ee 
We can most readily identify $X$ with the bundle $\calo_{\mathbb{P}^2}(-2)\oplus \calo_{\mathbb{P}^2}(-1)$. The $\mathbb{P}^2$ base can be covered by three local patches defined by $Z_1 \neq 0$, $Z_2 \neq 0$ and $Z_3 \neq 0$ respectively. The coordinates $Z_4$ and $Z_5$ describe the fibers. For example, in the patch where $Z_3 \neq 0$, the $\mathbb{P}^2$--base can be described by the complex coordinates
\be
\lambda_1\equiv\frac{Z_1}{Z_3} \,,\quad \lambda_2\equiv\frac{Z_2}{Z_3} \,,
\ee while
\be
\zeta_1 \equiv Z_4Z^2_3  \,,\quad \zeta_2 \equiv Z_5Z_3 \,,
\ee
can be used as fibral coordinates for $\calo_{\mathbb{P}^2}(-2)\oplus \calo_{\mathbb{P}^2}(-1)$.

The non-vanishing homology groups are $H_0(X;\mathbb{Z})\simeq H_2(X;\mathbb{Z})\simeq H_4(X;\mathbb{Z})=\mathbb{Z}$ and are generated by a point, $H\subset \cals$ and $\cals$ respectively, where $\cals\simeq \mathbb{P}^2$ is the zero-section of the vector bundle $\calo_{\mathbb{P}^2}(-2)\oplus \calo_{\mathbb{P}^2}(-1)$, while $H\simeq \mathbb{P}^1$ is its hyperplane divisor. Indeed, for the base Sasaki-Einstein $Y\simeq Y^{1,2}(\mathbb{P}^2)$ we have \cite{Martelli:2008rt}
\be
\begin{aligned}
	&H_0(Y;\mathbb{Z})=H_2(Y;\mathbb{Z})=H_7(Y;\mathbb{Z})=\mathbb{Z}\\
	&H_1(Y;\mathbb{Z})=H_3(Y;\mathbb{Z})=\mathbb{Z}_2\,,\quad H_5(Y;\mathbb{Z})=\mathbb{Z}\oplus \mathbb{Z}_2\\
	&H_4(Y;\mathbb{Z})=H_6(Y;\mathbb{Z})=0
\end{aligned}
\ee
By Poincar\'e duality we can then obtain $H^k(Y;\mathbb{Z})\simeq H_{7-k}(Y;\mathbb{Z})$.

In order to compute the metric, we first need the K\"ahler potential for the resolved cone $X$ over $Y^{1,2}(\mathbb{P}^2)$. Let us work in a local patch with complex coordinates $z^i\equiv (\lambda_1,\lambda_2,\zeta_1,\zeta_2)$. 
The $\mathbb{P}^2$ base is naturally endowed with the Fubini-Study K\"ahler potential
\be
k_{\mathbb{P}^2}=\frac{1}{2 \pi}\log(1+|\lambda_1|^2+|\lambda_2|^2)\,,
\ee
which we normalize such that 
\be
\label{jP2norm}
j_{\mathbb{P}^2}\equiv \ii\del\delbar k_{\mathbb{P}^2}=\frac1{6\pi}\calr_{\mathbb{P}^2}\,,\qquad \int_Hj_{\mathbb{P}^2}=1\,,
\ee
where $\calr_{\mathbb{P}^2}$ is the associated  Ricci form.

We may regard $e^{2\pi k_{\mathbb{P}^2}}$ as a metric for the line bundle $\calo_{\mathbb{P}^2}(-1)$ and then we can introduce the following globally defined radial coordinates over $X\simeq\calo_{\mathbb{P}^2}(-2)\oplus \calo_{\mathbb{P}^2}(-1)$:
\be\label{tdefY12b}
t_1\equiv |\zeta_1|^2e^{4\pi k_{\mathbb{P}^2}}\,,\quad t_2\equiv |\zeta_2|^2e^{2\pi k_{\mathbb{P}^2}}~.
\ee

The K\"ahler potential for the CY metric over $X$ can be written in the form
\be\label{kY12}
k_X=U(t_1,t_2;v)+\sigma k_{\mathbb{P}^2}(\lambda,\bar\lambda)
\ee
where $U(t_1,t_2;v)$ is a globally defined function which we will determine shortly and $\sigma$ is the unique K\"ahler modulus, which parametrizes the volume of  $H\simeq\mathbb{P}^1$ at the tip of the resolved cone. In fact 
\be
\sigma=\frac{2\pi}{\ell_{\text{\tiny P}}}\int_HJ_X
\ee
as it follows immediately from  \eqref{jP2norm}, and 
\be
J_X\equiv \frac{\ell_{\text{\tiny P}}}{2\pi} \ii\del\delbar k_X= \frac{\ell_{\text{\tiny P}}}{2\pi} \left(\ii\del\delbar U + \sigma j_{\mathbb{P}^2} \right)
\ee
We will show momentarily that the K\"ahler potential \eqref{kY12} leads to the same class of metrics as that found in \cite{Martelli:2007pv}.

In order to better characterize $U(t_1,t_2;v)$ we first introduce two new real coordinates\footnote{Notice the redefinition $x\rightarrow 1-x$ and $y\rightarrow 1-y$ with respect to \cite{Martelli:2007pv}.} 
\be\label{xyt2}
x=x(t_1,t_2;\sigma)\,,\quad y=y(t_1,t_2;\sigma)
\ee
obtained by inverting the following relations
\be\label{Y12t12va}
\begin{aligned}
	t_1&\equiv \sigma^{\frac{8}{3 u_+}}\exp\left[\frac{2}{u_+}\left(\int^x_{u_+}\d u\,\frac{u-u_+}{{\mathfrak F}(u)}+\int_y^{u_+}\d u\frac{u_+-u}{{\mathfrak F}(u)}\right)\right]\,,\\
	t_2&\equiv \sigma^{\frac{4}{3u_-}}\exp\left[\frac{1}{u_-}\left({\rm P}\int^x_{u_-}\d u\frac{u-u_-}{{\mathfrak F}(u)}+\int^y_{u_-}\d u\frac{u-u_-}{{\mathfrak F}(u)}\right)\right]\,.
\end{aligned}
\ee
The real coordinates $x$ and $y$ obey the consistency conditions \eqref{xycons}, and $u_-<u_+$ are the two real zeros of the function ${\mathfrak G}(u)$ defined in \eqref{Gmathfrak}. The function ${\mathfrak F }(u)$ is related to ${\mathfrak G}(u)$ in \eqref{Gmathfrak} as
\be\label{Fmathfraka}
{\mathfrak F}(u)\equiv \frac34 \frac{{\mathfrak G}(u)}{u^2} = \frac{3}{4}u^2-u+\frac{2\nu}{u^2}\,.
\ee 
By rewriting
\be\label{Fmathfraka2}
{\mathfrak F}(u) =\frac{3}{4u^2}(u-u_+)(u-u_-)(u-u_*)(u-\overline{u_*})
\ee
the radial coordinates \eqref{Y12t12va} can be more explicitly written in terms of the roots of ${\mathfrak G}$ as in \eqref{Y12t12v}. Other useful identities are $\frac{u_+(u_--1)}{u_+-u_-}=-\frac23$ and $\frac{u_-(u_+-1)}{u_+-u_-}=\frac13$.
Moreover, note that
$0<u_- < 1 < u_+$
and that ${\mathfrak F}(u)$ is negative for $u\in (u_-,u_+)$ and positive for $u\in (u_+,\infty)$. The condition \eqref{s+-fix} fixes $u_\pm,\nu$ to be irrational numbers. Furthermore 
\be\label{u_*a}
u_* = \frac{1}{3}\left(1-\eta +\ii(\eta-1)\sqrt{\frac{\eta+2}{\eta}} \right) \simeq -0.306 -\ii ~0.437
\ee
and its complex conjugate $\overline{u_*}$ are the two complex zeros of ${\mathfrak F}(u)$. Since ${\mathfrak F}(u)$ depends on $\nu$ or equivalently on $\eta$ according to $(\ref{nu_of_g})$, these zeros all depend on $\nu$ or $\eta$.

Finally, in terms of \eqref{Y12t12va} and the K\"ahler modulus $\sigma$, the function $U(t_1,t_2;\sigma)$ appearing in \eqref{kY12} reads
\be\label{GY12a}
\begin{aligned}
	U(t_1,t_2;\sigma)=&\,\frac{\sigma}{2\pi u_- u_+}\Big[\int^x_{u_-}\d u\frac{(u-u_-)(u-u_+)}{{\mathfrak F}(u)}+\int^y_{u_-}\d u\frac{(u-u_-)(u-u_+)}{{\mathfrak F}(u)}\\
	&-\frac43\Big(u_++u_--\frac43\Big)\log \sigma+c\Big]
\end{aligned}
\ee
where $c$ is an arbitrary constant and the logarithmic term in the second line is required to satisfy the asymptotic conditions \eqref{bcond}.

From the implicit definition \eqref{Y12t12va} of $x$ and $y$, one can compute the differentials
\be\label{delxy}
\begin{aligned}
	\d x&=\frac{{\mathfrak F}(x)}{2(u_+-u_-)(x-y)}\left[u_+(y-u_-)\d\log t_1-2u_-(y-u_+)\d\log t_2\right]\\
	\d y&= \frac{{\mathfrak F}(y)}{2(u_+-u_-)(y-x)}\left[u_+(x-u_-)\d\log t_1-2u_-(x-u_+)\d\log t_2\right]~.
\end{aligned}
\ee
It is also useful to introduce the globally defined $(1,0)$-forms
\be\label{1formseta}
\begin{aligned}
	\eta_1&\equiv \frac{u_--y}{3(u_--1)}\del\log t_1-\frac{u_+-y}{3(u_+-1)}\del\log t_2\\
	\eta_2&\equiv \frac{u_--x}{3(u_--1)}\del\log t_1-\frac{u_+-x}{3(u_+-1)}\del \log t_2
\end{aligned}
\ee
with $\del\log t_1=\d\log\zeta_1+4\del k_{\mathbb{P}^2}$ and $\del\log t_2=\d\log\zeta_1+2\del k_{\mathbb{P}^2}$. Hence, from the K\"ahler potential \eqref{kY12}, upon using \eqref{delxy} and \eqref{1formseta}, we can compute
\be\label{Y12v}
\begin{aligned}
	J_X=&\,\frac{\sigma}{u_- u_+\pi}\left[x y\, j_{\mathbb{P}^2}+\frac{\ii {\mathfrak F}(x)}{2(x-y)}\eta_1\wedge\bar\eta_1+\frac{\ii {\mathfrak F}(y)}{2(y-x)}\eta_2\wedge\bar\eta_2\right]~.
\end{aligned}
\ee
Finally, by introducing the angular variables $(\tau,\psi)$ defined by
\be
\begin{aligned}
	\Im\log\zeta_1\equiv2\psi+\frac{2\tau}{u_+}\,,\quad 
	\Im\log\zeta_2\equiv \psi+\frac{\tau}{u_-}
\end{aligned}
\ee
the  K\"ahler metric associated with \eqref{Y12v} 
can be written as
\be\label{Y12v2}
\begin{aligned}
	\d s^2_X=&\,\frac{\sigma}{u_- u_+ \pi}\Bigg[xy\,\d s^2_{\mathbb{P}^2}+\frac{x-y}{4}\left(\frac{\d x^2}{{\mathfrak F}(x)}-\frac{\d y^2}{{\mathfrak F}(y)}\right)\\
	&+\frac{{\mathfrak F}(x)}{x-y}\big(\d\tau+y(\d\psi+A)\big)^2+\frac{{\mathfrak F}(y)}{y-x}\big(\d\tau+x(\d\psi+A)\big)^2\Bigg]
\end{aligned}
\ee
where
\be
A=\pi \d^{\rm c}k_{\mathbb{P}^2}
\ee
with $\d^{\rm c}=-\ii(\del-\delbar)$, is such that $\d A=2\pi j_{\mathbb{P}^2}$. Up to an overall rescaling, this metric is in the form presented in \cite{Martelli:2007pv}.

The asymptotic conical region is obtained by taking  the limit $x\rightarrow\infty $. By identifying the asymptotic radial coordinate $r$ through
\be
r^2\simeq \frac{4\sigma}{3u_+u_-\pi}x
\ee 
we asymptotically have $\d s^2_X\simeq \d r^2+r^2\d s^2_{Y}$, where
\be
\begin{aligned}
	\d s^2_{Y}&\equiv\d s^2_{T}+\frac9{16}\left[\d\tau+y(\d\psi+A)\right]^2\\
	&\text{with}\quad\d s^2_{T}\equiv \frac34y\d s^2_{\mathbb{P}^2}-\frac{3}{16{\mathfrak F}(y)}\d y^2-\frac34{\mathfrak F}(y)(\d\psi+A)^2
\end{aligned}
\ee
is the Sasaki-Einstein metric over $Y\equiv Y^{1,2}(\mathbb{P}^2)$.

\subsection{Evaluation of integrals}

The integrals in (\ref{Y12t12va}) and (\ref{GY12a}) can then be performed explicitly. We find that the former are
\be\label{Y12t12expla}
\begin{split}
t_1^{\frac{3 u_+}{8}} &= \sigma   \left(\frac{(x-u_-) (y-u_-)}{(u_+-u_-)^2}\right)^{\frac{u_-^2}{|u_--u_*|^2}} 
\Bigg| \left(\frac{(x-u_*) (y-u_*)}{(u_+-u_*)^2}\right)^{\frac{u_*^2}{(u_*-u_-) (u_*-\overline{u_*})}}\Bigg|^2   \\
t_2^{\frac{3 u_-}{4}} &= \sigma   \left(\frac{(x-u_+) (u_+-y)}{(u_+-u_-)^2}\right)^{\frac{u_+^2}{|u_+-u_*|^2}} 
\Bigg| \left(\frac{(x-u_*) (y-u_*)}{(u_--u_*)^2}\right)^{\frac{u_*^2}{(u_*-u_+) (u_*-\overline{u_*})}}\Bigg|^2 
\end{split}
\ee 
and that the integrals appearing in (\ref{GY12a}) are given by
\be\label{integrals_simpa}
\begin{split}
&	\int^x_{u_-}\d u\frac{(u-u_-)(u-u_+)}{{\mathfrak F}(u)}+\int^y_{u_-}\d u\frac{(u-u_-)(u-u_+)}{{\mathfrak F}(u)}=\\
&\qquad\qquad= \frac{4}{3}\left(x+y-2u_- + \Im\left[{u_*^2 \log\frac{(x-u_*) (y-u_*)}{(u_--u_*)^2}}\right]/~\Im{u_*} \right)~.
\end{split}
\ee
Thus, from the K\"ahler potential \eqref{kY12}, we find the dual symplectic potential \eqref{sympot}
\begingroup
\allowdisplaybreaks
\begin{align}
F_X(l;\sigma) &= - \frac{1}{2\pi} \sum\limits_{i=1}^4 l_i \log l_i - \frac{1}{2\pi} s_3 \log s_3 \nonumber
\\
&\quad\,+ \frac{1}{2\pi} s_{P_{\rm can}} \log (2l_1+l_2+\sigma)- \frac{1}{2\pi} s_{P} \log \sigma \nonumber
\\
\begin{split}
&\quad\,-\frac{1}{2\pi} l_1 \log \frac{2l_1+l_2+\sigma}{l_1}-\frac{1}{2\pi} l_2 \log \frac{2l_1+l_2+\sigma}{l_2}
\\
&\quad\,- \frac{4 u_-^2}{3 \pi u_+ |u_--u_*|^2} l_1 \log \frac{l_1}{\sigma}-\frac{2 u_+^2}{3 \pi u_- |u_+-u_*|^2} l_2 \log \frac{l_2}{\sigma}
\end{split}
\\
&\quad\,- \frac{4}{3 \pi u_+} l_1 \log \left| \alpha^{\frac{u_*^2}{(u_*-u_-)(u_*-\bar u_*)}} \right|^2 - \frac{2}{3\pi u_-} l_2 \log \left| \alpha^{\frac{u_*^2}{(u_*-u_+)(u_*-\bar u_*)}} \right|^2 \nonumber
\\
&\quad\,+\frac{2\sigma}{3 \pi u_- u_+} \frac{\Im[u_*^2 \log \alpha]}{\Im u_*} \nonumber
\end{align}
\endgroup
where we have introduced
\begingroup
\allowdisplaybreaks
\begin{align}
s_3 &= 2 l_1 + l_2 - l_3- l_4 +\sigma
\nonumber\\
\begin{split}
s_{P_{\rm can}} &= 3 l_1+2 l_2 + \sigma
\\
s_{P} &= \frac{8}{3\,u_+}l_1+\frac{4}{3\,u_-}l_2+\frac{4(3u_++3u_--4)}{9\,u_-u_+}\sigma
\end{split}
\\
\alpha &= 2  u_- (u_+-u_*) \frac{l_1}{\sigma}+  u_+(u_--u_*) \frac{l_2}{\sigma} +(u_--u_*)(u_+-u_*)\nonumber
\end{align}
\endgroup


\subsection{\texorpdfstring{$P^A$}{PA}, superconformal \texorpdfstring{$R$}{R}-charges and volumes of 5-cycles} \label{subsubsec:Rvolumes}

We have seen in section \ref{sec:toric} that the constants $P^A$ introduced in (\ref{sP}) can be interpreted as twice the scaling dimensions (or superconformal $R$-charges) of the homogeneous chiral fields $Z_A$. The AdS/CFT correspondence relates the latter to the volumes of 5-cycles $\Pi^A=\partial \cald^A$ in the Sasaki-Einstein manifold $Y=\partial X$ which are bases of the toric divisors $\cald^A=\{Z_A=0\}$ in the Calabi-Yau cone $X$. Consistency therefore requires that $P^A$ in our formalism are related to the volumes of 5-cycles $\Pi^A$ as in (\ref{Pvolrel}). In this appendix we verify this equality in the case of $Y=Y^{1,2}(\bP^2)$, matching the quantities computed in section \ref{subsec:HEFT_Y12} in the holographic EFT with the volumes and Reeb vector computed by volume minimization \cite{MartelliSparksYau2006,MartelliSparksYau2008} in appendix A.1 of \cite{Cremonesi2011}. 

The components of the Reeb vector and the ratios of the volumes of the 5-cycles $\Pi^A$ ($A=1,\dots,5$) to the volume of the Sasaki-Einstein $Y=Y^{1,2}(\bP^2)$ are irrational numbers. The expressions computed in section \ref{subsec:HEFT_Y12} and those computed in \cite{Cremonesi2011} have different functional forms, and proving that they are equal seems hard at first sight. One can be excused to settle for a numerical comparison, which is easily done. In this appendix we will however prove these equalities, by exploiting the fact that volumes, components of the Reeb vector and $R$-charges are algebraic numbers \cite{MartelliSparksYau2006}. It turns out that the relevant quantities are each the only real root of a certain cubic polynomials with integer coefficients, which can be determined by standard algebraic methods (or by using \texttt{Mathematica} \cite{Mathematica}).

Let us first focus on the holographic EFT side. Starting with expression (\ref{g}) for $\eta$, and using the expressions (\ref{sigma+-_of_g}), (\ref{Y12Deltazrho}), (\ref{Y12bp}) and (\ref{Y12PA}), one finds that the following quantities are the only real roots of the following cubic polynomials: 
\be\label{table_HEFT}
\begin{tabular}{c|c}
 Quantity & Polynomial $P(z)$ \\ \hline
 $\eta$ & $2 z^3-2 z^2-3 z-1$\\
$u_+$ & $27 z^3-54 z^2+33 z-8$ \\
$u_-$ & $27 z^3-18 z^2+3z-4$ \\ 
$\Delta(e^{2\pi z^1})$ & $2 z^3-11 z^2+24 z-16$\\
$\Delta(e^{2\pi z^2})$ & $2 z^3-z^2+4 z-4$\\
$\frac{1}{N}\Delta(e^{-2\pi \rho})$ & $2 z^3+3 z^2+16 z-8$ \\
$b_1$ & $z^3-11 z^2+48 z-64 $ \\
$b_2$ & $z^3-z^2+8 z-16 $ \\
$p$ & $z^3+3 z^2+32 z-32 $ \\ 
$\frac{1}{2}P^1=\frac{1}{2}P^2=\frac{1}{2}P^3$ & $2 z^3+3 z^2+16 z-8$ \\ 
$\frac{1}{2}P^4$ & $2 z^3-17 z^2+72 z-16$ \\ 
$\frac{1}{2}P^5$ & $z^3-2 z^2+20 z-8$  
 \end{tabular}
\ee

For the formulae of appendix A.1.1 of \cite{Cremonesi2011} for the Reeb vector and the superconformal $R$-charges $R^A\equiv \frac{\pi \text{vol}(\Pi^A)}{3\text{vol}(Y)}$ computed from volumes of 5-cycles, the same method gives
\footnote{The homogeneous coordinates $(Z_1, Z_2, Z_3, Z_4, Z_5)$ of (\ref{Y12glsm}) were denoted as $(b_0,d_0,a_2,a_1,c_0)$ in \cite{Cremonesi2011} and were identified with the following operators in the flavored ABJM theory on the worldvolume of a single M2-brane: $(B_1,B_2,\tilde{T},T,A_2)$.}
\be\label{table_volumes}
\begin{tabular}{c|c}
	Quantity & Polynomial $P(z)$ \\ \hline
	$x$ & $z^3-z^2+8 z-16$\\
	$4-x$ & $z^3-11 z^2+48 z-64$ \\
	$R^1=R^2=R^3$ & $2 z^3+3 z^2+16 z-8$ \\ 
	$R^4$ & $2 z^3-17 z^2+72 z-16$ \\ 
	$R^5$ & $z^3-2 z^2+20 z-8$  
\end{tabular}
\ee

Comparing (\ref{table_volumes}) with (\ref{table_HEFT}), we see that the nontrivial components of the Reeb vector are identified as $4-x=b_1$ and $x=b_2$, and that $P^A=2 R^A$ in agreement with the relation (\ref{Pvolrel}).

\subsection{Counting holomorphic sections and chiral operators}\label{appsub:counting_Y12}

In this section we compare the geometric count of holomorphic sections with the count of monopole operators in the worldvolume theory on M2-branes. We will start with the case of $N=1$ M2-brane before the S operation, and then generalise the results in two ways: acting with the S operation and taking $N>1$. We are eventually interested in $N\gg 1$ M2-branes to make contact with the holographic EFT.

\subsubsection{\texorpdfstring{$N=1$}{N=1} count before the \texorpdfstring{$S$}{S} operation}

Let us start from the geometric side and look at the generating function of holomorphic sections of the line bundle $\calo_X(n D)$, where $X$ is the cone over $Y^{1,2}(\bP^2)$, as in \cite{Butti:2006au,Butti:2007jv}. $D$ is the toric divisor associated to a GLSM field of charge $1$, up to linear equivalence. In other words, we count the effective chiral operators of section \ref{sec:chiralop} with fixed charge $n$ in the formal limit $N=1$.

From the point of view of the GLSM \eqref{Y12glsm2}, we count operators of electric charge $n$ using the Molien integral 
\begin{equation}\label{Y12count_1}
\begin{split}
    g_{1,n}(T) &=\oint \frac{dw}{2\pi i w} w^{-n} \PE[(T_1+T_2+T_3)w+T_4 w^{-2} +T_5 w^{-1}]\\
    &= \sum_{r_1,\dots,r_5 \in \bZ_{\ge 0}} \delta_{r_1+r_2+r_3-2r_4-r_5,n}\prod_{A=1}^5 T_A^{r_A} 
\end{split}
\end{equation}
where $\PE$ denotes the plethystic exponential
\footnote{If $x_i$ are variables and $n_i$ integers, the plethystic exponential is given by $\PE[\sum_i n_i x_i]=\prod_i (1-x_i)^{-n_i}$.}
and $T_A$ are fugacities associated to the homogeneous coordinates $Z_A$. We assume $|T_A|<1$ so that we can Taylor expand the $\PE$. The Molien integral projects to gauge invariant states, but we have inserted a background electric charge $-n$ under the gauge $U(1)$ which therefore needs to be compensated by a total charge of $n$ carried by GLSM fields.  
The Kronecker delta in the second line of \eqref{Y12count_1} is trivialised by setting $r_A \equiv s_A(\mathbf{m},n):= \langle \mathbf{m}, \mathbf{v}_A \rangle + M_A n$ as in \eqref{chargecond}, where the lattice 4-vector $\mathbf{m}$ takes value in the quantized Delzant polytope $\mathcal{P}^\bZ_n:=\mathcal{P}_n \cap M_\bZ$, see \eqref{poly} for the continuous version. Here $\mathbf{v}_A$ are the coordinates of the generators of the toric fan \eqref{Y12vec} and we take $M_A=\delta_{A,3}$. Then 
\begin{equation}
    \prod_{A=1}^5 T_A^{r_A} = (T_3^2 T_4)^{m_1} (T_3 T_5)^{m_2} (T_1 T_3^{-1})^{m_3} (T_2 T_3^{-1})^{m_4} (T_3)^n \equiv \omega^n \prod_{i=1}^4 t_i^{m_i} 
\end{equation}
where 
\begin{equation}
    t_i = \prod_{A} T_A^{v^i_A}
\end{equation}
are fugacities for the toric $U(1)^4$ symmetries and 
\begin{equation}
    \omega = \prod_{A} T_A^{M_A}
\end{equation}
is a fugacity for the Betti $U(1)$ symmetry, so we can write the generating function counting operators of charge $n$ as 
\begin{equation}\label{Y12count_2}
    g_{1,n}(t,\omega)= \omega^n \sum_{\mathbf{m} \in \calp_n^\bZ} \prod_{i=1}^4 t_i^{m_i}~.
\end{equation}

It is possible to calculate \eqref{Y12count_2} in closed form for $n\ge 0$ and $n \le 0$, but the result is not particularly illuminating so we omit the details here. Instead, following appendix F of \cite{CremonesiMekareeyaZaffaroni2016}, we will show that the geometric formula using the Molien integral \eqref{Y12count_1} reproduces the formula for the Hilbert series that counts monopole operators in the $U(1)_{3/2} \times U(1)_{-3/2}$ flavoured Chern-Simons quiver gauge theory of section \ref{sec:Y12quiver}, with a fixed background magnetic charge $n$ for the `baryonic' gauge $U(1)$. To see that, it is convenient to introduce the $U(1)_M$ symmetry in the GLSM \eqref{Y12glsm0}, with a fugacity $\gamma$ that we integrate over and an electric charge $m$ that we sum over. We can rewrite \eqref{Y12count_1} as 
\begin{equation}\label{Y12count_3}
    g_{1,n}(T) =\sum_{m \in \bZ}\oint \frac{dw}{2\pi i w} w^{-n} \oint \frac{d\gamma}{2\pi i \gamma} \gamma^{-m} \PE[(T_1+T_2)w+T_3 w \gamma+T_4 w^{-2}\gamma^{-1} +T_5 w^{-1}]~.
\end{equation}
This is ineffectual, since performing the sum over $m$ leads to a Dirac delta function that sets $\gamma=1$ and returns the original Molien integral \eqref{Y12count_1}. Instead, changing integration variable $\gamma=u w^{-1}$ and performing the integral over $u$ yields
\begingroup
\allowdisplaybreaks
\begin{align}
    g_{1,n}(T) &=\sum_{m \in \bZ}\oint \frac{dw}{2\pi i w} w^{-(n-m)} \oint \frac{du}{2\pi i u} u^{-m} \PE[(T_1+T_2)w+T_3 u+T_4 w^{-1}u^{-1} +T_5 w^{-1}]~\nonumber\\
    &=\sum_{m \in \bZ}\oint \frac{dw}{2\pi i w} w^{-(n-m)} \PE[(T_1+T_2)w+(T_3 T_4 +T_5) w^{-1}] T_3^{\frac{|m|+m}{2}}(T_4 w^{-1})^{\frac{|m|-
    m}{2}} \nonumber\\
    &= \sum_{m \in \bZ} \left(\frac{T_3}{T_4}\right)^{\frac{m}{2}}\left(T_3 T_4\right)^{\frac{|m|}{2}}   \oint \frac{dw}{2\pi i w} w^{-(n-\frac{3}{2}m+\frac{1}{2}|m|
    )} \PE[(T_1+T_2)w+(T_3 T_4 +T_5) w^{-1}]~, \label{Y12count_4}
\end{align}
\endgroup
see equation (F.9) of \cite{CremonesiMekareeyaZaffaroni2016} for details of the manipulations. Finally, letting
\begin{equation}\label{fug_map_Y12}
    T_1=ty~, \quad T_2=t/y~,\quad T_3=t^{1/2}z~, \quad T_4=t^{1/2}x/z~,\quad T_5=t/x
\end{equation}
leads to the final expression 
\begin{equation}\label{Y12count_6}
\begin{split}
    g_{1,n}(t,x,y,z) &=\sum_{m \in \bZ} z^m t^{\frac{|m|}{2}} x^{\frac{|m|-m}{2}}
    \oint \frac{dw}{2\pi i w} w^{-(n-\frac{3}{2}m+\frac{1}{2}|m|
    )} \PE\big[tw(y+\frac{1}{y})+\frac{t}{w}(x+\frac{1}{x})\big]~.
\end{split}
\end{equation}

Equation \eqref{Y12count_6} precisely takes the form of the Hilbert series that counts dressed monopole operators in the abelian $U(1)_{3/2}\times U(1)_{-3/2}$ Chern-Simons quiver of section \ref{sec:Y12quiver} (before the S operation), computed as in \cite{CremonesiMekareeyaZaffaroni2016}. In the UV quiver theory $n$ is interpreted as a background magnetic charge for the baryonic gauge $U(1)$; in other words, we turn on background magnetic charges $n_1=-n_2=n$ for the topological symmetries of the two $U(1)$ gauge groups. The integral over the fugacity for the decoupled diagonal gauge $U(1)$ sets the dynamical gauge magnetic charges of the two $U(1)$ gauge groups to be equal: $m_1=m_2\equiv m$, where $m$ is the magnetic charge for the diagonal gauge $U(1)$, which we sum over, weighted by the topological fugacity $z=z_1 z_2$ for the diagonal gauge $U(1)$. $w^{-1}$ is the fugacity for the `baryonic' gauge $U(1)$, which is integrated over. $t^2$ is the fugacity for the $R$-symmetry, which assigns $R$-charge $1/2$ to all bifundamentals. $y$ is the fugacity for an $SU(2)$ global symmetry that rotates $b_1$ and $b_2$, whereas $x$ is the fugacity for a $U(1)$ global symmetry that acts on $a_1$, $a_2$ and the fundamental flavours. (There is also a mixed bare Chern-Simons term at level $1/2$ between the latter $U(1)$ symmetry and the diagonal gauge $U(1)$.) Taking into account the quantum corrections to the charges of monopole operators as explained in \cite{CremonesiMekareeyaZaffaroni2016}, one precisely finds  formula \eqref{Y12count_6} for the Hilbert series. 

The fugacity map \eqref{fug_map_Y12} identifies the homogeneous coordinates of the GLSM with gauge variant operators in the abelian quiver Chern-Simons theory as follows:
\begin{equation}
    Z_1 \leftrightarrow b_1~,\quad Z_2 \leftrightarrow b_2~, \quad Z_3 \leftrightarrow t~, \quad Z_4 \leftrightarrow \tilde t~, \quad Z_5 \leftrightarrow a_2~,
\end{equation}
(recall that $a_1=t \tilde t$) \cite{BeniniClossetCremonesi2010}. The toric coordinates $e^{-2\pi z^i}$ are instead identified with
\begin{equation}
    e^{-2\pi z^1} \leftrightarrow t^2 \tilde t= t a_1~,\quad e^{-2\pi z^2} \leftrightarrow t a_2~,\quad e^{-2\pi z^3} \leftrightarrow \frac{b_1}{t} ~,\quad  e^{-2\pi z^4} \leftrightarrow \frac{b_2}{t} ~, 
\end{equation}
which are gauge invariant but not globally defined chiral operators. 
The holomorphic operators of `Betti charge' $n$ counted by formula \eqref{Y12count_2} can then be written as 
\begin{equation}\label{effective_chiral_operators_Y12}
    b_1^{m_3} b_2^{m_4} t^{2m_1+m_2-m_3-m_4+n} \tilde t^{m_1} a_2^{m_2}
\end{equation}
in terms of the independent holomorphic variables $b_1, b_2, t, \tilde t, a_2$ of the abelian quiver Chern-Simons theory. The operators \eqref{effective_chiral_operators_Y12} are genuine globally defined chiral operators if and only if they are holomorphic in $b_1, b_2, t, \tilde t, a_2$, that is if and only if the powers in \eqref{effective_chiral_operators_Y12} are all non-negative. This is precisely the condition that $\mathbf{m} \in \calp_n^\bZ$.

\subsubsection{\texorpdfstring{$N=1$}{N=1} count after the \texorpdfstring{$S$}{S} operation}

We now apply the S operation to the topological symmetry associated to the `baryonic' $U(1)$ gauge symmetry. On the geometry side, using \eqref{Y12count_1} the generating function becomes 
\begin{equation}\label{g_1 Y12}
\begin{split}
   g_1(T,b) &\equiv \sum_{n \in \bZ} b^n g_{1,n}(T)= \sum_{n \in \bZ} b^n  \sum_{r_1,\dots,r_5 \in \bZ_{\ge 0}} \delta_{r_1+r_2+r_3-2r_4-r_5,n}\prod_{A=1}^5 T_A^{r_A} \\
   &= \sum_{r_1,\dots,r_5 \in \bZ_{\ge 0}} b^{r_1+r_2+r_3-2r_4-r_5} \prod_{A=1}^5 T_A^{r_A}\\
   &= \PE \big[b(T_1+T_2+T_3)+b^{-2}T_4+b^{-1} T_5 \big]~,
\end{split}
\end{equation}
which is the Hilbert series of the extended geometric moduli space $\calm_{N=1}^{\mathrm{ext}}=\bC^{4+b_2(Y^{1,2}(\bP^2))}=\bC^5$. This corresponds to adding to the GLSM \eqref{Y12glsm2} a homogeneous coordinate $X \in \bC^*$ of charge $-1$ which can be used to soak up the charge of the other homogeneous coordinates, as explained after equation \eqref{projsec}. The basic gauge invariants of this extended GLSM are then 
\begin{equation}\label{extended_glsm_inv_Y12}
    X Z_1~, \quad X Z_2~, \quad X Z_3~, \quad X^{-2} Z_4~, \quad X^{-1} Z_5~, \quad  
\end{equation}
in one-to-one correspondence with the $5$ homogeneous coordinates of the original GLSM.

On the quiver Chern-Simons theory side, we follow the prescription of appendix B of \cite{CremonesiMekareeyaZaffaroni2016} to apply the S operation and use formula \eqref{Y12count_6} to obtain  
\begin{equation}\label{g_1 Y12 quiver}
\begin{split}
   g_1&(t,x,y,z,b) \equiv \sum_{n \in \bZ} b^n  g_{1,n}(t,x,y,z) \\
   &=\sum_{n,m \in \bZ} b^n z^m t^{\frac{|m|}{2}} x^{\frac{|m|-m}{2}}
    \oint \frac{dw}{2\pi i w} w^{-(n-\frac{3}{2}m+\frac{1}{2}|m|
    )} \PE\big[tw(y+\frac{1}{y})+\frac{t}{w}(x+\frac{1}{x})\big]\\
    &=\sum_{m \in \bZ} z^m t^{\frac{|m|}{2}} x^{\frac{|m|-m}{2}} b^{-(-\frac{3}{2}m+\frac{1}{2}|m|
    )} \PE\big[tb(y+\frac{1}{y})+\frac{t}{b}(x+\frac{1}{x})\big]\\
    &= \PE\big[ zt^{\frac{1}{2}}b +z^{-1} t^{\frac{1}{2}}x b^{-2} - t b^{-1} x + t b(y+y^{-1})+ tb^{-1}(x+x^{-1})   \big]  \\
    &= \PE\big[ t b y+ t b y^{-1}+  zt^{\frac{1}{2}}b +z^{-1} t^{\frac{1}{2}}x b^{-2} + tb^{-1} x^{-1}   \big]~.
\end{split}
\end{equation}
A contribution of magnetic charge $(n,m)$ in the second line corresponds to a dressing of the bare monopole operator $\calt_{n}T_m$, where $T_m$ is a monopole operator of charge $m$ under the diagonal gauge $U(1)$ in $U(1)_{3/2}\times U(1)_{-3/2}$ (so that $T_1\equiv T$ and $T_{-1}\equiv \tilde T$) and $\calt_n$ is a monopole operator of magnetic charge $n$ under the newly gauged topological symmetry for the baryonic gauge $U(1)$. Note that $\calt_n$ carries charge $n$ under the `baryonic' gauge $U(1)$. Since no matter fields are charged under the newly gauged $U(1)$, the classical relations $\calt_i \calt_j = \calt_{i+j}$ hold, with $\calt_0$ the identity operator. Hence $\calt_n= \calt^n$, and we can identify $\calt\equiv\calt_1$ with the extra homogeneous coordinate $X$ in the extended GLSM. 
For the same reason, a monopole operator of magnetic charges $(n,m)$ factorises into the product of two monopole operators for the quiver gauge group and for the extra $U(1)$ gauge factor. To go from the second to the third line of \eqref{g_1 Y12 quiver} we have swapped the integral over $w$ with the sum over $n$, which produces a delta function $2\pi i w\delta(w-b)$. Finally, we have summed over $m$ to get from the third to the fourth line. 

 In the order in which they are written in the last line of \eqref{g_1 Y12 quiver}, the generators of the extended moduli space in the last line of \eqref{g_1 Y12 quiver} are the gauge invariant operators 
\begin{equation}
    \calt b_1~, \quad  \calt b_2~, \quad  \calt t~, \quad  \calt^{-2} \tilde t~, \quad  \calt^{-1} a_2~, \quad  
\end{equation}
which precisely correspond to the invariants \eqref{extended_glsm_inv_Y12}.

\subsubsection{Count for \texorpdfstring{$N>1$}{N>1}}

The generalisation to multiple membranes is straightforward and does not introduce any conceptual novelty. 

On the geometric (or GLSM) side, we have $N$ sets of  homogeneous coordinates $Z_A^I$ labelled by $I=1,\dots,N$. The gauge group of the GLSM is the semidirect product of $U(1)^N$, with $N$ copies of the same $U(1)$ each acting on $Z_A^I$ with a different $I$, and the symmetric group $S_N$ which permutes the $I$ indices (since the membranes are indistinguishable). The gauge invariant operators of Betti charge $0$ take precisely the form \eqref{Y12mesonic}, and the Hilbert series which counts these operators is given by the symmetric product formula
\be
g_{N,0}(T) = \mathrm{Sym}^N g_{1,0}(T) = \frac{1}{N!} \left(\frac{\partial}{\partial \nu} \right)^N \PE[\nu g_{1,0}(T)]\bigg|_{\nu=0}~.
\ee

Acting with the S operation leads to a sum over Betti charges $n$, and the operators take the form
\be
\calo_{{\bf m},n} = X^n \left(\prod_{I}\left[(Z^I_1)^{m_3^I}(Z^I_2)^{m_4^I}(Z^I_3)^{2m^I_1+m^I_2-m_3^I-m_4^I+n}(Z^I_4)^{m^I_1}(Z^I_5)^{m^I_2}\right]\right)_{\rm sym}~
\ee
or equivalently \eqref{Y12chiral}. The generating function which counts these operators at fixed $n$ is given by the symmetric product formula \cite{Butti:2007jv}
\be\label{sym_line_bundles}
g_{N,Nn}(T) = \mathrm{Sym}^N g_{1,n}(T) = \frac{1}{N!} \left(\frac{\partial}{\partial \nu} \right)^N \PE[\nu g_{1,n}(T)]\bigg|_{\nu=0}~,
\ee
and summing over $n$ we obtain
\be
g_N(T,b) = \sum_{n \in \bZ} b^n g_{N,Nn}(T)~.
\ee

On the quiver gauge theory side, we have a $U(N)_{3/2} \times U(N)_{-3/2}$ quiver Chern-Simons theory. We need to count dressed monopole operators of charge $Nn$ under the off-diagonal combination of the $U(1)\times U(1)$ center of the $U(N) \times U(N)$ gauge  group. These are bare monopole operators of magnetic charges $(m_1,\dots,m_N;m_1,\dots,m_N)$, dressed by bifundamental chiral operators of the residual gauge theory which describes the massless degree of freedom in the monopole background, and symmetrized over the Weyl group. For instance, starting with $N=5$ and turning on a magnetic charge $(2,2,2,1,1;2,2,2,1,1)$, the residual theory factorizes into a $U(3)\times U(3)$ quiver and a $U(2)\times U(2)$ quiver. We refer to \cite{CremonesiMekareeyaZaffaroni2016} for the details of the construction. Taking into account the bare Chern-Simons levels and one-loop corrections in the monopole background, the above monopole has electric charge $(-4,-4,-4,-2,-2;4,4,4,2,2)$ under the Cartan of the gauge group. To get a gauge invariant operator of the $(U(3)\times U(3)) \times (U(2)\times U(2))$ residual gauge theory, the above bare monopole operator must be dressed by a dibaryonic operator of the $SU\times SU$ quiver of opposite electric charge. The minimal option is to dress the bare monopole with the dibaryon $(\det\nolimits_3 A^{(3)})^4 (\det\nolimits_2 A^{(2)})^2$, where we used the notation introduced in section \ref{sec:FT} and omitted global $SU(2)$ indices to avoid clutter. Averaging over the broken Weyl group, one obtains a gauge invariant operator in the full theory. 

For our purposes, the discussion can be drastically simplified if we assume the validity of a conjecture of \cite{Butti:2007jv}, which has been tested in many examples using Hilbert series and is expected from string theory considerations: this states that the generating function of mesonic (respectively dibaryonic) operators in the $N>1$ theory can be obtained from a symmetric product of the generating functions of mesonic (resp. dibaryonic) operators in the $N=1$ theory. The matching between the field-theoretic and the geometric count for $N>1$ then follows from the matching at $N=1$ reviewed in the previous subsections of the appendix, along with this conjecture. 

A few examples of gauge invariant operators in the $U(N) \times U(N)$ gauge theory (before the S operation) and in the $U(N) \times U(N) \times U(1)$ gauge theory (after the S operation) are discussed in section \ref{sec:FT} and at the end of section \ref{sec:Y12model}.

\section{The resolved cone over \texorpdfstring{$Q^{111}$}{Q111}}
\label{app:Q111}

In section \ref{sec:Q111model} we computed the holographic EFT of the moduli when the internal manifold is the resolved cone over $Q^{111}$. Here, following \cite{Benishti:2010jn}, we recall some basic features of this geometry and provide explicit expressions for the K\"ahler form and the resolved metric.

The homogeneous space $Q^{111}$ is defined by the quotient
	\be
	Q = \frac{SU(2)_1\times SU(2)_2 \times SU(2)_3 \times U(1)_\calr}{U(1)_1 \times U(1)_2 \times U(1)_3}\,.
	\ee
Given a local patch, the metric with isometry group $SU(2)_1\times SU(2)_2 \times SU(2)_3 \times U(1)_\calr$ is
	\be\label{Q111_Metric_Cone}
	\begin{aligned}
	\d s^2_{Q^{111}} &= \frac{1}{16} \left(\d \psi + \sum\limits_{i=1}^3\cos \theta_i\d \phi_i \right)^2 + \frac{1}{8}\sum\limits_{i=1}^3 \left(\d \theta_i^2 + \sin^2 \theta_i \d \phi_i^2\right)
	\end{aligned}
	\ee
where $\psi$ has period $4\pi$ and $\theta_i \in [0,\pi]$, $\phi_i \in [0,2\pi]$. Such a metric is regular and manifestly portrays $Q^{111}$ as a $U(1)$ bundle over the $\mathbb{P}^1\times\mathbb{P}^1\times\mathbb{P}^1$ base, with connection $\sum_i \cos \theta_i \d \phi_i$.

We can then construct a cone $\calc(Q^{111})$ over the $Q^{111}$ base, whose metric is
	\be
	\d s^2_{\calc(Q^{111})}= \d r^2 + r^2 \d s^2_{Q^{111}}\,.
	\ee
The singularity at the tip of the cone can be resolved by substituting the apex of the cone with two $\mathbb{P}^1$s.  The resulting resolved geometry $X$ is the bundle $\calo(-1,-1)_B\oplus\calo(-1,-1)_B$ over the base $B \simeq \mathbb{P}^1_1 \times \mathbb{P}^1_2$. This structure can be more easily described by interpreting the resolved geometry as the moduli space of a gauged linear sigma model with six chiral fields $Z_A$, $A=1,\ldots,6$ charged under the gauge group $U(1)^2$. The charges of the fields are chosen as
\be\label{Q111_GLSMb}
\begin{array}{ c | c c c c c c | c }
	& Z_1 & Z_2 & Z_3 & Z_4 & Z_5 & Z_6 & \mathrm{FI}\\\hline
	Q^{1A} & 1 & 1 & 0 & 0 & -1 & -1 & \sigma^1\\
	Q^{2A} & 0 & 0 & 1 & 1 & -1 & -1 & \sigma^2\\
\end{array}
\ee
where $\sigma^1$ and $\sigma^2$ are  FI parameters. The coordinates over the resolved cone $X$ then satisfy the D-flatness conditions
\be
\begin{aligned}
 &|Z_1|^2+ |Z_2|^2-|Z_5|^2-|Z_6|^2=\ell_{\text{\tiny P}}\sigma^1\,,
 \\
 &|Z_3|^2+ |Z_4|^2-|Z_5|^2-|Z_6|^2=\ell_{\text{\tiny P}}\sigma^2\,.
\end{aligned}
\ee
Let us assume that the parameters $\sigma^1$ and $\sigma^2$ belong to the K\"ahler cone I in \eqref{KcQ111}, that is $\sigma^1 \geq 0$ and $\sigma^2 \geq 0$. Hence, $\{Z_1, Z_2\}$ and $\{Z_3, Z_4\}$ are the homogeneous coordinates of the two $\mathbb{P}^1$ in the base. Choosing charts where $Z_1 \neq 0$ and $Z_3 \neq 0$, the base can be parametrised by the gauge invariant coordinates
\be
\label{Q111_coorda}
\lambda_1 = \frac{Z_2}{Z_1}\,,\quad \lambda_2 = \frac{Z_4}{Z_3}\,
\ee
and the fiber $\calo(-1,-1)_B\oplus\calo(-1,-1)_B$ by
\be
\label{Q111_coordb}
\zeta_1 =Z_1Z_3Z_5\,,\quad \zeta_2 = Z_1Z_3Z_6\,.
\ee
Moreover, we can introduce the globally defined radial coordinate
\be
t = (|\zeta_1|^2+|\zeta_2|^2) (1+ |\lambda_1|^2)(1+ |\lambda_2|^2)\,.
\ee
Compatibility with the $SU(2)^3 \times U(1)$ isometry of the $Q^{111}$ base requires the K\"ahler potential to be of the form
\be
\label{Q111_ka}
k_X(z,\bar z; \sigma) = U(t; \sigma) + \sigma^1 k_{(1)}+ \sigma^2 k_{(2)}\,.
\ee
The first contribution $U(t; \sigma)$, besides carrying a dependence on the $\mathbb{P}^1$--volume moduli $\sigma^a$, depends solely on the radial coordinate $t$. The last two contributions, $k_{(1)}$ and $k_{(2)}$, which are the canonical K\"ahler potentials on $\mathbb{P}^1_{(1)}$ and $\mathbb{P}^1_{(2)}$ defined in \eqref{Q111_kP}, are related to the resolution of the singularity.

From \eqref{Q111_ka} we immediately obtain the K\"ahler form 
\be
\label{Q111_J}
J = \frac{\ell_{\text{\tiny P}}}{2\pi} \left(\ii \del\delbar U(t;\sigma) + \sigma^1 j_{(1)}+ \sigma^2 j_{(2)} \right)
\ee
with
\be
j_{(a)} = \frac{\ii}{(1+|\lambda_a|^2)^2} \d \lambda_a \wedge \d \bar{\lambda}_a~.
\ee
By introducing the holomorphic forms
\be\label{eq:Q111_newbasis}
	\eta \equiv \frac{\del \rho^2}{\rho}, \quad \xi \equiv  \frac{\zeta_2 \del \zeta_1 - \zeta_1 \del \zeta_2}{|\zeta_1|^2+|\zeta_2|^2}~,
\ee
with $\rho = \sqrt{t}$, we can explicitly write the K\"ahler form \eqref{Q111_J} as
\be\label{eq:Q111_JBs}
\begin{split}
    \frac{2\pi}{\ell_{\text{\tiny P}}} J &=[\sigma^1+t\, U'(t;\sigma)]j_{(1)}+[\sigma^2+t\, U'(t;\sigma)]j_{(2)} 
    \\
    &\quad\,+\ii[t\, U''(t;\sigma) + U'(t;\sigma)] \eta \wedge \bar\eta + \ii U'(t;\sigma) \xi \wedge \bar\xi
\end{split}
\ee
From \eqref{Q111Udef} one can show that ${\mathfrak G}_0 (\sqrt{t};\sigma) = 2\pi t\, U(t;\sigma)$, allowing us to rewrite \eqref{eq:Q111_JBs} as 
\be\label{eq:Q111_JBG}
\begin{split}
    \frac{2\pi}{\ell_{\text{\tiny P}}}  J &=\left[\sigma^1+\frac{{\mathfrak G}_0 (\sqrt{t};\sigma)}{2\pi}\right]j_{(1)}+\left[\sigma^2+\frac{{\mathfrak G}_0 (\sqrt{t};\sigma)}{2\pi}\right]j_{(2)} 
    \\
    &\quad\,+\frac{\ii}{2\pi} \frac{\del {\mathfrak G}_0 (\sqrt{t};\sigma)}{\del t}  \eta \wedge \bar\eta + \ii \frac{{\mathfrak G}_0 (\sqrt{t};\sigma)}{2 \pi t} \xi \wedge \bar\xi\,.
\end{split}
\ee
To make contact with the real basis used in \cite{Benishti:2010jn}, let us perform the change of coordinates
\be
\begin{aligned}
    r^2 &\equiv \frac{1}{\pi}{\mathfrak G}_0 (\sqrt{t};\sigma)\,,
    \\ 
    \zeta_1 &\equiv \sqrt{t} e^{\frac{i}{2}(\psi+\phi_1+\phi_2+\phi_3)}\cos \frac{\theta_1}{2}\cos \frac{\theta_2}{2}\cos \frac{\theta_3}{2}\,,
    \\
    \zeta_2 &\equiv \sqrt{t} e^{\frac{i}{2}(\psi-\phi_1+\phi_2+\phi_3)}\sin \frac{\theta_1}{2}\cos \frac{\theta_2}{2}\cos \frac{\theta_3}{2}\,,
    \\
    \lambda_1 &\equiv e^{-\ii \phi_2} \tan \frac{\theta_2}{2}\,,
    \\
    \lambda_2 &\equiv e^{-\ii \phi_3} \tan \frac{\theta_3}{2}~,
\end{aligned}
\ee
where $\psi \in [0,4\pi]$, $\theta_i \in [0,\pi]$, $\phi_i \in [0,2\pi]$, so that the K\"ahler form becomes
\be\label{Q111_Real_Kahler}
	\begin{aligned}
	\frac{2\pi}{\ell_{\text{\tiny P}}}  J = &\frac{r}{2}\d r \wedge \left(\d \psi + \sum\limits_{i=1}^3 \cos \theta_i \d \phi_i\right) - \frac{r^2}{4} \d \theta_1 \wedge \sin \theta_1\d \phi_1  \\
	&-\frac{2\sigma^1+r^2}{4} \d \theta_2 \wedge \sin \theta_2\d \phi_2- \frac{2\sigma^2+r^2}{4}\d \theta_3 \wedge \sin \theta_3\d \phi_3\,.
	\end{aligned}
	\ee
and the associated metric is 
\be\label{Q111_Metric_Resolved}
	\begin{aligned}
	\d s^2_{\calc(Q^{111}), \textrm{res}} &=  \frac{\d r^2}{\kappa(r)} + \kappa(r) \frac{r^2}{16} \left(\d \psi + \sum\limits_{i=1}^3\cos \theta_i\d \phi_i \right)^2 + \frac{r^2}{8} \left(\d \theta_1^2 + \sin^2 \theta_1 \d \phi_1^2\right)\\
	&+\frac{(2a+r^2)}{8} \left(\d \theta_2^2 + \sin^2 \theta_2 \d \phi_2^2\right)+ \frac{(2b+r^2)}{8} \left(\d \theta_3^2 + \sin^2 \theta_3 \d \phi_3^2\right)\,,
	\end{aligned}
	\ee
with $\sigma^1,\sigma^2 >0$ and
	\be
	\kappa(r)= \frac{(2A_-+r^2)(2A_++r^2)}{(2\sigma^1+r^2)(2\sigma^2+r^2)}
	\ee
where
	\be
	A_\pm = \frac{1}{3}\left(2\sigma^1+2\sigma^2\pm \sqrt{4(\sigma^1)^2-10\sigma^1\sigma^2+4(\sigma^2)^2}\right)
	\ee
Note that $A_\pm$ are complex when $\frac{\sigma^2}{2}<\sigma^1<2\sigma^2$, but $\kappa$ is real. The metric \eqref{Q111_Metric_Resolved} coincides with the one presented in \cite{Benishti:2010jn} and, in the limit $\sigma^1, \sigma^2 \to 0$, the metric \eqref{Q111_Metric_Cone} is recovered.

\subsection{Counting holomorphic sections and chiral operators}\label{appsub:counting_Q111}

We will be brief in this appendix, since it follows the same logic as appendix \ref{appsub:counting_Y12}, but now applied to $Q^{111}$. We restrict our attention to $N=1$, since the generalization to $N>1$ goes as in appendix \ref{appsub:counting_Y12}.

\subsubsection{\texorpdfstring{$N=1$}{N=1} count before the \texorpdfstring{$S$}{S} operation}

We start from the geometric side and calculate the generating function that counts holomorphic sections of the line bundle $\calo_X(n^a M_{Aa} D_A)=\calo_X(n^1 D_1 + n^2 D_3)$, where $X$ is the cone over $Q^{111}$, and $D_A$ is the toric divisor associated to the field $Z_A$ in the GLSM \eqref{Q111_GLSMb}, up to linear equivalence (we have taken  $M_{A1}=\delta_{A,1}$, $M_{A2}=\delta_{A,3}$). In other words, we count the effective chiral operators of section \ref{sec:chiralop} with fixed charges ${\bf n}=(n^1, n^2)$ in the formal limit $N=1$.

Using the GLSM \eqref{Q111_GLSMb}, the generating function is given by the Molien integral 
\begin{equation}\label{Q111count_1}
\begin{split}
    g_{1,{\bf n}}(T) &= \prod_{a=1}^2 \oint \frac{dw_a}{2\pi i w_a} w_a^{-n^a}  \PE[(T_1+T_2)w_1 + (T_3+T_4)w_2 + (T_5+T_6) w_1^{-1} w_2^{-1}\\
    &= \sum_{r_1,\dots,r_6 \in \bZ_{\ge 0}} \delta_{r_1+r_2-r_5-r_6,n^1}\delta_{r_3+r_4-r_5-r_6,n^2}\prod_{A=1}^6 T_A^{r_A}~.
\end{split}
\end{equation}

The Kronecker delta function constraints are solved by setting $r_A = s_A(\mathbf{m},\mathbf{n}):= \langle \mathbf{m}, \mathbf{v}_A \rangle + M_{Aa} n^a$ as in \eqref{chargecond}, where $\mathbf{m}$ takes value in the quantized Delzant polytope $\mathcal{P}^\bZ_\mathbf{n}:=\mathcal{P}_\mathbf{n} \cap M_\bZ$, see \eqref{poly} for the continuous version, and $\mathbf{v}_A$ are the coordinates of the generators of the toric fan  \eqref{Q111_TFan}. Then 
\begin{equation}\label{operator_count_Q111_fixed_n}
    \prod_{A=1}^6 T_A^{r_A} = T_1^{n^1} T_3^{n^2} (T_1 T_3 T_5)^{m_1} (T_1 T_3 T_6)^{m_2} (T_1^{-1} T_2)^{m_3} (T_3^{-1} T_4)^{m_4}  \equiv \prod_{a=1}^2\omega_a^{n^a} \prod_{i=1}^4 t_i^{m_i} 
\end{equation}
where 
\begin{equation}
    t_i = \prod_{A} T_A^{v^i_A}~, \qquad \qquad  \omega_a = \prod_{A} T_A^{M_{Aa}}
\end{equation}
are fugacities for the toric $U(1)^4$ symmetries and the Betti $U(1)^2$ symmetries respectively. The generating function counting operators of charge $\bf n$ is then 
\begin{equation}\label{Q111count_2}
    g_{1,{\bf n}}(t,\omega)= \prod_{a=1}^2 \omega_a^{n^a} \sum_{\mathbf{m} \in \calp_{\bf n}^\bZ} \prod_{i=1}^4 t_i^{m_i}~.
\end{equation}

It is possible to calculate \eqref{Y12count_2} in closed form for $\bf n$ in each of the three chambers \eqref{KcQ111} (with $\sigma^a$ replaced by $n^a$), see appendix F of \cite{CremonesiMekareeyaZaffaroni2016} for details. Here, following the same appendix, we will content ourselves with showing that the geometric formula using the Molien integral \eqref{Q111count_1} reproduces the formula for the Hilbert series that counts monopole operators in the $U(1)_{0} \times U(1)_{0}$ flavoured Chern-Simons quiver gauge theory of section \ref{sec:Q111quiver}, with a fixed background magnetic charge $n^1$ for the `baryonic' gauge $U(1)$ and a background magnetic charge $n^2$ for the flavour $U(1)_F$. We omit the details of the manipulations,  which are analogous to those of appendix \ref{appsub:counting_Y12}, and simply state the result: 
\begin{equation}\label{Q111count_4}
\begin{split}
    g_{1,{\bf n}}(T) 
    &= \sum_{m \in \bZ} \oint \frac{dw}{2\pi i w} w^{-n^1} \left(\frac{T_3w}{T_6}\right)^{\frac{m}{2}}
    \left(\frac{T_5}{T_4w}\right)^{\frac{m-n^2}{2}}
    \left(\frac{T_3 T_6}{w}\right)^{\frac{|m|}{2}} 
    \left(\frac{T_4 T_5}{w} \right)^{\frac{|m-n^2|}{2}}  \\
    & \quad\qquad \times \PE[(T_1+T_2)w+(T_3 T_6 + T_4T_5) w^{-1}]~.
\end{split}
\end{equation}
Finally, letting
\begin{equation}\label{fug_map_Q111}
    T_1=ty~, \quad T_2=\frac{t}{y}~,\quad T_3= \frac{t^{1/2} z}{x}~, \quad T_4=t^{1/2}~,\quad T_5=t^{1/2} x ~,\quad T_6=\frac{t^{1/2}}{z}
\end{equation}
leads to the final expression  
\begin{equation}\label{Q111count_6}
\begin{split}
    g_{1,{\bf n}}(t,x,y,z) &=\sum_{m \in \bZ} z^m t^{\frac{1}{2}(|m|+|m-n^2|)} x^{\frac{1}{2}(-n^2-|m|+|m-n^2|)} \\
    &\times \oint \frac{dw}{2\pi i w} w^{-(n^1-\frac{n^2}{2}+\frac{|m|}{2}+\frac{|m-n^2|}{2}
    )} \PE[tw(y+y^{-1})+t w^{-1}(x+x^{-1})]
    ~,
\end{split}
\end{equation}
which is precisely the Hilbert series that counts dressed monopole operators in the abelian $U(1)_{0}\times U(1)_{0}$ flavoured Chern-Simons quiver of section \ref{sec:Q111quiver}, with suitable mixed Chern-Simons levels, computed as in \cite{CremonesiMekareeyaZaffaroni2016}.

The fugacity map \eqref{fug_map_Q111} identifies the homogeneous coordinates of the GLSM with gauge variant operators in the abelian quiver Chern-Simons theory as follows:
\begin{equation}\label{identification_1_Q111}
    Z_1 \leftrightarrow b_1~,\quad Z_2 \leftrightarrow b_2~, \quad Z_3 \leftrightarrow r~, \quad Z_4 \leftrightarrow \tilde r~, \quad Z_5 \leftrightarrow s~, \quad Z_6 \leftrightarrow \tilde s~,
\end{equation}
where $r$, $\tilde r$, $s$, $\tilde s$ are effective monopole operators which satisfy 
\begin{equation}\label{abttilde_from_rsrtildestilde}
    \begin{split}
        \tilde r s &= a_1~, \quad\qquad r \tilde s = a_2\\
        r s &= t~, \quad\qquad~~ \tilde r \tilde s = \tilde t~.
    \end{split}
\end{equation}
The effective monopole operators $\tilde r, s$ (respectively $\tilde s, r$) are needed to describe the pinching of the circle parametrized by the dual photon at $s=\sigma^2$ (resp. $s=0$), where the flavours $p_1, q_1$ (resp. $p_2, q_2$) become massless.

The toric coordinates $e^{-2\pi z^i}$ are instead identified with
\begin{equation}
    e^{-2\pi z^1} \leftrightarrow b_1 r s = t b_1~,\quad e^{-2\pi z^2} \leftrightarrow b_1 r \tilde s=a_2 b_1~,\quad e^{-2\pi z^3} \leftrightarrow \frac{b_2}{b_1} ~,\quad  e^{-2\pi z^4} \leftrightarrow \frac{\tilde r}{r} ~, 
\end{equation}
which are gauge invariant but not globally defined chiral operators. Indeed the last two operators are only defined  locally in a patch of the two blown-up $\bP^1$'s. The holomorphic operators of Betti charge ${\bf n}=(n^1, n^2)$ counted by formula \eqref{Q111count_2} can then be written as 
\begin{equation}\label{effective_chiral_operators_Q111}
    b_1^{m_1+m_2-m_3+n^1} b_2^{m_3} r^{m_1+m_2-m_3+n^2} \tilde r^{m_4} s^{m_1} \tilde s^{m_2}~.
\end{equation}

\subsubsection{\texorpdfstring{$N=1$}{N=1} count after the \texorpdfstring{$S$}{S} operation}

Now we apply the S operation to the two $U(1)$ symmetries that we turned on a background magnetic charge for in the previous subsection. 

On the geometry side, using \eqref{Q111count_1} the generating function becomes 
\begin{equation}\label{g_1 Q111}
\begin{split}
   g_1(T,b) &\equiv \sum_{n^1,n^2 \in \bZ} b_1^{n^1} b_2^{n^2} g_{1,\mathbf{n}}(T)\\
   &= \PE \big[b_1(T_1+T_2)+b_2(T_3+T_4)+ b_1^{-1}b_2^{-1}(T_5+T_6) \big]~,
\end{split}
\end{equation}
which is the Hilbert series of the extended geometric moduli space $\calm_{N=1}^{\mathrm{ext}}=\bC^{4+b_2(Q^{111})}=\bC^6$. This corresponds to adding to the GLSM \eqref{Q111_GLSM0} two homogeneous coordinate $X_a \in \bC^*$ ($a=1,2$)  of charge $-\delta_{ab}$ under the $b$-th $U(1)$ gauge factor, which can be used to soak up the charge of the other homogeneous coordinates, as explained after equation \eqref{projsec}. The basic gauge invariants of this extended GLSM are then 
\begin{equation}\label{extended_glsm_inv_Q111}
    X_1 Z_1~, \quad X_1 Z_2~, \quad X_2 Z_3~, \quad X_2 Z_4~, \quad X_1^{-1} X_2^{-1} Z_5~, \quad  X_1^{-1} X_2^{-1} Z_6~,
\end{equation}
in one-to-one correspondence with the $6$ homogeneous coordinates of the original GLSM.

On the quiver Chern-Simons theory side, we gauge the topological $U(1)$ of the gauged baryonic $U(1)$, as well as the $U(1)_F$ flavour symmetry. Following appendix B of \cite{CremonesiMekareeyaZaffaroni2016} and using formula \eqref{Q111count_6} we find  
\begin{equation}\label{g_1 Q111 quiver}
\begin{split}
   g_1&(t,x,y,z,b) \equiv \sum_{n^1,n^2 \in \bZ} b_1^{n^1} b_2^{n^2}  g_{1,{\bf n}}(t,x,y,z) \\
   &= \PE\big[ b_1 t y+ b_1 t y^{-1}+  
   b_2 t^{\frac{1}{2}} z x^{-1} + b_2 t^{\frac{1}{2}} + b_1^{-1}b_2^{-1} t^{\frac{1}{2}} x + b_1^{-1}b_2^{-1} t^{\frac{1}{2}} z^{-1} \big]~,
\end{split}
\end{equation}
which precisely reproduces \eqref{g_1 Q111}, with the generators written in the same order. Next, we identify these generators as dressed gauge invariant monopole operators. Let $t_{m;n^1,n^2}$ be the monopole operator of magnetic charges $(m,m)$ under the $U(1)\times U(1)$ gauge group of the original quiver, and $(n^1,n^2)$ under the newly gauged $U(1)^2$ introduced in the double S operation. Since no matter fields are charged under the gauged topological symmetry of the gauged baryonic $U(1)$, monopole operators for this $U(1)$ gauge symmetry, which have magnetic charge $n^1\neq 0$, can be factored out: for all $m, n^1, n^2$  
\begin{equation}\label{factor_Q111}
    t_{m;n^1,n^2}= t_{0;n^1,0}t_{m;0,n^2}=(\calt_1)^{n^1}t_{m;0,n^2}
\end{equation}
where $\calt_{1}\equiv t_{0;1,0}$ is invertible. The generators in the second line of \eqref{g_1 Q111 quiver}, in the order in which they are written, are then the gauge invariant operators
\begin{equation}
    \calt_1 b_1~, \quad  \calt_1 b_2~, \quad  t_{1;0,1}~, \quad  t_{0;0,1}~, \quad (\calt_1)^{-1} t_{0;0,-1} ~,\quad (\calt_1)^{-1} t_{-1;0,-1}~. \quad  
\end{equation}
Comparing this with \eqref{extended_glsm_inv_Q111}, we find the further identifications $X_1 \leftrightarrow \calt_1$ and
\begin{equation}
    \quad X_2 Z_3 \leftrightarrow t_{1;0,1} ~, \quad X_2 Z_4 \leftrightarrow t_{0;0,1} ~, \quad X_2^{-1} Z_5 \leftrightarrow t_{0;0,-1} ~, \quad X_2^{-1} Z_6 \leftrightarrow t_{-1;0,-1}~.
\end{equation}

\end{appendix}




\providecommand{\href}[2]{#2}\begingroup\raggedright\endgroup


\end{document}